\documentclass[12pt]{article}
\pdfoutput=1
\usepackage[colorlinks,linkcolor=Blue,citecolor=Blue,bookmarks,bookmarksnumbered]{hyperref}
\usepackage[scaled=0.85]{helvet}
\usepackage{amsmath,amssymb,accents,mathrsfs,pHEDR0N_arXiv/XoohmE}
\usepackage{graphicx,color}
\graphicspath{{Figures/}}
\usepackage{booktabs}
\usepackage{multirow}
\usepackage{placeins}
\usepackage{amsmath}
\usepackage{subfigure}
\usepackage{multirow}
\usepackage{lipsum}  
\newcommand\numberthis{\addtocounter{equation}{1}\tag{\theequation}}
\usepackage{float}
\usepackage{pHEDR0N_arXiv/XoohmE}
\newenvironment{lcverbatim}
 {\SaveVerbatim{cverb}}
 {\endSaveVerbatim
  \flushleft\fboxrule=0pt\fboxsep=.5em
  \colorbox{cverbbg}{%
    \makebox[\dimexpr\linewidth-2\fboxsep][l]{\BUseVerbatim{cverb}}%
  }
  \endflushleft
}
\usepackage{colortbl}
\definecolor{cverbbg}{gray}{0.93}

\usepackage[table,x11names]{xcolor}
\usepackage{fancyvrb,newverbs,xcolor}
\usepackage{lipsum}
\usepackage{colortbl}
\definecolor{cverbbg}{gray}{0.93}
\newverbcommand{\cverb}
  {\setbox\verbbox\hbox\bgroup}
  {\egroup\colorbox{cverbbg}{\box\verbbox}}

\definecolor{Green}  {rgb}{0.10,0.70,0.10} 
\definecolor{Orange} {rgb}{1.00,0.50,0.15} 
\definecolor{Red}    {rgb}{0.90,0.00,0.12} 
\definecolor{Purple} {rgb}{0.50,0.25,0.55} 
\definecolor{Turque} {rgb}{0.00,0.65,0.85} 
\definecolor{Blue}   {rgb}{0.00,0.00,1.00} 
\definecolor{Magenta}{rgb}{1.00,0.00,1.00} 
\definecolor{Gold}   {rgb}{1.00,0.75,0.25} 
\definecolor{Seaweed}{rgb}{0.01,0.24,0.09} 
\definecolor{Brown}  {rgb}{0.43,0.26,0.32} 
\definecolor{grey1}  {rgb}{0.20,0.20,0.20} 
\definecolor{grey2}  {rgb}{0.40,0.40,0.40} 
\definecolor{grey3}  {rgb}{0.60,0.60,0.60} 
\definecolor{grey4}  {rgb}{0.80,0.80,0.80} 
\definecolor{grey5}  {rgb}{0.90,0.90,0.90} 
\def\C#1#2{{\ifcase#1\or
             \color{Green}\or \color{Orange}\or \color{Red}\or
              \color{Purple}\or \color{Turque}\or \color{Blue}\or
               \color{Magenta}\or \color{Gold}\or \color{Seaweed}\or
                \color{Brown}\or\color{grey1}\or\color{grey2}\or
                 \color{grey3}\else\color{grey4}\fi#2}}

\definecolor{Slate} {rgb}{0.00,0.45,0.55}

\usepackage{color} 
\usepackage{listings} 
\usepackage{setspace} 
 
\definecolor{Code}{rgb}{0,0,0} 
\definecolor{Decorators}{rgb}{0.5,0.5,0.5} 
\definecolor{Numbers}{rgb}{0.5,0,0} 
\definecolor{MatchingBrackets}{rgb}{0.25,0.5,0.5} 
\definecolor{Keywords}{rgb}{0,0,1} 
\definecolor{self}{rgb}{0,0,0} 
\definecolor{Strings}{rgb}{0,0.63,0} 
\definecolor{Comments}{rgb}{0,0.63,1} 
\definecolor{Backquotes}{rgb}{0,0,0} 
\definecolor{Classname}{rgb}{0,0,0} 
\definecolor{FunctionName}{rgb}{0,0,0} 
\definecolor{Operators}{rgb}{0,0,0} 
\definecolor{Background}{rgb}{0.98,0.98,0.98}

\lstdefinelanguage{Python}{ 
numbers=left, 
numberstyle=\footnotesize, 
numbersep=1em, 
xleftmargin=1em, 
framextopmargin=2em, 
framexbottommargin=2em, 
showspaces=false, 
showtabs=false, 
showstringspaces=false, 
frame=l, 
tabsize=4, 
basicstyle=\ttfamily\small\setstretch{1}, 
backgroundcolor=\color{Background}, 
commentstyle=\color{Comments}\slshape, 
stringstyle=\color{Strings}, 
morecomment=[s][\color{Strings}]{"""}{"""}, 
morecomment=[s][\color{Strings}]{'''}{'''}, 
morekeywords={import,from,class,def,for,while,if,is,in,elif,else,not,and,or,print,break,continue,return,True,False,None,access,as,,del,except,exec,finally,global,import,lambda,pass,print,raise,try,assert}, 
keywordstyle={\color{Keywords}\bfseries}, 
morekeywords={[2]@invariant,pylab,numpy,np,scipy}, 
keywordstyle={[2]\color{Decorators}\slshape}, 
emph={self}, 
emphstyle={\color{self}\slshape}, 
}

\def\rI{{\rm I}}
\def\rJ{{\rm J}}
\def\rK{{\rm K}}
\def\rL{{\rm L}}
\def\rR{{\rm R}}

\def\hi{{\hat\imath}}

\def\hk{{\hat{k}}}

\def\fracm#1#2{\hbox{\large{${\frac{{#1}}{{#2}}}$}}}

\def\vCent#1{\vcenter{\hbox{\hss#1\hss}}}
\def\be{\begin{equation}}
\def\ee{\end{equation}}
\newcommand{\bea}{\begin{eqnarray}}
\newcommand{\eea}{\end{eqnarray}}
\newcommand{\ena}{\end{eqnarray}}

\def\pp{{\mathchoice
          {
              \kern 1pt%
              \raise 1pt
              \vbox{\hrule width5pt height0.4pt depth0pt
                    \kern -2pt
                    \hbox{\kern 2.3pt
                          \vrule width0.4pt height6pt depth0pt
                          }
                    \kern -2pt
                    \hrule width5pt height0.4pt depth0pt}%
                    \kern 1pt
           }
            {
              \kern 1pt%
              \raise 1pt
              \vbox{\hrule width4.3pt height0.4pt depth0pt
                    \kern -1.8pt
                    \hbox{\kern 1.95pt
                          \vrule width0.4pt height5.4pt depth0pt
                          }
                    \kern -1.8pt
                    \hrule width4.3pt height0.4pt depth0pt}%
                    \kern 1pt
            }
            {
              \kern 0.5pt%
              \raise 1pt
              \vbox{\hrule width4.0pt height0.3pt depth0pt
                    \kern -1.9pt  
                    \hbox{\kern 1.85pt
                          \vrule width0.3pt height5.7pt depth0pt
                          }
                    \kern -1.9pt
                    \hrule width4.0pt height0.3pt depth0pt}%
                    \kern 0.5pt
            }
            {
              \kern 0.5pt%
              \raise 1pt
              \vbox{\hrule width3.6pt height0.3pt depth0pt
                    \kern -1.5pt
                    \hbox{\kern 1.65pt
                          \vrule width0.3pt height4.5pt depth0pt
                          }
                    \kern -1.5pt
                    \hrule width3.6pt height0.3pt depth0pt}%
                    \kern 0.5pt
            }
        }}

\def\mm{{\mathchoice
                       {
                             \kern 1pt
               \raise 1pt    \vbox{\hrule width5pt height0.4pt depth0pt
                                  \kern 2pt
                                  \hrule width5pt height0.4pt depth0pt}
                             \kern 1pt}
                       {
                            \kern 1pt
               \raise 1pt \vbox{\hrule width4.3pt height0.4pt depth0pt
                                  \kern 1.8pt
                                  \hrule width4.3pt height0.4pt depth0pt}
                             \kern 1pt}
                       {
                            \kern 0.5pt
               \raise 1pt
                            \vbox{\hrule width4.0pt height0.3pt depth0pt
                                  \kern 1.9pt
                                  \hrule width4.0pt height0.3pt depth0pt}
                            \kern 1pt}
                       {
                           \kern 0.5pt
             \raise 1pt  \vbox{\hrule width3.6pt height0.3pt depth0pt
                                  \kern 1.5pt
                                  \hrule width3.6pt height0.3pt depth0pt}
                           \kern 0.5pt}
                       }}

\def\ad{{\kern0.5pt
                   \alpha \kern-5.05pt \raise5.8pt\hbox{$\textstyle.$}\kern
0.5pt}}

\def\bd{{\kern0.5pt
                   \beta \kern-5.05pt \raise5.8pt\hbox{$\textstyle.$}\kern
0.5pt}}

\def\qd{{\kern0.5pt
                   q \kern-5.05pt \raise5.8pt\hbox{$\textstyle.$}\kern
0.5pt}}
\def\Dot#1{{\kern0.5pt
     {#1} \kern-5.05pt \raise5.8pt\hbox{$\textstyle.$}\kern
0.5pt}}

\catcode`@=11
\def\un#1{\relax\ifmmode\@@underline#1\else
        $\@@underline{\hbox{#1}}$\relax\fi}
\catcode`@=12

\def\a{\alpha}
\def\b{\beta}
\def\c{\chi}
\def\d{\delta}

\def\g{\gamma}

\def\i{\iota}
\def\j{\psi}
\def\k{\kappa}
\def\l{\lambda}

\def\o{\omega}

\def\r{\rho}

\def\t{\tau}

\def\L{\Lambda}
\def\O{\Omega}
\def\P{\Pi}

\def\S{\Sigma}

\def\ca{{\cal A}}

\def\cf{{\cal F}}

\def\dslash{\not{\hbox{\kern-2pt $\partial$}}}
\def\Dslash{\not{\hbox{\kern-4pt $D$}}}
\def\pslash{\not{\hbox{\kern-2.3pt $p$}}}
 \newtoks\slashfraction
 \slashfraction={.13}
 \def\slash#1{\setbox0\hbox{$ #1 $}
 \setbox0\hbox to \the\slashfraction\wd0{\hss \box0}/\box0 }
 
\def\kcr{{\hbox{\ro \char'170}}}                
\def\ktl{{\hbox{\ro \char'170}}}        
\def\ktr{{\hbox{\ro \char'170}}}        
\def\kbl{{\hbox{\ro \char'170}}}        
\def\kbr{{\hbox{\ro \char'170}}}        

\def\plpl{\raise-2pt\hbox{$\raise3pt\hbox{$_+$}\hskip-6.67pt\raise0.0pt
\hbox{$^+$}\hskip 0.01pt$}}
\def\mimi{\raise-2pt\hbox{$\raise3pt\hbox{$_-$}\hskip-6.67pt\raise0.0pt
\hbox{$^-$}\hskip 0.01pt$}} 

\def\bo{{\raise.15ex\hbox{\large$\Box$}}}               
\def\TH{{\raise.2ex\hbox{$\displaystyle \bigodot$}\mskip-4.7mu \llap H \;}}
\def\face{{\raise.2ex\hbox{$\displaystyle \bigodot$}\mskip-2.2mu \llap {$\ddot
        \smile$}}}                                      

\def\dt#1{\on{\hbox{\bf .}}{#1}}                
\def\Dot#1{\dt{#1}}

\def\Tilde#1{\widetilde{#1}}                    
\def\leftrightarrowfill{$\mathsurround=0pt \mathord\leftarrow \mkern-6mu
        \cleaders\hbox{$\mkern-2mu \mathord- \mkern-2mu$}\hfill
        \mkern-6mu \mathord\rightarrow$}
\def\dvec#1{\vbox{\ialign{##\crcr
        \leftrightarrowfill\crcr\noalign{\kern-1pt\nointerlineskip}
        $\hfil\displaystyle{#1}\hfil$\crcr}}}           
\def\dt#1{{\buildrel {\hbox{\LARGE .}} \over {#1}}}     

\def\fracm#1#2{\hbox{\large{${\frac{{#1}}{{#2}}}$}}}
\def\sfrac#1#2{{\vphantom1\smash{\lower.5ex\hbox{\small$#1$}}\over
        \vphantom1\smash{\raise.4ex\hbox{\small$#2$}}}} 
\def\bfrac#1#2{{\vphantom1\smash{\lower.5ex\hbox{$#1$}}\over
        \vphantom1\smash{\raise.3ex\hbox{$#2$}}}}       
\def\afrac#1#2{{\vphantom1\smash{\lower.5ex\hbox{$#1$}}\over#2}}    

\let\bm\relax
\newcommand{\bm}[1]{{\boldsymbol{#1}}}

\def\ad{{\dot{\alpha}}}
\def\bd{{\dot{\beta}}}

 \font\rOpe=cmsy10                        
 \def\ktl{{\hbox{\rOpe\char'170}}}        
 \def\kbl{{\hbox{\rOpe\char'170}}}        
 \def\kcr{{\reflectbox{\rOpe\char'170}}}        
 \def\ktr{{\reflectbox{\rOpe\char'170}}}        
 \def\kbr{{\reflectbox{\rOpe\char'170}}}        
 \def\Border{\vbox{\hsize0pt
        \setlength{\unitlength}{1mm}
        \newcount\xco
        \newcount\yco
        \xco=-21
        \yco=12
        \begin{picture}(0,0)(-7.5,0)
        \put(\xco,\yco){$\ktl$}
        \advance\yco by-1
        {\loop
        \put(\xco,\yco){$\kcr$}
        \advance\yco by-2
        \ifnum\yco>-240
        \repeat
        \put(\xco,\yco){$\kbl$}}
        \xco=170
        \yco=12
        \put(\xco,\yco){$\ktr$}
        \advance\yco by-1
        {\loop
        \put(\xco,\yco){$\kcr$}
        \advance\yco by-2
        \ifnum\yco>-240
        \repeat
        \put(\xco,\yco){$\kbr$}}
        \put(-19.5,13){\scalebox{.6065}{%
         University of Maryland Center for String and Particle  Theory \&\ Physics Department%
        |University of Maryland Center for String and Particle  Theory \&\ Physics Department}}
        \put(-19.5,-241.5){\scalebox{.5835}{%
         ****University of Maryland * Center for String and
         Particle  Theory* Physics Department****University of Maryland *Center
        for String and Particle  Theory* Physics Department}}
        \end{picture}
        \par\vskip-8mm}}
\definecolor{UMred}{rgb}{.9,.05,.2}
\definecolor{HUblue}{rgb}{.0,.3,.7}

\definecolor{Red}    {rgb}{0.90,0.00,0.12} 
\definecolor{Blue}   {rgb}{0.00,0.00,1.00} 
\definecolor{Green}  {rgb}{0.10,0.70,0.10} 
\definecolor{Turque} {rgb}{0.00,0.65,0.85} 
\definecolor{Orange} {rgb}{1.00,0.50,0.15} 
\definecolor{Magenta}{rgb}{1.00,0.00,1.00} 
\definecolor{Gold}   {rgb}{1.00,0.75,0.25} 
\definecolor{Seaweed}{rgb}{0.01,0.24,0.09} 
\definecolor{Purple} {rgb}{0.50,0.25,0.55} 
\definecolor{Brown}  {rgb}{0.43,0.26,0.32} 
\definecolor{grey1}  {rgb}{0.20,0.20,0.20} 
\definecolor{grey2}  {rgb}{0.40,0.40,0.40} 
\definecolor{grey3}  {rgb}{0.60,0.60,0.60} 
\definecolor{grey4}  {rgb}{0.80,0.80,0.80} 
\definecolor{grey5}  {rgb}{0.90,0.90,0.90} 
\def\C#1#2{{\ifcase#1\or
             \color{Red}\or \color{Green}\or \color{Blue}\or\
              \color{Turque}\or \color{Orange}\or \color{Magenta}\or 
               \color{Gold}\or \color{Seaweed}\or \color{Purple}\or
                \color{Brown}\or\color{grey1}\or\color{grey2}\or
                 \color{grey3}\else\color{grey4}\fi#2}}

\definecolor{Slate} {rgb}{0.00,0.45,0.55}

\newdimen\parshift\parshift=\parindent
\catcode`@=11
 \long\def\@footnotetext#1{\insert\footins{\reset@font\footnotesize
           \interlinepenalty\interfootnotelinepenalty\splittopskip%
            \footnotesep\splitmaxdepth\dp\strutbox\floatingpenalty\@MM%
             \hsize\columnwidth\addtolength{\hsize}{-2\parindent}
              \@parboxrestore\protected@edef\@currentlabel%
              {\csname p@footnote\endcsname\@thefnmark}%
                \color@begingroup%
                 \@makefntext{\rule\z@\footnotesep\ignorespaces#1%
                  \@finalstrut\strutbox}%
                \color@endgroup}}
 \long\def\@makefntext#1{\hglue\parshift%
           \vbox{\noindent\baselineskip=11pt plus.5pt minus.5pt\hb@xt@0em{\hss\@makefnmark\kern1pt}#1}}
\catcode`@=12

\newskip\humongous \humongous=0pt plus 1000pt minus 1000pt
\def\caja{\mathsurround=0pt}
\def\eqalign#1{\,\vcenter{\openup2\jot \caja
        \ialign{\strut \hfil$\displaystyle{##}$&$
        \displaystyle{{}##}$\hfil\crcr#1\crcr}}\,}
\newif\ifdtup

\makeatletter
\def\section{\@startsection{section}{1}{\z@}
        {3ex plus-1ex minus-.2ex}{1pt plus1pt}{\large\sf\bfseries\boldmath}}
\def\subsection{\@startsection{subsection}{2}{\z@}
         {1.5ex plus-1ex minus-.2ex}{0.01pt plus1pt}{\sf\slshape}}
\def\subsubsection{\@startsection{subsubsection}{3}{\z@}
          {1.5ex plus-1ex minus-.2ex}{0.01pt plus0.2pt}{\sf\boldmath}}
\def\paragraph{\@startsection{paragraph}{4}{\z@}
           {.75ex \@plus.5ex \@minus.2ex}{-2mm}{\sf\bfseries\boldmath}}
\makeatother

 \allowdisplaybreaks
 \seceq

\usepackage{lipsum}
\usepackage{listings}
\definecolor{MyDarkGreen}{rgb}{0.0,0.4,0.0} 
\usepackage[enableskew,vcentermath]{youngtab}

\definecolor{Hey}{rgb}{.9,.05,.4}
\definecolor{orange}{rgb}{1,.5,0}
\definecolor{plum}{rgb}{.4,0,.6}
\definecolor{R}{rgb}{1,0,0}
\definecolor{G}{rgb}{0.1,0.7,0}
\definecolor{B}{rgb}{0,0,1}
\begin{document}

\thispagestyle{empty}
\noindent{\small
\hfill{  \\ 
$~~~~~~~~~~~~~~~~~~~~~~~~~~~~~~~~~~~~~~~~~~~~~~~~~~~~~~~~~~~~~~~~~$
$~~~~~~~~~~~~~~~~~~~~~~~~~~~~~~~~~~~~~~~~~~~~~~~~~~~~~~~~~~~~~~~~~$
{}
}}
\vspace*{0mm}
\begin{center}
{\large \bf
$ ~~$ A  Pr\' ecis: Minimal Four Color 
Holoraumy and  
$~$\\[2pt]
Wolfram's ``New Kind of Science" Paradigm}   \vskip0.3in
{\large {
$~~~~~~~~~~~~~$
S.\ James Gates, Jr.\footnote{gatess@umd.edu}$^{,a,b}$, and 
Youngik (Tom) Lee\footnote{youngik${}_-$lee@almuni.brown.edu}
$~~~~~~$
\newline
}}
\\*[8mm]
\emph{
\centering
$^{a} $Department of Physics, 1117 Toll Hall,
\\[1pt]
University of Maryland, College Park, MD 20742-4111, USA,
\\[4pt] and \\[4pt]
$^{b}$School of Public Policy, 2023
Thurgood Marshall Hall,
\\[1pt]
University of Maryland, College Park, MD 20742-4111, USA,
\\[12pt]
} $~$
 \\*[45mm]
{ ABSTRACT}\\[4mm]
\parbox{142mm}{\parindent=2pc\indent\baselineskip=14pt plus1pt
Adinkras are graphical representations of the gauge invariant field components in supersymmetric theories and their orbits under the action of supersymmetry (SUSY) generators in the
context of supermultiplets. A discussion is given that provides
a thorough review of the concepts of holoraumy, permutahedra,
and gadgets.  One consequence of these concepts, the additional concept of hopper operators, is discussed.  These play a particularly
important role that ignites the processes needed for the study of adinkras related to minimal 4D, $\cal N$ = 1 supermultiplets (chiral, vector, tensor and complex linear supermultiplets) through the prism of very simple cellular automata following Wolfram's `New Kind of Science' paradigm.
}

 \end{center}
\vfill
\noindent PACS: 11.30.Pb, 12.60.Jv\\
Keywords: supersymmetry, Adinkra, supermultiplet 
\vfill
\clearpage
%

%
\tableofcontents
\newpage
\section{Introduction}

In 2002, Stephen Wolfram published a book entitled, {\it {A New Kind of
Science}} \cite{18}, a treatise containing examples and formal theory of the application of computational algorithms, including cellular automata.  From the viewpoint of a
journeyman theoretical physicist, this work can be viewed as the introduction of
a paradigm for mathematical models of complex systems.  We will thus refer to the
``Wolfram Paradigm'' in our subsequent discussion.

The adinkra construction \cite{Adnk1} is a graphical presentation paradigm of supersymmetric representation theory. Recently, there has
appeared very convincing and strong arguments that show  linkages to algebraic geometry
\cite{Reager}. This builds upon previously discovered and surprising connections to
classical error-correction\cite{codes1,codes2,codes3}, Riemann surfaces\cite{GEO1,GEO2,MON},
critical groups\cite{IIGn}, and Clifford Algebras\cite{CLIFF}.
There is another such linkage to the permutahedron concept\cite{pHR0n1,pHR0n2,pHR0n3,pHR0n4} as a useful component \cite{Adnk300}
in analyzing the structure of adinkra graphs.
This all highlights the power of the 
adinkra construct as a unifying tool for exploring the connections between distinct areas of physics and 
mathematics.

As advocated by Wolfram, his paradigm can be viewed as a discrete version of the continuum limit of physics, while the adinkra construction is a geometric representation of the supersymmetry algebra. In fact, the adinkra graphs arise from another limit of spacetime. These images emerge in the `Carroll Limit,' (a Wigner-In\" on\"u contraction of the spacetime Lorentz group) proposed by mathematicians L\'evy-Leblond and Gupta in the 1960's \cite{CARR1,CARR2}.  Therefore, adinkras provides a natural framework for studying supersymmetric theories in a discrete setting, which is complementary to the approach of the Wolfram Paradigm\footnote{In 2010, SJG and SW had a short discussion about
this possibility at a gathering in Concord, MA.  See also the video accessed via the link
https://www.youtube.com/watch?v=go9XxI1Igsk on-line}.

In the analysis of dynamical systems, the notion of an ``attractor'' \cite{X4} is a well-known concept. Given a 
wide variety of initial conditions describing a state space, the evolution of the these tends to ``accummulate'' 
at some special points in state space.  In systems containing a sufficiency of conserved quantities, a 
Hamiltonian serves the role of providing the operator that evolves the system.  The Wolfram Paradigm raises 
the issue of whether adinkras possess recursive behaviors that are reminiscent of these phenomena.  To answer 
this question is one primary purpose of this work.

The outline of our paper goes as follows.

In chapter two, there is a discussion of the adinkra related concepts of ``holoraumy" and ``gadgets"
\cite{KIAS,K2,K3,K4,K5,K4g,K5g}.  At
first this may seem as a non sequitur.  However, it will be shown the existence of a holoraumy matrix 
is a sufficient first step to implement the recursive process that begins the implementation and study of the Wolfram Paradigm in the context of adinkras.  

The first appearance in the literature of holoraumy is taken as the starting point.  The existence of two such tensors, the bosonic and fermionic holoraumy types, is reviewed. Operators called ``hoppers" and ``jumpers" are shown to 
exist as consequences of the properties of the bosonic holoraumy tensor\footnote{The fermionic holoraumy tensor leads to similar left-operating matrices on the R-matrices, 
 but as these $~~~~$ are not needed in our discussion, we do not dwell on them.}.  These operators will act from 
 the left on the adinkra L-matrices that describe the three minimal 4D, $\cal N$ = 1 supermultiplets
 are introduced.  
 
Next the concept of the permutahedron is reviewed for the convenience of the reader and used to define the `permutahedronic projection' that maps sets of four 
L-matrices onto permutahedron vertices. The concept of a ``weight" for permutation operators is defined and
 a new basis for the L-matrices called the ``weight-based representation" is introduced. This basis provides a
 uniform way to compare all L-matrix descriptions and provides a tool that can be used as a `definitional'
convention.  The hopper operators are explicitly
 calculated and a hemispherical projection operator is introduced based on one such operator.

The final section of this chapter reviews the notion of
gadgets \cite{K2,K3,K4,K5,K4g,K5g,6}.  The first gadget 
provides a map from the holoraumy matrices into the
real numbers and thus defines a metric on the space of
holoraumy matrices.  The second gadget defines a `wedge'
product for this space.
 
 In a short chapter three, there is presented a simplified description of the Wolfram Paradigm that is needed for the
 subsequent calculations.  The concept of ``sequential substitution" or ``rewriting rules" is introduced
 and graphical illustrations of it are presented.
 
The fourth chapter takes on the tasks of the implementation of the recursive processes indicated to study the interaction of 
the Wolfram Paradigm with the adinkra graphs for the CM, VM, and TM supermultiplets.  The analysis includes studies of both the L-matrices and the associated R-matrices for each system of 
 adinkras.  This chapter has three subsections, one dedicated to each of the representations.  For each of these multiplets, it is found that the process `converges' in three steps to 
 a state that is reminiscent of the presence of an attractor in
 the recursion.
 
In chapter five, the complex linear system is studied recursively in the same manner as in the preceding 
chapter.  However, a different outcome is found.  In particular, and in a stark contrast to the minimal cases, 
we do not end with a convergence at three steps.

In chapters six and seven, we will present the technical computer coding algorithms based on chapters 
two to five that graphically construct the adinkra.

\newpage\section{From Holoraumy to Hopper Operators}
In the work of \cite{KIAS}, the following equation was introduced
\begin{equation}
 {
\eqalign{
\left[ \,
  (\,{\rm R}^{(\cal R)}_\rI)_\hi{}^j\>(\, {\rm L}^{(\cal R)}_\rJ)_j{}^\hk - (\,{\rm R}^{(\cal 
 R)}_\rJ)_\hi{}^j\>(\,{\rm L}^{(\cal R)}_\rI)_j{}^\hk \, \right]
  &= 2\Big[ \,\ell^{({\cal R})1}_{\rI\rJ}\, (\g^2 \g^3){}_{\hi}{}^\hk
   +  \ell^{({\cal R})2}_{\rI\rJ}\, (\g^3 \g^1){}_{\hi}{}^\hk
   + \ell^{({\cal R})3}_{\rI\rJ}\, (\g^1 \g^2){}_{\hi}{}^\hk   \, \Big] 
   \cr
  &~~
  ~+~  2\Big[ \,  i\,  {{\Tilde \ell}^{(\cal R)}}_{
 \rI\rJ}{}^{1}\, (\g^5){}_{\hi}{}^\hk  \,+\,
  i\, {{\Tilde \ell}^{(\cal R)}}_{\rI\rJ}{}^{2}\, 
 (\g^0 \g^5){}_{\hi}{}^\hk  \,-\,  {{\Tilde \ell}^{(\cal R)}}_{\rI\rJ}{}^{3}\, 
 (\g^0){}_{\hi}{}^\hk 
  \, \Big]  ~~,
} } \label{GarDNAdnk}
\end{equation}
that relates the adjacency matrices (L-matrices \& 
R-matrices) in minimal 4-color adinkras to a set of
$\gamma$-matrices in four dimensional spacetime.
The label $\cal R$ denotes the three distinct supermultiplets, 
the chiral, vector, and tensor ones respectively. We use labels 
CM, VM, and TM to denote each one individually.
For the CM, VM, and TM representations, the quantities
$\ell$, and ${\Tilde{\ell}}$ are non-vanishing
constants that take on values of +1, 0, or -1.  The result shown in Eq.\ (\ref{GarDNAdnk}) pointed toward the beginning of discussions of the 
`holoraumy' and the `gadget' concepts \cite{K2,K3,K4,K5,K4g,K5g,6} in the literature. 

In particular, dividing both sides of Eq.\ (\ref{GarDNAdnk}) by a factor of two, yields on its LHS,
the `fermionic holoraumy tensor,' denoted by the symbol $ (\,{\cal F}^{(\cal R)}_{\rI \, \rJ})_\hi{}^\hk$. 
The works cited at the end of the last paragraph also noted there is a corresponding `bosonic holoraumy tensor,' denoted by the symbol $ (\,{\cal B}^{(\cal R)}_{\rI \, \rJ})_i{}^k$ and defined by the equation
\be
 (\,{\cal B}^{(\cal R)}_{\rI \, \rJ})_i{}^j ~=~ \fracm 12   \, \left[ \,
   (\,{\rm L}^{(\cal R)}_\rI)_\i{}^\hk\>(\, {\rm R}^{(\cal R)}_\rJ)_\hk{}^j - 
  (\,{\rm L}^{(\cal R)}_\rJ)_\i{}^\hk\>(\,{\rm R}^{(\cal R)}_\rI)_\hk{}^j \, \right]  ~=~ \fracm 12   \, \left[ \,
  (\,{\rm L}^{(\cal R)}_{[\, \rI})_\i{}^\hk\>(\, {\rm R}^{(\cal R)}_{\rJ ]})_\hk{}^j   \, \right]  ~~~.
  \label{e:BR1}
\ee
There does not exist any obvious relationship for the  $ (\,{\cal B}^{(\cal R)}_{\rI \, \rJ})_i{}^k$
that is similar to the form of Eq.\ (\ref{GarDNAdnk}). In the work of \cite{K2}, it was noted that as
matrices $ (\,{\cal F}^{(\cal R)}_{\rI \, \rJ})$ and $ (\,{\cal B}^{(\cal R)}_{\rI \, \rJ})$ generate two 
distinct representations of Lie groups isomorphic to subgroups of Spin($\cal N$).  Furthermore and obviously
both of these matrices vanish when the relationship I = J is satisfied.

In the inaugural lecture of the `BTPC Idea' seminar 
series\footnote{https://www.youtube.com/watch?v=TGnG2JQXn1k},
the concept of `hopper operators' for acting on the vertices
of the permutahedron as it relates to adinkras was presented.
In 2021, it was brought to the attention of one of the
authors (SJG)\footnote{Brown University M.S. 
degree student, Lily Anderson, completed a draft research paper including the concept of `jumpers.'}, the concept of  `jumpers' could be introduced.   In a 
much later work, \cite{Adnk300}, the concept of `hoppers' was formally introduced into
 the literature.
In the remainder of this section, we will explain the relationships between holoraumy,
jumpers, and hoppers. 

We matrix multiply $ (\,{\cal B}^{(\cal R)}_{\rI \, \rJ})$ by $ (\,{\rm L}^{(\cal R)}_\rK)$ where I = K 
for fixed values and I $\ne$ J to find
 \be{ \eqalign{
 (\,{\cal B}^{(\cal R)}_{\rI \, \rJ})_i{}^j \, (\,{\rm L}^{(\cal R)}_\rK){}_j{}^\hi 
 ~=&~ \fracm 12   \, \left[ \, (\,{\rm L}^{(\cal R)}_\rI)_\i{}^\hk\>(\, 
 {\rm R}^{(\cal R)}_\rJ)_\hk{}^j -   (\,{\rm L}^{(\cal R)}_\rJ)_\i{}^\hk\>(\,{\rm R}^{(\cal R)}_\rI)_\hk{}^j \, \right] \, (\,{\rm L}^{(\cal R)}_\rK){}_j{}^\hi   \cr
~=&~ -\,  \left[ \, 
(\,{\rm L}^{(\cal R)}_\rJ)_\i{}^\hk\>(\,{\rm R}^{(\cal R)}_\rI)_\hk{}^j \, \right] \, (\,{\rm L}^{(\cal R)}_\rK){}_j{}^\hi 
 }}
  \label{e:BR2}
\ee
by use of the first of the fundamental definitions and then the second of the `Garden Algebra' L-matrices 
and R-matrices seen in
Eq.\ (\ref{GA0}) 
 \be{ \eqalign{
 & (\,{\rm L}^{(\cal R)}_\rI)_\i{}^\hk\>(\, {\rm R}^{(\cal R)}_\rJ)_\hk{}^j + 
  (\,{\rm L}^{(\cal R)}_\rJ)_\i{}^\hk\>(\,{\rm R}^{(\cal R)}_\rI)_\hk{}^j  ~=~  2 \, \d_{\rI \, \rJ } 
  \, \d{}_\i{}^j  ~~~,  \cr
 &(\,{\rm R}^{(\cal R)}_\rI)_\hi{}^j\>(\, {\rm L}^{(\cal R)}_\rJ)_j{}^\hk + (\,{\rm R}^{(\cal R)}_\rJ)_\hi{}^j\>(\,{\rm L}^{(\cal R)}_\rI)_j{}^\hk ~=~   2 \, \d_{\rI \, \rJ } \, \d{}_\hi{}^\hk ~~~.  \cr
  }}
  \label{GA0}
\ee

Thus, the result Eq. (\ref{e:BR2}) can be rewritten in the form
 \be{ 
 (\,{\rm L}^{(\cal R)}_\rJ
 )_\i{}^\hi\> ~=~ -\,  (\,{\cal B}^{(\cal R)}_{\rK \, \rJ})_i{}^j \, (\,{\rm L}^{(\cal R)}_\rK){}_j{}^\hi
 ~=~ \,  (\,{\cal B}^{(\cal R)}_{\rJ \, \rK})_i{}^j \, (\,{\rm L}^{(\cal R)}_\rK){}_j{}^\hi~~~,
 }
  \label{e:BR4}
\ee
when J and K are both set to fixed values.
This equation is the key to understanding `hopper' and `jumper' operators. 
 Jumpers are the simpler of 
the two sets of operators to understand. \cite{36} For fixed values of the K index in  Eq.\ (\ref{e:BR4}), implies the
${\cal B}^{({\cal R)}}$-matrix is a jumper operator that can be used to `jump' any particular L-matrix with 
the fixed-index K to any L-matrix associated with the fixed-index J in the set. 

To define hopper operators requires first the introduction of a `weight,' i.e. 
a real number to be assigned to each signed permutation
matrix L.  This means if we take the absolute values of the L-matrices,
we are left purely with permutation matrices.  However, permutations
have an intrinsic isomorphism to the real numbers.  This is seen via
the following argument.

In the work of \cite{3}
, the results of Table \ref{t:0brane0}

\begin{table}[H]
\setlength\extrarowheight{2pt}
$$
    \begin{array}{|c|c|c|c|c|c|c|} 
   \hline
 & {\bm {\rL}}_1 & {\bm {\rL}}_2 & {\bm {\rL}}_3 & {\bm {\rL}}_4  \\  \hline
{\rm VM} & \vev{2\bar41\bar3} & \vev{13\bar2\bar4} & \vev{4231} & \vev{3\bar1\bar42}
\\ \hline
{\rm CM} & \vev{1\bar42\bar3} & \vev{23\bar1\bar4} &  \vev{3\bar2\bar41} & \vev{4132} 
\\ \hline
{\rm TM} & \vev{1\bar3\bar4\bar2} & \vev{24\bar31} & \vev{312\bar4} & \vev{4\bar213}
\\ \hline
    \end{array}
$$
    \caption{Signed Permutation Element Decomposition of L-matrices}
    \label{t:0brane0}
\end{table}  \noindent
appeared using the concise definition of each L-matrix as a product of a Boolean
factor\footnote{See \cite{permutadnk} for the definition of the Boolean Factors.} times
a permutation written in `bracket notation.'

Next taking the absolute values of the entries in each L-matrix acts to `strip away'
all of the Boolean Factors yielding the ones shown 
in Table. \ref{t:0braneX}.

\begin{table}[H]
$$
\begin{array}{|c|c|c|c|c|} 
\hline
 & |{\bm {\rL}}_1 | & |{\bm {\rL}}_2| & |{\bm {\rL}}_3 |& | {\bm {\rL}}_4 |  \\  \hline
{\rm VM} &  \vev{2413} &  \vev{1324} &   \vev{4231} &  \vev{3142} 
\\ \hline
{\rm CM} &  \vev{1423} &   \vev{2314} &  \vev{3241} &  \vev{4132} 
\\ \hline
{\rm TM} &  
\vev{1342} &  \vev{2431} & \vev{3124} &  \vev{4213}
\\ \hline
    \end{array}
$$
\caption{Absolute Value Element Decomposition of L-matrices}
\label{t:0braneX}
\end{table} \noindent

It is very efficient to use the three color permutahedron to show the relations of the supermultiplets to sets of 
permutation matrices.  Thus, here we first give a brief introduction to the 3-color permutahedron\footnote{This
image is adapted from one at the Wikipedia  URL- https://en.wikipedia.org/wiki/Permutohedron} shown in Fig. 1.  
To the left side of the image is the actual permutahedron.  Each of its nodes represent one of the 4! permutations. 
On each of the nodes is written an address in the form of a four digit number.  This is where the isomorphism
enters.
$$
\vCent
{\setlength{\unitlength}{1mm}
\begin{picture}(-20,0)
\put(-46,-62){\includegraphics[width=3.7in]{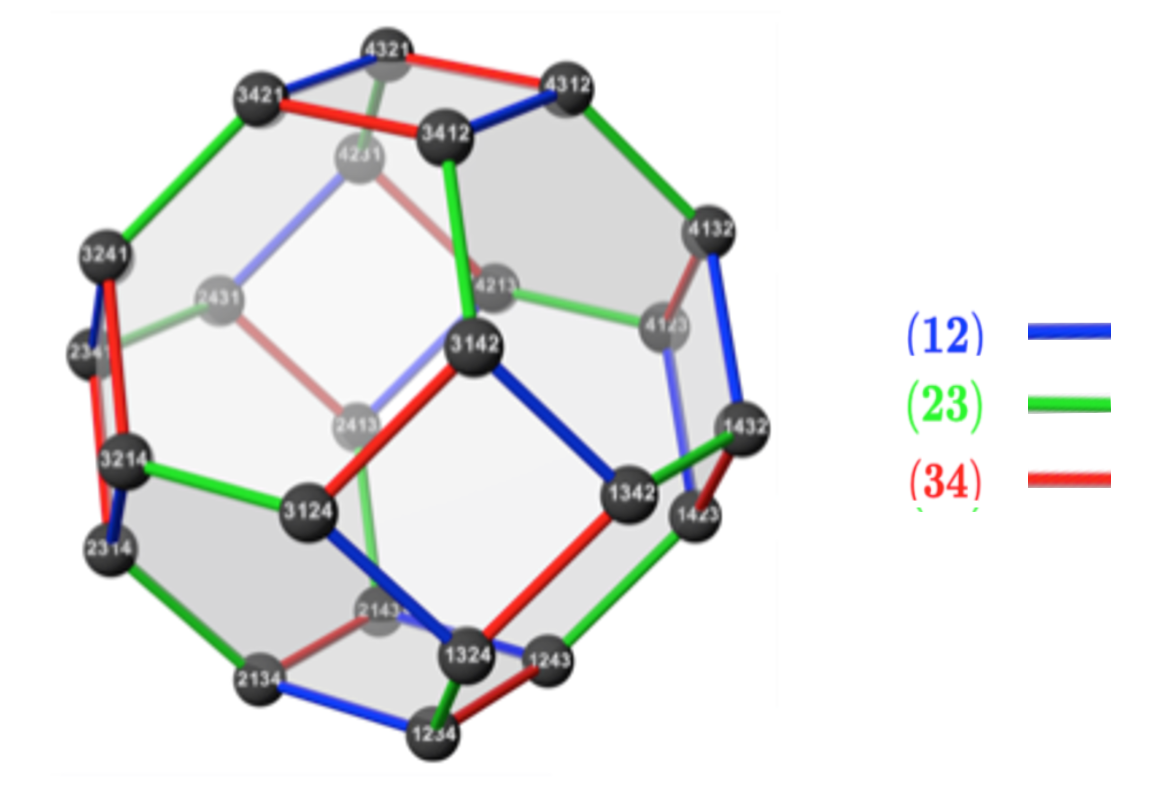}}
\put(-46,-67){$\bm {\rm {Figure ~1:}}$}
\put(-26,-67){ {\rm {The Three Color Permutahedron}}}
\end{picture}}
\nonumber
$$ \noindent
\vskip2.4in
\noindent
Using the `bracket notation' $\langle n_1 \,  n_2 \,  n_3 \,  n_4 \, ... \rangle $, any permutation matrix $\cal P$ is described by a set of 
digits and the address of each permutation ${\cal P}$ can be 
defined by the digit $ n_1 \,  n_2 \,  n_3 \,  n_4, \, ...   $.  Hence each address on 
the nodes specifies one of the 24 permutation elements.
 
On the right, three 2-cycle permutations, described by adjacent digits, are specified and each is `color coded.'  The permutahedron 
is a visual
representation of a multiplication table, using Bruhat ordering \cite{Bruhat}, showing the 2-cycles acting from the left on all 24 permutation elements.  
Via the use of a computer code in the work of \cite{permutadnk} all of the solutions to the Eq.(\ref{e:BR2}) were investigated.  This resulted in the discovery of three
additional classes of solutions given the respective. names of $\rm{VM}_1$, $\rm{VM}_2$, and $\rm{VM}_3$.  These are shown in Table \ref{t:0braneXzX}.

\begin{table}[H]
$$
\begin{array}{|c|c|c|c|c|} 
\hline
 & |{\bm {\rL}}_1 | & |{\bm {\rL}}_2| & |{\bm {\rL}}_3 |& | {\bm {\rL}}_4 |  \\  \hline
{\rm VM}{}_1 &  \vev{1432} &  \vev{2341} &   \vev{3214} &  \vev{4123} 
\\ \hline
{\rm VM}{}_2 &  \vev{1243} &   \vev{2134} &  \vev{3412} &  \vev{4312} 
\\ \hline
{\rm VM}{}_3 &  
\vev{1234} &  \vev{2143} & \vev{3412} &  \vev{4321}
\\ \hline
    \end{array}
$$
\caption{Absolute Value Element Decomposition of L-matrices}
\label{t:0braneXzX}
\end{table}

It is perhaps useful to emphasize the distinction between the information displayed in Table \ref{t:0brane0} 
versus that displayed in Table \ref{t:0braneXzX}.  The former table's content was derived by calculating the 
``0-brane'' or ``Carroll'' limits of 4D, $\cal N$ = 1 supermultiplets.  While the latter's content was derived
solely from seeking all mathematical solutions to the equations in Eq.  (\ref{GA0}).  With this preamble, 
we are now able to show the relations of the three CM, VM, and TM supermultiplets to sets of permutation 
elements in the permutahedron. Each supermultiplet is shown as a set of color highlighted nodes within 
the permutahedron.  We use the color schema 
$$
\vCent
{\setlength{\unitlength}{1mm}
\begin{picture}(-20,0)
\put(-46,-4.4){\includegraphics[width=0.43in]{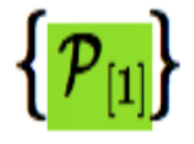}}
\put(-60,-3) {\includegraphics[width=0.4in]{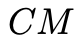}}
\put(4,-4){\includegraphics[width=0.4in]{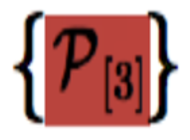}}
\put(-10,-3) {\includegraphics[width=0.4in]{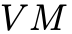}}
\put(44,-4){\includegraphics[width=0.4in]{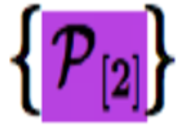}}
\put(30,-3) {\includegraphics[width=0.4in]{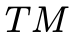}}
\end{picture}}
\nonumber
$$
\noindent
and thus we arrive at the Permutahedronic Projections shown below.
\vskip0.02in
$~~~~~~~~~~~~~~~~~CM~~~~~~~~~~~~~~~~~~~~~~~~~~~~~~~~~~~~~VM
~~~~~~~~~~~~~~~~~~~~~~~~~~~~~~~~~~~~~TM$
$$
\vCent
{\setlength{\unitlength}{1mm}
\begin{picture}(-20,0)
\put(-86,-52){\includegraphics[width=6.8in]{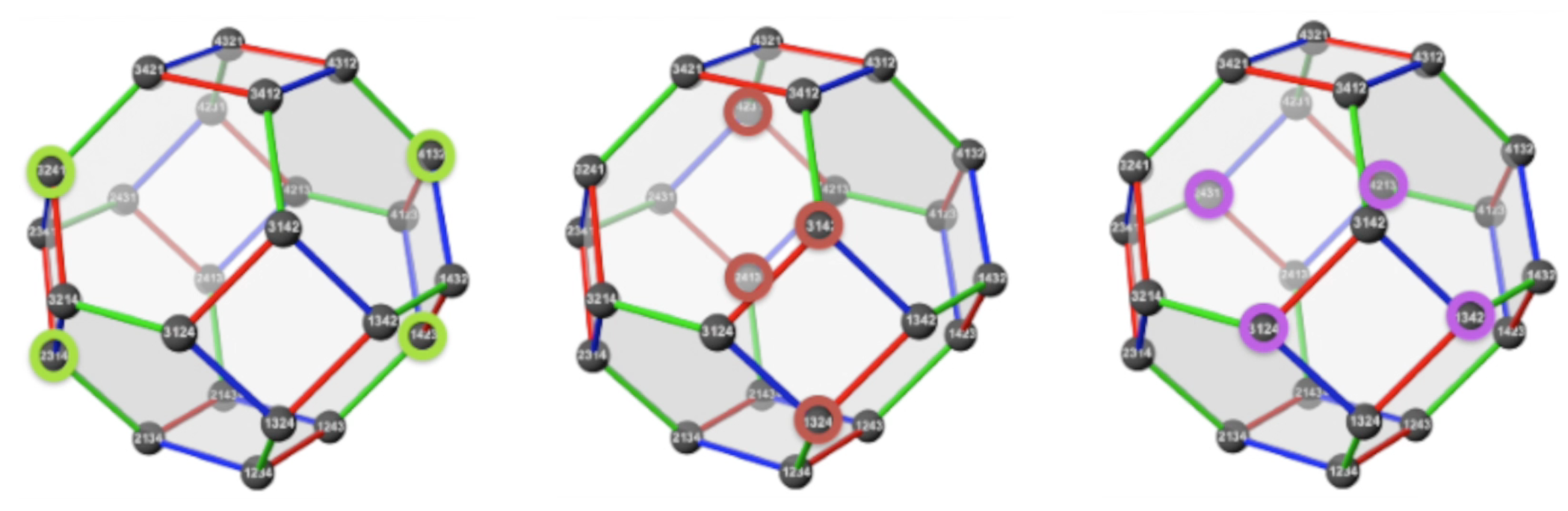}}
\put(-84,-59){$\bm {\rm {Figure ~2:}}$}
\put(-64,-59){ {\rm {The 3-color Permutahedronic Projection of the CM, VM, and TM supermultiplets.}}}
\end{picture}}
\nonumber
$$ \noindent
\vskip2.3in
\noindent
$~$
In a parallel consideration of the $\rm{VM}_1$, $\rm{VM}_2$, and $\rm{VM}_3$ representations, their 
relations to sets of permutation elements in the permutahedron can also be found.  For these we use the
color schema
$$
\vCent
{\setlength{\unitlength}{1mm}
\begin{picture}(-20,0)
\put(-46,-4.4){\includegraphics[width=0.43in]{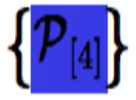}}
\put(-60,-3) {\includegraphics[width=0.36in]{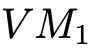}}
\put(4,-4.5){\includegraphics[width=0.44in]{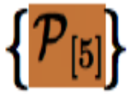}}
\put(-10,-3) {\includegraphics[width=0.4in]{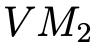}}
\put(44,-4.9){\includegraphics[width=0.48in]{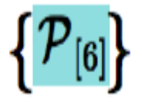}}
\put(30,-3) {\includegraphics[width=0.4in]{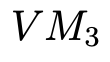}}
\end{picture}}
\nonumber
$$
\noindent
and thus we arrive at the Permutahedronic Projections shown Fig. 3 below.
\vskip0.02in
$~~~~~~~~~~~~~~~~~VM_1~~~~~~~~~~~~~~~~~~~~~~~~~~~~~~~~~~~VM_2
~~~~~~~~~~~~~~~~~~~~~~~~~~~~~~~~~~~VM_3$
$$
\vCent
{\setlength{\unitlength}{1mm}
\begin{picture}(-20,0)
\put(-83,-52){\includegraphics[width=6.8in]{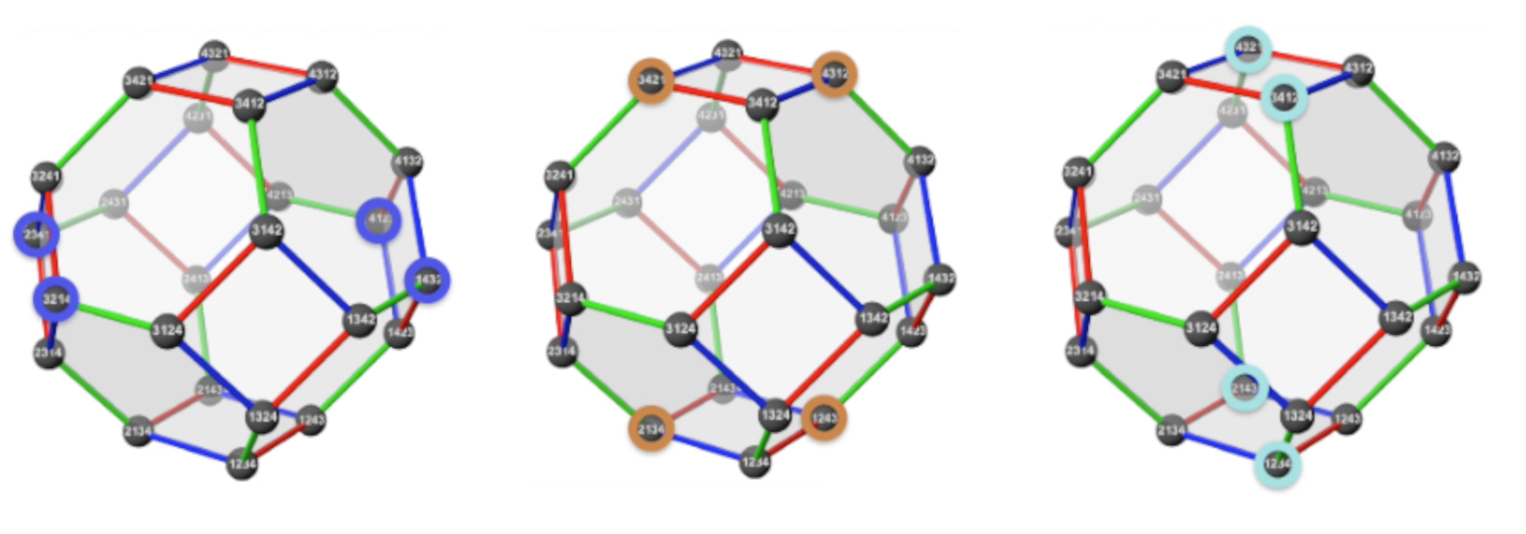}}
\put(-84,-59){$\bm{\rm{Figure ~3:}}$}
\put(-64,-59){ {\rm {The 3-color Permutahedronic Projection of the CM, VM, and TM supermultiplets.}}}
\end{picture}}
\nonumber
$$
\vskip2.25in

The final prerequisite required for defining hopper operators is to introduce a concept of the `weight' for the 
permutation matrices.  A ``weight function" $w({\cal P})$ can be introduced by simply interpreting the addresses 
as real numbers.  Examples of this can be seen by demonstrating this concept for the permutations associated 
with the three supermultiplets.

\begin{table}[!h]
$$
\begin{array}{|c|c|c|c|c|} 
\hline
 &w( |{\bm {\rL}}_1  |) &w( |{\bm {\rL}}_2 |) &w( |{\bm {\rL}}_3 |) & w( |{\bm {\rL}}_4 | ) \\  \hline
${\rm{VM}}$ &   2,413 &   1,324 &    4,231 &    3,142 \\ \hline
${\rm{TM}}$ &   1,342 &   2,431 &    3,124 & 4,213
\\ \hline
${\rm{CM}}$ &   1,423&   2,314 &   3,241  &    4,132 
\\ \hline
    \end{array}
$$
\caption{Weight Values of L-matrices}
\label{t:0braneY}
\end{table} 
\noindent
This table clearly demonstrates that each set has an element of lightest weight.  Furthermore, each 
set can now be re-ordered from its lightest element to its heaviest element.  In fact, this isomorphism
is so powerful that a new set of L-matrices denoted by ${\bm {\rL}}_{w_1}$, ${\bm {\rL}}_{w_2}$, ${\bm 
{\rL}}_{w_3}$, and ${\bm {\rL}}_{w_4}$ can be defined\footnote{Such a reordering would only effect the 
VM system by exchanging the first entry with the second one and simultaneously exchanging the third entry with the fourth one}.  
\begin{table}[H]
\setlength\extrarowheight{2pt}
$$
\begin{array}{|c|c|c|c|c|c|c|} 
\hline
 &  {\bm {{\bm {\rL}}}}_1  & {\bm {{\bm {\rL}}}}_2  & {\bm {{\bm {\rL}}}}_3  &  {\bm {{\bm {\rL}}}}_4   \\  \hline
{\rm TM} & \vev{1\bar3\bar4\bar2} & \vev{24\bar31} & \vev{312\bar4} & \vev{4\bar213}
\\ \hline
{\rm CM} & \vev{1\bar42\bar3} & \vev{23\bar1\bar4} &  \vev{3\bar2\bar41} & \vev{4132} 
\\ \hline
{\rm VM} & \vev{13\bar2\bar4}& \vev{2\bar41\bar3} & \vev{3\bar1\bar42} & \vev{4231} 
\\ \hline
{\rm VM}{}_1 &  \vev{1\bar4\bar32} &  \vev{\bar2\bar3\bar41} &   \vev{3\bar21\bar4} &  \vev{4123} 
\\ \hline
{\rm VM}{}_2 &  \vev{12\bar4\bar3} &   \vev{\bar213\bar4} &  \vev{3412} &  \vev{4\bar31\bar2} 
\\ \hline
{\rm VM}{}_3 &  
\vev{12\bar3\bar4} &  \vev{\bar21\bar43} & \vev{3412} &  \vev{4\bar3\bar21}
\\ \hline
    \end{array}
$$
\caption{Weighted Ordered Permutation Element Decomposition of L-matrices}
\label{t:0braneZ}
\end{table} 
\noindent

The $w_1$, $w_2$, $w_3$, and $w_4$ values are four integers such
that $w_1$ $<$ $w_2$ $<$ $w_3$ $<$ $w_4$ and the conditions,
\begin{equation}
w({{\bm {\rL}}_{w_1}}) ~<~ w({{\bm {\rL}}_{w_2}}) ~<~ w({{\bm {\rL}}_{w_3}}) ~<~ 
w({{\bm {\rL}}_{w_4}}) ~~~.
\end{equation}
are satisfied.  We note this type of `sorting' can be carried out for any
number of L-matrices and used in any basis.  So for example, the result in Eq.\
(\ref{GarDNAdnk}) becomes
\be
\eqalign{
\left[ \,  ({\rm R}^{(\cal R)}_{w_1})_\hi{}^j  \, 
 ({\rm L}^{(\cal R)}_{w_2})_j{}^\hk ~-~ ({\rm R}^{(\cal R)}_{w_2})_\hi{}^j  \, 
 ({\rm L}^{(\cal R)}_{w_1})_j{}^\hk
 \, \right]    = 2&\Big[ \,\ell^{({\cal R})1}_{{w_1}{w_2}}\, (\g^2 \g^3){}_{\hi}{}^\hk
   +  \ell^{({\cal R})2}_{{w_1}{w_2}}\, (\g^3 \g^1){}_{\hi}{}^\hk
   + \ell^{({\cal R})3}_{{w_1}{w_2}}\, (\g^1 \g^2){}_{\hi}{}^\hk   \, \Big] \cr
  ~~
  ~+&~ 2 \Big[ \,  i\,  {{\Tilde \ell}^{(\cal R)}}_{
 {w_1}{w_2}}{}^{1}\, (\g^5){}_{\hi}{}^\hk  \,+\,
  i\, {{\Tilde \ell}^{(\cal R)}}_{{w_1}{w_2}}{}^{2}\, 
 (\g^0 \g^5){}_{\hi}{}^\hk  \,-\,  {{\Tilde \ell}^{(\cal R)}}_{{w_1}{w_2}}{}^{3}\, 
 (\g^0){}_{\hi}{}^\hk 
  \, \Big]  ~~,  {~~~~}   \cr
  &~~
}\label{GarDNAdnkX}
\ee
in the weighted L-matrix basis and the results for 
the $\ell$ and $\tilde {\ell}$ parameters become
 \be
\begin{array}{cccccc}
 \ell_{{w_1}{w_2}}^{(CM)2}=1 & \ell _{{w_1}{w_3}}^{(CM)3}=1 & \ell _{{w_1}{w_4}}^{(CM)1}=1 & \ell _{{w_2}{w_3}}^{(
 CM)1}=1 & \ell _{{w_2}{w_4}}^{(CM)3}=-1 & \ell_{{w_3}{w_4}}^{(CM)2}=1 \\
\Tilde{\ell }_{{w_1}{w_2}}^{(VM)3}= 1 &
\Tilde{\ell }_{{w_1}{w_3}}^{(VM)2}=1 &
\Tilde{\ell }_{{w_1}{w_4}}^{(VM)1}=1 & 
\Tilde{\ell }_{{w_2}{w_3}}^{(VM)1}=-1 & 
\Tilde{\ell }_{{w_2}{w_4}}^{(VM)2}=1 & 
\Tilde{\ell }_{{w_3}{w_4}}^{(VM)3}=-1 \\
 \Tilde{\ell }_{{w_1}{w_2}}^{(TM)3}=1 & \Tilde{\ell }_{{w_1}{w_3}}^{(TM)2}=1 & \Tilde{\ell
 }_{{w_1}{w_4}}^{(TM)1}=1 & \Tilde{\ell }_{{w_2}{w_3}}^{(TM)1}=-1 &
\Tilde{\ell }_{{w_2}{w_4}}^{(TM)2}=1 & \Tilde{\ell }_{{w_3}{w_4}}^{(TM)3}=-1
\end{array}  
\label{e:ls4W}
\ee

We can now define hopper operators using the results in Eq.\ (\ref{e:BR4}).  This equation
can be calculated in the basis of the weighted L-matrices to yield
 \be{ 
 (\,{\rm L}^{(\cal R)}_{w_j}
 )_i{}^\hi\> ~=~ 
   (\,{\cal B}^{(\cal R)}_{{w_j} \, {w_i}})_{i}{}^j \, (\,{\rm L}^{(\cal R)}_{w_i}){}_j{}^\hi~~~.
 }
  \label{e:BR4xx}
\ee

Hopper operators are defined by using only the lightest weighted L-matrix, ${\bm {\rm L}}_{{w_1}}$ on the right with all $w$-indices at fixed values.
 \be{ 
 (\,{\rm L}^{(\cal R)}_{w_j}
 )_i{}^\hi\> ~=~ 
   (\,{\cal B}^{(\cal R)}_{{w_j} \, {w_1}})_{i}{}^j \, (\,{\rm L}^{(\cal R)}_{w_1}){}_j{}^\hi~~~.
 }
  \label{e:BR4xZx}
\ee
in this equation.  Since there is only one remaining free index to specify which other weighted L-matrix 
appears here, we simplify our notation by the introduction of the symbol $\cal H$ (for hopper) that carries 
a free index $\ell$ and write Eq.\ (\ref{e:BR4xZx}) in the form

 \be{ 
 (\,{ {\rm L}}^{(\cal R)}_{w_{\ell}}
 )_i{}^\hi\> ~=~ 
   (\,{\cal H}_{{\ell}-1}^{(\cal R)} ){}_i{}^j   \, (\,{\rm L}^{(\cal R)}_{w_1}){}_j{}^\hi~~~.
 }
  \label{e:BR4xYx}
\ee
which leads to the results shown in Fig. 4.
\newline $~$ \newline $~$ \newline
$$
\vCent
{\setlength{\unitlength}{1mm}
\begin{picture}(-20,0)
\put(-36,-12){\includegraphics[width=3.2in]{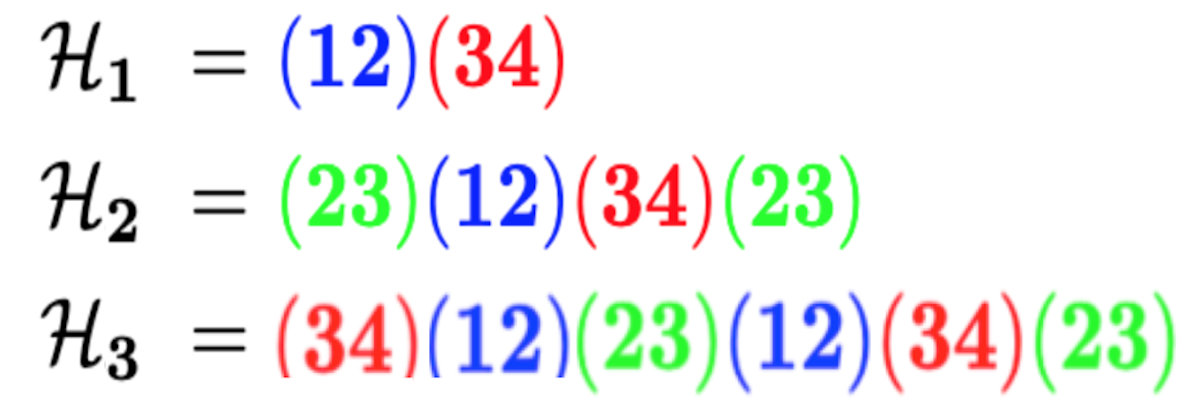}}
\put(-36,-18){${\bm {\rm {Figure ~4:}}}$}
\put(-16,-18){{ {\rm {Hopper Operators}}}}
\end{picture}} 
\nonumber
\label{f:PX2ZZ}
$$
\vskip0.7in

It is of note that this result is independent of which supermultiplet CM, VM, or TM is used as the
starting point.  It is often convenient to add the identity to the list of hopper operators by making
the definition ${\bm {\cal H}}_0$ $\define$ $()$.  Together with the identity operator, the hoppers 
form the vierergruppe \cite{permutadnk}.

Using ${\bm {\cal H}}_3$, two projection operators are shown in the results listed as Eq. (\ref{f:PX4}).
\be
\eqalign{
&\mathcal{P}^-=\frac{1}{2}[{\rm\mathbf{I}}-\mathcal{H}_3],\quad
\mathcal{P}^+=\frac{1}{2}[{\rm\mathbf{I}}+\mathcal{H}_3]\cr
&\mathcal{P}^+\mathcal{P}^-=
\mathcal{P}^-\mathcal{P}^+=0,\quad
\mathcal{P}^++\mathcal{P}^-={\rm\mathbf{I}}\cr
}
\label{f:PX4}
\ee

Looking back at the Table \ref{t:0braneY}, it is clear that taking the absolute value of the first column leads the
following identifications
\be
~~~~
 {{\bm {\rL}}}^{(VM)}_{w_1}  ~=~ \langle 1324 \rangle 
 ~~~~,~~~~{{\bm {\rL}}}^{(TM)}_{w_1}  ~=~ \langle 1342\rangle  ~~~~,~~~~
 {{\bm {\rL}}}^{(CM)}_{w_1} ~=~ \langle 1423\rangle  ~~~~,
\label{e:VcMs}
\ee
and looking at the Table \ref{t:0braneXzX}, a set of lowest weight states can also be identified
according to
\be
~~~~
 {{\bm {\rL}}}^{({VM}_3)}_{w_1}  ~=~ \langle 1234 \rangle ~~~~,~~~~
  {{\bm {\rL}}}^{({VM}_2)}_{w_1}  ~=~ \langle 1243 \rangle ~~~~,~~~~
   {{\bm {\rL}}}^{({VM}_1)}_{w_1}  ~=~ \langle 1432 \rangle 
  ~~~~.
\label{e:VcMs2}
\ee
in a similar manner. These look very similar to a set of `ground states,' analogous to what one would encounter in 
a QM system with ladder operators. In the cases of both Eq. (\ref{e:VcMs}) and Eq. (\ref{e:VcMs2}) 
the order of the ladder operators is given by the hopper operators ${\bm {\cal H}}_0$, ${\bm {\cal H}}_1$,
${\bm {\cal H}}_2$ and ${\bm {\cal H}}_3$. ¬†Moreover the highest weight states are the antipodes of 
the lowestweight states for each representation in the permutahedron. ¬†

Further insight into the ground states follows by looking at their locations in the 3-color permutahedron. ¬†
All of the square and hexagonal faces can be assigned a weight as the sum of the weights at their 
respective vertices. ¬†The lightest weight hexagonal face contains the trivial permutation, and its weight is 9,998.
There exists a set of hopper operators that `move around' this lightest weight hexagonal face.
We will denote these by the symbols
${\widehat {\bm {\cal H}}}_0$, ${\widehat {\bm {\cal H}}}_1$,...,
and ${\widehat {\bm {\cal H}}}_7$.
Here ${\widehat {\bm {\cal H}}}_0$ = $()$ as before with
${ \bm{\cal H}}_0$. The explicit forms of the remaining
six is given by
$~~$
$$
\vCent
{\setlength{\unitlength}{1mm}
\begin{picture}(-20,0)
\put(-73,-33){\includegraphics[width=2.2in]{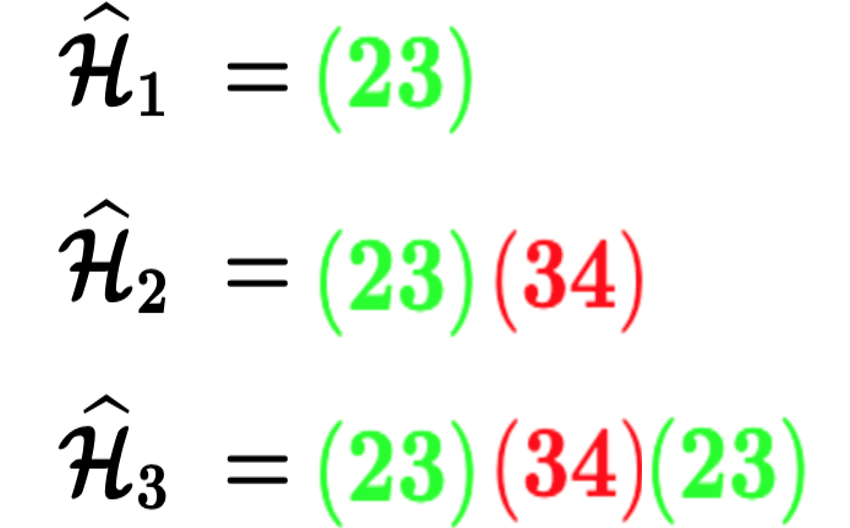}}
\put(-8,-33.5){\includegraphics[width=3.25in]{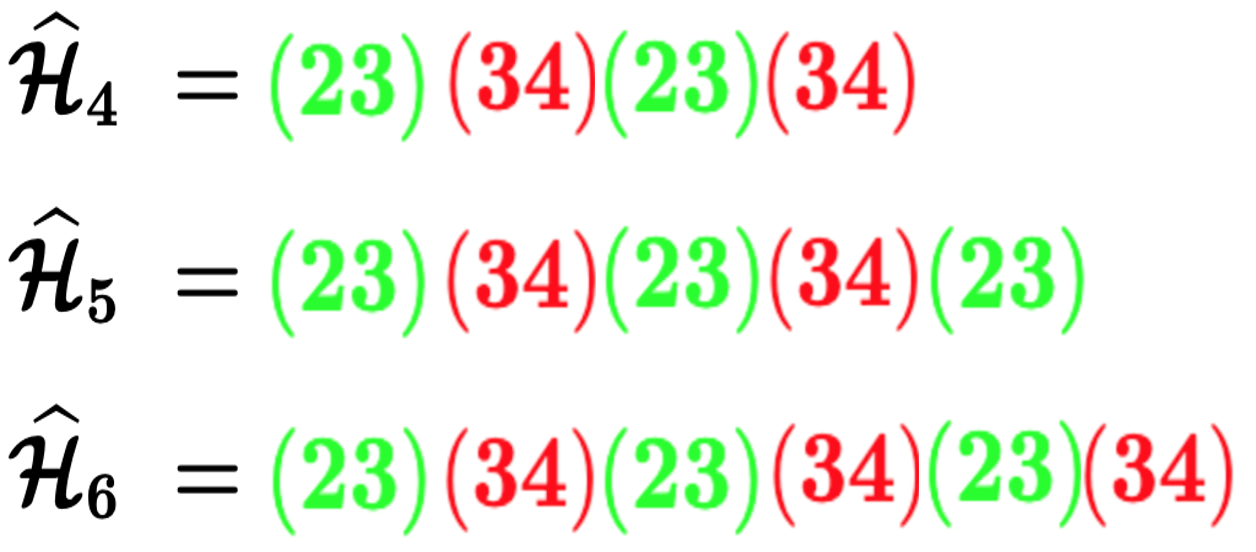}}
\put(-50,-44){${\bm {\rm {Figure ~5:}}}$}
\put(-30,-44){{ {\rm {Lightest `Benzene Ring' Hopper Operators}}}}
\end{picture}}
\nonumber
\label{f:PX2ZZ}
$$
$$~~$$
$$~~$$
$$~~$$
$$~~$$
$$~~$$

By looking at the image in Fig. 6, however, it is clear there exist another set of hexagonal face hopper operators.
We can denotes these by the symbols ${{\widehat{\bm {\cal H}}}}_0^{w_1}$, ${\widehat {\bm {\cal H}}}_1^{w_1}$,...,
and ${\widehat {\bm {\cal H}}}_6^{w_1}$. here the superscript $w_1$ here denotes the weight of the hexagonal 
face which is given by 13,344.  These hexagonal face hopper operators are obtained by replacing the green links 
by blue links and the red links by green links in Fig. 5.
$~~$

$$
\vCent
{\setlength{\unitlength}{1mm}
\begin{picture}(-20,0)
\put(-56,-55){\includegraphics[width=5in, height=2.3in]{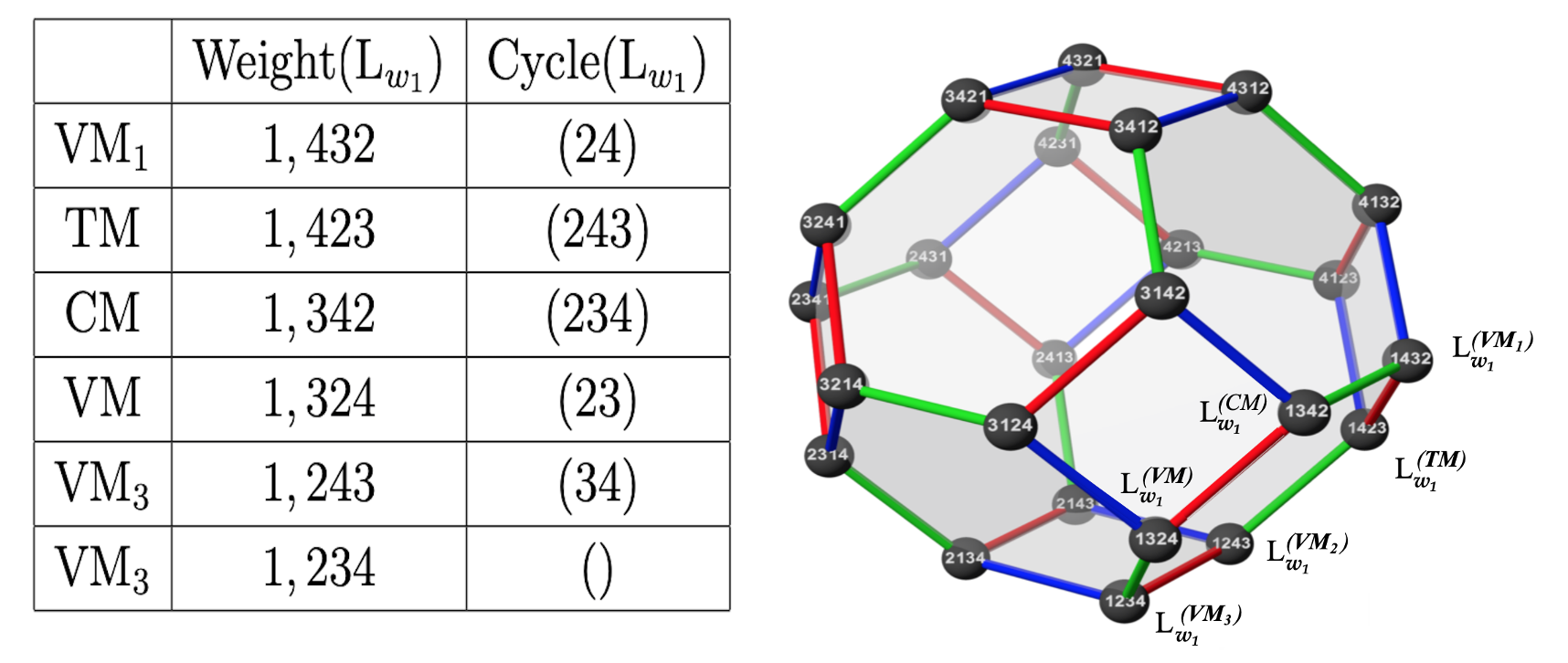}}
\put(-56,-60){$\bm {\rm {Figure ~6:}}$}
\put(-36,-60){ {\rm {The Three Color Permutahedron Ground States}}}
\end{picture}}
\nonumber
$$ 
\vskip2.4in
Before we continue in the following discussion to calculate holoraumy matrices, it is convenient to recall work 
from \cite{3} in which two sets of matrices denoted by the symbols ${\bm {\a{}}}$ and ${\bm \b}$ matrices were 
defined.  These emerged from the reduction of 4D, $\cal N$ = 1 supermultiplets.  We have two different but 
equivalent ways to denote these using either Bar-Bracket or Boolean-Factor-Cycle notations.  Explicitly, we have
\be  \eqalign{
&{\bm \a}_1 ~=~ -\, i \,\vev{43{\bar2}{\bar1}} ~=~ -\, i \, (12)_b \, (14)(23)~~,~~ \cr
&{\bm \a}_2 ~=~ -\, i \,\vev{2{\bar1}4{\bar3}} ~=~ -\, i \, (10)_b \, (12)(34) ~~,~~ \cr
&{\bm \a}_3 ~=~ -\, i \,\vev{3{\bar4}{\bar1}2}  ~=~ -\, i \, (6)_b \, (13)(24)  ~~, \cr
&{\bm \b}_1 ~=~ -\, i \,\vev{4{\bar3}2{\bar1}}  ~=~ -\, i \, (10)_b \, (14)(23)  ~~, \cr
&{\bm \b}_2 ~=~ -\, i \,\vev{34{\bar1}{\bar2}}  ~=~ -\, i \, (12)_b \, (13)(24)  ~~, \cr
&{\bm \b}_3 ~=~ -\, i \,\vev{2{\bar1}{\bar4}3}  ~=~ -\, i \, (6)_b \, (12)(34) ~~,
} \ee
and these describe two commuting representations of su(2).  Now we can also use the weighted L-matrices to 
calculate their holoraumy matrices and express the results in terms of the ${\bm {\a{}}}$ and ${\bm \b}$ matrices.
\begin{table}[H]
\setlength\extrarowheight{2pt}
$$
\begin{array}{|c|c|c|c|c|c|c|c|c|} 
\hline
(\cal R)
 & {\bm {\cal B}}^{({\cal R})}_{{w_1} \, {w_2} } & {\bm {\cal B}}^{({\cal R})}_{{w_1} \, {w_3} }  & {\bm {\cal B}}^{({\cal R})}_{{w_1} \, {w_4} } 
 & {\bm {\cal B}}^{({\cal R})}_{{w_2} \, {w_3} }  & {\bm {\cal B}}^{({\cal R})}_{{w_2} \, {w_4} } & {\bm {\cal B}}^{({\cal R})}_{{w_3} \, {w_4} } \\ \hline  
{\rm VM}{}_1 & i\, {\bm \b}_{1}   &   i \, {\bm \b}_{2}  &    i \, {\bm \b}_{3}   &   i \, {\bm \b}_{3}  &  -\, i \, {\bm \b}_{2}  &   i \, {\bm \b}_{1}  \\ \hline
{\rm CM} & -\, i\, {\bm \a}_{3}   &   i \, {\bm \a}_{1}  &    i \, {\bm \a}_{2}   & -\.  i \, {\bm \a}_{2}  &  i \, {\bm \a}_{1}  &   i \, {\bm \a}_{3}  \\ \hline
{\rm TM} & i\, {\bm \a}_{1}   &   i \, {\bm \a}_{2}  &    i \, {\bm \a}_{3}   &   i \, {\bm \a}_{3}  & -\, i \, {\bm \a}_{2}  &   i \, {\bm \a}_{1}  \\ \hline
{\rm VM} & -\, i\, {\bm \a}_{3}   &   i \, {\bm \a}_{2}  &    i \, {\bm \a}_{1}   &   i \, {\bm \a}_{1}  & -\, i \, {\bm \a}_{2}  &  -\, i \, {\bm \a}_{3}  \\ \hline
{\rm VM}{}_2 & i\, {\bm \a}_{2}   &   i \, {\bm \a}_{1}  &    i \, {\bm \a}_{3}   & -\,  i \, {\bm \a}_{3}  &  i \, {\bm \a}_{1}  &  -\, i \, {\bm \a}_{2}  \\ \hline
{\rm VM}{}_3 & i\, {\bm \b}_{3}   &   i \, {\bm \b}_{2}  &    i \, {\bm \b}_{1}   & -\,  i \, {\bm \b}_{1}  &  i \, {\bm \b}_{2}  &  -\, i \, {\bm \b}_{3}  \\ \hline
\end{array}
$$    
\caption{Weighted Ordered Holoraumy of Mininal Reps}
\label{t:0braneXZX}
\end{table} 

Given any explicit set of ${\bm \g}^{\mu}$ matrices, in a Majorana representation, for a four dimensional Minkowski space 
and their covering algebra $\{ \,{\rm I}, \, {\bm \g}^{\mu}, \, {\bm \g}^{\mu} \wedge  {\bm \g}^{\nu}, \, {\bm \g}^{\mu} \wedge  
{\bm \g}^{\nu} \wedge {\bm \g}^{\rho} , \, {\bm \g}^{\mu} \wedge  {\bm \g}^{\nu} \wedge {\bm \g}^{\rho} \wedge  {\bm \g}^{\sigma}, 
\, \}$, of necessity the two subsets $\{ \, {\bm \g}^{1} \wedge  {\bm \g}^{2}, \, {\bm \g}^{2} \wedge  {\bm \g}^{3}, \, {\bm \g}^{3} 
\wedge  {\bm \g}^{1} \, \}$ and $\{ \, {\bm \g}^{0} , i\, {\bm \g}^{5}, i\, {\bm \g}^{0} {\bm \g}^{5}  \, \}$ necessarily form two 
commuting representations of su(2).  This implies there must exist an isomorphism between elements of the holoraumy 
matrices and the elements in the subsets containing $\{ \, {\bm \g}^{1} \wedge  {\bm \g}^{2}, \, {\bm \g}^{2} \wedge  {\bm 
\g}^{3}, \, {\bm \g}^{3} \wedge  {\bm \g}^{1} \, \}$ and $\{ \, {\bm \g}^{0} , i\, {\bm \g}^{5}, \, i\, {\bm \g}^{0} {\bm \g}^{5}  \, \}$.

The works in \cite{K4g,K5g} introduced two mapping operations from pairs of supermultiplet representations $({\cal R})$ 
and $({\cal R}^{\prime})$ into the real numbers.  These operations were given the names of `gadgets', and respectively
denoted by the symbols, $ { {\cal G}}{}_{(1)}\left[  ({\cal R}) \, , \,  ({\cal R}^{\prime}) \right]$ and $ { {\cal G}}{}_{(2)}\left[  
({\cal R}) \, , \,  ({\cal R}^{\prime}) \right]$.  With some improvements in definitions we can write (using no summation of 
repeated $ w$ indices on the B-matrices),
\be
\eqalign{
{ {\cal G}}{}_{(1)} \left[  ({\cal R}) \, , \,  ({\cal R}^{\prime}) \right] ~&\equiv ~~~ -\, \frc{1}{24} \, {\rm {Tr}}
\, {\Big [} ~ 
 {\bm {\cal B}}_{{w_1} \,{w_2}}^{\,({\cal R})  } \,  
 {\bm {\cal B}}_{{w_1} \,{w_2}}^{\, ({\cal R}^{\prime})  }  ~+~ 
 {\bm {\cal B}}_{{w_1} \,{w_3}}^{\,({\cal R})  } \,  
 {\bm {\cal B}}_{ {w_1} \,{w_3}}^{\, ({\cal R}^{\prime})  } ~+~
 {\bm {\cal B}}_{{w_1} \,{w_4}}^{\,({\cal R})  } \,  
 {\bm {\cal B}}_{{w_1} \,{w_4}}^{\, ({\cal R}^{\prime})  }    ~+~ \cr
&{~~~~~~~~~~~~~~~~~~~~~}      
 {\bm {\cal B}}_{{w_2} \,{w_3}}^{\, ({\cal R})  }  
 {\bm {\cal B}}_{{w_2} \,{w_3}}^{\, ({\cal R}^{\prime})  } \, ~+~ 
 {\bm {\cal B}}_{{w_2} \,{w_4}}^{\,({\cal R})  } \,  
 {\bm {\cal B}}_{ {w_2} \,{w_4}}^{\, ({\cal R}^{\prime})  } ~+~
 {\bm {\cal B}}_{{w_3} \,{w_4}}^{\,({\cal R})  } \,  
 {\bm {\cal B}}_{{w_3} \,{w_4}}^{\, ({\cal R}^{\prime})  }  ~ {\Big ]} 
~~~,~~~~~~
}    \label{Gdgt1}
\ee 
and 
\be
\eqalign{
{ {\cal G}}{}_{(2)} \left[  ({\cal R}) \, , \,  ({\cal R}^{\prime}) \right] ~&\equiv ~~~ -\, \frc{1}{24} \, {\rm {Tr}}
\, {\Big [} ~ 
 {\bm {\cal B}}_{{w_1} \,{w_2}}^{\,({\cal R})  } \,  
 {\bm {\cal B}}_{{w_1} \,{w_2}}^{\, ({\cal R}^{\prime})  }  ~-~ 
 {\bm {\cal B}}_{{w_1} \,{w_3}}^{\,({\cal R})  } \,  
 {\bm {\cal B}}_{ {w_1} \,{w_3}}^{\, ({\cal R}^{\prime})  } ~+~
 {\bm {\cal B}}_{{w_1} \,{w_4}}^{\,({\cal R})  } \,  
 {\bm {\cal B}}_{{w_1} \,{w_4}}^{\, ({\cal R}^{\prime})  }    ~+~ \cr
&{~~~~~~~~~~~~~~~~~~~~~}      
 {\bm {\cal B}}_{{w_2} \,{w_3}}^{\, ({\cal R})  }  
 {\bm {\cal B}}_{{w_2} \,{w_3}}^{\, ({\cal R}^{\prime})  } \, ~-~ 
 {\bm {\cal B}}_{{w_2} \,{w_4}}^{\,({\cal R})  } \,  
 {\bm {\cal B}}_{ {w_2} \,{w_4}}^{\, ({\cal R}^{\prime})  } ~+~
 {\bm {\cal B}}_{{w_3} \,{w_4}}^{\,({\cal R})  } \,  
 {\bm {\cal B}}_{{w_3} \,{w_4}}^{\, ({\cal R}^{\prime})  }  ~ {\Big ]} 
~~~~.~~~~~
}    \label{Gdgt2}
\ee
Since its definition, Eq. (\ref{Gdgt1}), implies $ {\cal G}{}_{(1)} [ \, ( {\cal R}) , ({\cal R}^{\prime})) \, ]$ can 
only takes values (-1 , -2/3, -1/3, 0 , 1/3, 2/3, 1), it defines an `angle' between two representations by
$
cos \left\{ \theta [({\cal R})  , \, ( {\cal R}^{\prime} )]  \right\}$ = ${\cal G}{}_{(1)}[({\cal R})  , \, ( {\cal R}^{\prime} )]$.
\vskip0.15in
\indent
$~~~~~$ ${\bm {\rm Conjecture}}$: 
\newline \indent $~~~~~$
The work in \cite{Reager} revealed that the valise adinkras
correspond to maximal 
\newline \indent $~~~~~$
Cohen-Macaulay modules in algebraic geometry.  We conjecture that the
\newline \indent $~~~~~$
values of the two gadgets provide digits that
classify these modules.
\vskip0.15in

The weighted L-matrix basis has a remarkable distinction in comparison to the basis used in \cite{K4g,K5g}. In 
the weighted L-matrix basis the three representations $(\rm{VM})$, $(\rm{TM})$, and $(\rm{CM})$ form a 
right-handed {\it {orthogonal}} triad of unit vectors when taken in proper weight order.  This is shown in Fig. 7 below.
$$
\vCent
{\setlength{\unitlength}{1mm}
\begin{picture}(-20,0)
\put(-38,-56){\includegraphics[width=3.2in]{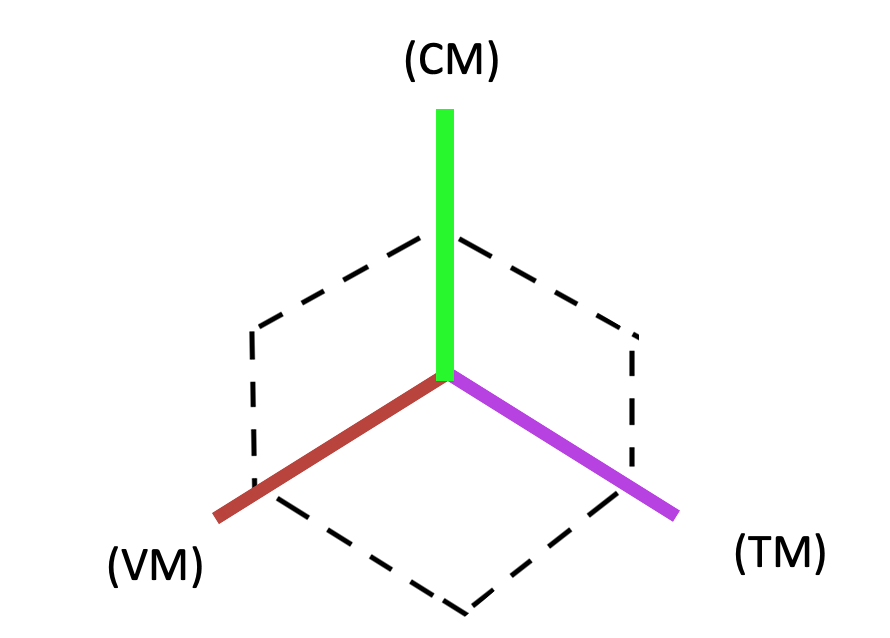}}
\put(-47,-66){${\bm {\rm {Figure ~7:}}}$}
\put(-28,-66){{ {\rm {  Triad of Minimal 4-Color Adinkra Reps}}}}
\put(-16,-18){{ {\rm {}}}}
\end{picture}}
\nonumber
\label{f:TRIAD}
$$ 
\vskip2.5in
\noindent
In fact, use of the definition above Eq. (\ref{Gdgt1}) on the six minimal representations leads to the conclusion that all the
representations lie at $90^o$ angles with respect to one another with a ``length'' of unity.

Although we did not use the fact that the ${\bm {\rm L}}$- and ${\bm {\rm R}}$-matrices related to adinkras
can be used to construct a Euclidean Clifford algebra via
the a definition
\be
{\bm\g}^{(\cal R)}_\rI ~=~ 
\left[\begin{array}{cc}
0~~~ & {\bm {\rm L}}^{(\cal R)}_\rI \\
{\bm {\rm R}}^{(\cal R)}_\rI & 0  
\end{array}\right] 
 ~\define ~ 
{\widehat {\cal F}}_\rI
\left( \, {\bm {\rm L}}, \, {\bm {\rm R}}  \, \right) ~~~,
\label{EuCliff}
\ee
this has long been known.

Now let us return to the isomorphism noted below Table \ref{t:0braneY}.  This is the origin of the 
results shown in Eq. (\ref{t:0brane0}) and is the mechanism by which the ${\bm {\rm L}}$-matrices 
and ${\bm {\rm R}}$-matrices define a set of Minkowski space $\gamma$ matrices. This equation 
implies there exist a function ${\cal F}^0$ and three functions ${\cal F}^{1}$, ${\cal F}^{2}$, and 
${\cal F}^{3}$, such that
\be 
{\bm {\gamma}}^0 ~=~ {\cal F}^0 \left( \, {\bm {\rm L}}, \, {\bm {\rm R}}  \, \right) ~,~ 
{\bm {\gamma}}^1 ~=~ {\cal F}^1 \left( \, {\bm {\rm L}}, \, {\bm {\rm R}}  \, \right) ~,~ 
{\bm {\gamma}}^2 ~=~ {\cal F}^2 \left( \, {\bm {\rm L}}, \, {\bm {\rm R}}  \, \right) ~,~ 
{\bm {\gamma}}^3 ~=~ {\cal F}^3 \left( \, {\bm {\rm L}}, \, {\bm {\rm R}}  \, \right) ~,~ 
\label{MnCliff}
\ee
where ${\cal F}^0$ is linear in the holoraumy and ${\cal F}^1$, ${\cal F}^2$, 
and ${\cal F}^3$ are quadratic in the holoraumy.

Thus some pairs of sets, of the adinkra $\bm {\rm L}$-matrices, simultaneously define the two
Clifford algebras in Eq. (\ref{EuCliff}) and Eq. (\ref{MnCliff}). The former is a Euclidean Clifford 
Algebra on the worldline and the latter is a Minkowski Clifford Algebra on the spacetime manifold.  
However, in the 4-color case, two of the minimal adinkra representations are required to satisfy a 
certain condition \cite{Spn+ChLR} to construct the spacetime Clifford algebra.

This is the essential mechanism by which adinkras encode 4D Lorentz representation data and is the 
ultimate enabler of the existence of SUSY Holography \cite{4}.

\newpage 
\section{A Simplified Introduction to Wolfram Paradigm}
The Wolfram Paradigm is a discretized space-time model that uses a graph generated from mathematical objects known as ``sequential substitution" or ``rewriting rules" \cite{18}. 
The aim is to find a correspondence between discretized physics on the graph and classical physics in the continuum limit. The rewriting rules play an important role in the Wolfram Paradigm as they are a collection of elements that contain information about the connections between them \cite{17}. 

One simple example of a rewriting rule is

\begin{equation}
\{\{x, y\}\} \rightarrow \{\{x, y\}, \{y, z\}\}
\end{equation}

Here $x$, $y$, and $z$ stand for any elements\footnote{The interested reader can refer the more available examples and explanations of rewriting rules at \cite{17}. Chapter 2, 109-111 pp.}. This rule states that wherever a relation that matches \{$x$,$y$\} appears, one should replace those pair as \{\{$x$,$y$\},\{$y$,$z$\}\}, where $z$ is a new element. 
For example if the initial state is given as \{\{1,2\}\}, then the rule will produce \{\{1,2\},\{2,$k$\}\} where $k$ is a new element. ($k$ can be any element that is not 1 or 2, but for convenience let us set k as 3.)
And if we apply the rule one more time, we can get \{\{1, 2\}, \{2, 4\}, \{2, 3\}, \{3, 5\}\}.

Through repeated application of such rules, complex structures can be generated, such as those found in cellular automata and other discrete systems.
And we can express this relation by graph as,

\setcounter{figure}{7}
\begin{figure}[H]
    \centering
    \includegraphics[scale=0.6]{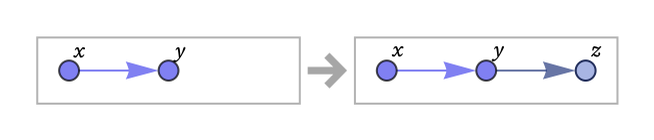}
    \caption{Graphical representation of rewriting rule\protect\cite{18}}
    \label{fig8-1}
\end{figure}

\begin{figure}[H]
    \centering
    \includegraphics[scale=0.3]{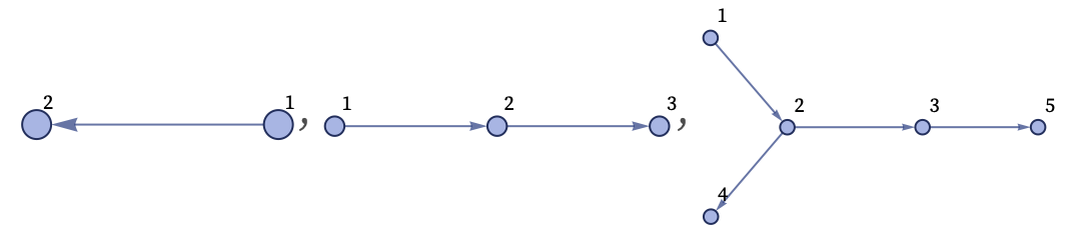}
    \caption{Graphical representation of development of rewriting rule\protect\cite{18}}
    \label{fig8-2}
\end{figure}

This is a simple example of graph generation but depending on the various rewriting rules there are numerous different ways of developing graphs by applying rewriting rules many times \cite{17}. 

In this work, we will explore the properties of adinkras under use of a rewriting rule in the Wolfram Paradigm. The Paradigm is a discretized space-time model that utilizes a graph generated from rewriting rules. While the example we presented is a simple illustration of graph generation, there are numerous ways to generate graphs depending on the rewriting rules applied. Therefore, our aim in this chapter is to investigate what impact the rewriting rules have on, firstly minimal adinkras and one non-minimal one.

\newpage
\section{4D $\mathcal{N}$=1 Minimal Multiplets}

In this chapter, based on our interpretation of the Wolfram Paradigm, we will explicitly present the relevant 
features of the paradigm to adinkras.  The general algorithm suggested in this paper is using hopping operator 
as a ingredient for rewriting rules. First by using the $\bm\rL$- (or $\bm\rR$- as necessary) matrices.
we define a
set of four operators denoted by $\bm{\rm X}$.   On these we apply the same logic as shown in Eq. (\ref{eq3})
to define four $\bm{\rm Y}$-matrices.  Next on the four $\bm{\rm Y}$-matrices, we once more use an equation
of the same form as in in Eq. (\ref{eq3}) to define four $\bm{\rm Z}$-matrices. etc. 
We repeat this process until 
we obtain the simplest form of hopping operators that can be written as  $\pm \mathbf{I}_n$.  At this stage, we
have reached a convergence as this process will never lead to other matrices.

The summary of our results are shown in the six tables ( Table \ref{tab1}, \ref{tab2}, \ref{tab3a}, \ref{tab4a}, \ref{tab5a}, and \ref{tab6a}), and a final Table \ref{tabZ} for the ``$\mathbf{W}$-matrices."

\begin{table}[H]
\setlength\extrarowheight{2pt}
\begin{center}
\begin{tabular}{|m{2.8em}|m{3.2cm}|m{3.2cm}|m{3.2cm}|m{3.2cm}|}
\hline
Chiral &  1&  2& 3& 4\\ 
\hline
$[{\bm \rL}_n]_{i}{}^{\hat j}$ & 
$\vev{14\bar2\bar3}$& 
$\vev{23\bar1\bar4}$& 
$\vev{3\bar2\bar41}$& 
$\vev{4132}$\\   
\hline
$[{\bm{\rm X}}_n]_{i}{}^{j}$ &  
$\vev{3\bar4\bar12}$&
$\vev{2\bar14\bar3}$&
$\vev{\bar341\bar2}$&
$\vev{2\bar14\bar3}$\\ 
\hline
$[{\bm{\rm Y}}_n]_{i}{}^{j}$ &  
$\vev{43\bar2\bar1}$&  
$\vev{43\bar2\bar1}$&   
$\vev{\bar4\bar321}$& 
$\vev{\bar4\bar321}$
\\ 
\hline
$[{\bm{\rm Z}}_n]_{i}{}^{j}$ &  
$\vev{1234}$&  
$\vev{\bar1\bar2\bar3\bar4}$&  
$\vev{1234}$&
$\vev{\bar1\bar2\bar3\bar4}$\\ 
\hline
$[{\bm{\rm W}}_n]_{i}{}^{j}$ &  
$\vev{\bar1\bar2\bar3\bar4}$&  
$\vev{\bar1\bar2\bar3\bar4}$&  
$\vev{\bar1\bar2\bar3\bar4}$&
$\vev{\bar1\bar2\bar3\bar4}$\\ 
\hline
\end{tabular}
\end{center}
\caption{$\bm\rL$-matrix summary table for Chiral supermultiplet.}
\label{tab1}
\end{table}

\begin{table}[H]\setlength\extrarowheight{2pt}
\begin{center}
\begin{tabular}{|m{2.8em}|m{3.2cm}|m{3.2cm}|m{3.2cm}|m{3.2cm}|}
\hline
Chiral &  1&  2& 3& 4\\ 
\hline
$[{\bm{\rm R}}_n]_{\hat i}{}^{k}$ & 
$\vev{13\bar4\bar2}$ 
&$\vev{\bar312\bar4}$
& $\vev{4\bar21\bar3}$
& $\vev{2431}$\\ 
\hline
$[{\bm{\rm X'}}_n]_{\hat i}{}^{\hat j}$ &  
$\vev{\bar21\bar43}$&
$\vev{\bar4\bar321}$&
$\vev{\bar21\bar43}$& 
$\vev{43\bar2\bar1}$\\ 
\hline
$[{\bm{\rm Y'}}_n]_{\hat i}{}^{\hat j}$ &  
$\vev{3\bar4\bar12}$& 
$\vev{\bar341\bar2}$&  
$\vev{\bar341\bar2}$&
$\vev{3\bar4\bar12}$
\\ 
\hline
$[{\bm{\rm Z'}}_n]_{\hat i}{}^{\hat j}$ &  
$\vev{\bar1\bar2\bar3\bar4}$&  
$\vev{1234}$&  
$\vev{\bar1\bar2\bar3\bar4}$&
$\vev{1234}$\\ 
\hline
$[{\bm{\rm W'}}_n]_{i}{}^{j}$ &  
$\vev{\bar1\bar2\bar3\bar4}$&  
$\vev{\bar1\bar2\bar3\bar4}$&  
$\vev{\bar1\bar2\bar3\bar4}$&
$\vev{\bar1\bar2\bar3\bar4}$\\ 
\hline
\end{tabular}
\end{center}
\caption{$\bm\rR$-matrix summary table for Chiral supermultiplet.}
\label{tab2}
\end{table}

\begin{table}[H]
\setlength\extrarowheight{2pt}
\begin{center}
\begin{tabular}{|m{2.8em}|m{3.2cm}|m{3.2cm}|m{3.2cm}|m{3.2cm}|}
\hline
Vector &  1&  2& 3& 4\\ 
\hline
$[{\bm \rL}_n]_{i}{}^{\hat j}$& 
$\vev{2\bar41\bar3}$&
$\vev{13\bar2\bar4}$&
$\vev{4231}$&
$\vev{3\bar1\bar42}$\\
\hline
$[{\bm{\rm X}}_n]_{i}{}^{j}$ &  
$\vev{3\bar4\bar12}$&
$\vev{\bar4\bar321}$&
$\vev{3\bar4\bar12}$& 
$\vev{43\bar2\bar1}$\\ 
\hline
$[{\bm{\rm Y}}_n]_{i}{}^{j}$ &  
$\vev{2\bar14\bar3}$&  
$\vev{\bar21\bar43}$&  
$\vev{\bar21\bar43}$&
$\vev{2\bar14\bar3}$\\ 
\hline
$[{\bm{\rm Z}}_n]_{i}{}^{j}$&  
$\vev{\bar1\bar2\bar3\bar4}$&  
$\vev{1234}$&  
$\vev{\bar1\bar2\bar3\bar4}$&
$\vev{1234}$\\ 
\hline
$[{\bm{\rm W}}_n]_{i}{}^{j}$ &  
$\vev{\bar1\bar2\bar3\bar4}$&  
$\vev{\bar1\bar2\bar3\bar4}$&  
$\vev{\bar1\bar2\bar3\bar4}$&
$\vev{\bar1\bar2\bar3\bar4}$\\ 
\hline
\end{tabular}
\end{center}
\caption{$\bm\rL$-matrix summary table for Vector supermultiplets}
\label{tab3a}
\end{table}

\begin{table}[H]
\setlength\extrarowheight{2pt}
\begin{center}
\begin{tabular}{|m{2.8em}|m{3.2cm}|m{3.2cm}|m{3.2cm}|m{3.2cm}|}
\hline
Vector &  1&  2& 3& 4\\ 
\hline
$[{\bm{\rm R}}_n]_{\hat i}{}^{j}$& 
$\vev{31\bar4\bar2}$& 
$\vev{1\bar32\bar4}$& 
$\vev{4231}$& 
$\vev{\bar241\bar3}$\\   
\hline
$[{\bm{\rm X}}'_n]_{\hat i}{}^{\hat j}$ &  
$\vev{2\bar1\bar43}$&
$\vev{\bar43\bar21}$&
$\vev{\bar214\bar3}$& 
$\vev{\bar43\bar21}$\\
\hline
$[{\bm{\rm Y}}'_n]_{\hat i}{}^{\hat j}$ &  
$\vev{34\bar1\bar2}$&  
$\vev{34\bar1\bar2}$&  
$\vev{\bar3\bar412}$& 
$\vev{\bar3\bar412}$\\ 
\hline
$[{\bm{\rm Z}}'_n]_{\hat i}{}^{\hat j}$ &  
$\vev{1234}$&  
$\vev{\bar1\bar2\bar3\bar4}$&  
$\vev{1234}$&
$\vev{\bar1\bar2\bar3\bar4}$\\ 
\hline
$[{\bm{\rm W}}'_n]_{i}{}^{j}$ &  
$\vev{\bar1\bar2\bar3\bar4}$&  
$\vev{\bar1\bar2\bar3\bar4}$&  
$\vev{\bar1\bar2\bar3\bar4}$&
$\vev{\bar1\bar2\bar3\bar4}$\\ 
\hline
\end{tabular}
\end{center}
\caption{$\bm\rR$-matrix summary table for Vector supermultiplets}
\label{tab4a}
\end{table}

\begin{table}[H]
\setlength\extrarowheight{2pt}
\begin{center}
\begin{tabular}{|m{2.8em}|m{3.4cm}|m{3.1cm}|m{3.1cm}|m{3.2cm}|}
\hline
Tensor &  1&  2& 3& 4\\ 
\hline
$[{\bm \rL}_n]_{i}{}^{\hat j}$ & 
$\vev{1\bar3\bar4\bar2}$& 
$\vev{24\bar31}$& 
$\vev{312\bar4}$& 
$\vev{4\bar213}$\\   
\hline
$[{\bm{\rm X}}_n]_{i}{}^{j}$&  
$\vev{\bar4\bar321}$&
$\vev{\bar341\bar2}$&
$\vev{\bar4\bar321}$& 
$\vev{3\bar4\bar12}$\\ 
\hline
$[{\bm{\rm Y}}_n]_{i}{}^{j}$
&  
$\vev{2\bar14\bar3}$&  
$\vev{\bar21\bar43}$&  
$\vev{\bar21\bar43}$& 
$\vev{2\bar14\bar3}$\\ 
\hline
$[{\bm{\rm Z}}_n]_{i}{}^{j}$ &  
$\vev{\bar1\bar2\bar3\bar4}$&  
$\vev{1234}$&  
$\vev{\bar1\bar2\bar3\bar4}$&
$\vev{1234}$\\ 
\hline
$[{\bm{\rm W}}_n]_{i}{}^{j}$ &  
$\vev{\bar1\bar2\bar3\bar4}$&  
$\vev{\bar1\bar2\bar3\bar4}$&  
$\vev{\bar1\bar2\bar3\bar4}$&
$\vev{\bar1\bar2\bar3\bar4}$\\ 
\hline
\end{tabular}
\end{center}
\caption{$\bm\rL$-matrix summary table for Tensor supermultiplet}
\label{tab5a}
\end{table}

\begin{table}[H]
\setlength\extrarowheight{2pt}
\begin{center}
\begin{tabular}{|m{2.8em}|m{3.4cm}|m{3.1cm}|m{3.1cm}|m{3.2cm}|}
\hline
Tensor &  1&  2& 3& 4\\ 
\hline
$[{\bm{\rm R}}_n]_{\hat i}{}^{j}$ & 
$\vev{1\bar4\bar2\bar3}$& 
$\vev{41\bar32}$& 
$\vev{231\bar4}$& 
$\vev{3\bar243}$\\   
\hline
$[{\bm{\rm X}}'_n]_{\hat i}{}^{\hat j}$ &  
$\vev{\bar214\bar3}$&  
$\vev{4\bar32\bar1}$&  
$\vev{2\bar1\bar43}$& 
$\vev{4\bar32\bar1}$\\ 
\hline
$[{\bm{\rm Y}}'_n]_{\hat i}{}^{\hat j}$ &  
$\vev{34\bar1\bar2}$&  
$\vev{34\bar1\bar2}$&  
$\vev{\bar3\bar412}$& 
$\vev{\bar3\bar412}$\\ 
\hline
$[{\bm{\rm Z}}'_n]_{\hat i}{}^{\hat j}$ &  
$\vev{1234}$&  
$\vev{\bar1\bar2\bar3\bar4}$&  
$\vev{1234}$&
$\vev{\bar1\bar2\bar3\bar4}$\\ 
\hline
$[{\bm{\rm W}}'_n]_{i}{}^{j}$ &  
$\vev{\bar1\bar2\bar3\bar4}$&  
$\vev{\bar1\bar2\bar3\bar4}$&  
$\vev{\bar1\bar2\bar3\bar4}$&
$\vev{\bar1\bar2\bar3\bar4}$\\ 
\hline
\end{tabular}
\end{center}
\caption{$\bm\rR$-matrix summary table for Tensor supermultiplet}
\label{tab6a}
\end{table}

\begin{table}[!ht]
\begin{center}
\begin{tabular}{|m{6.2em}|m{2.2cm}|m{2.2cm}|m{2.2cm}|m{2.2cm}|}
\hline
Multiplet $~~~~~$ &  $n=1$& $n=2$& $n=3$& $n=4$\\ 
\hline
${\rm {Chiral}} ~[{\bm{\rm W}}_n]_{i}{}^{j}$&  
-$\mathbf{I}_4$&  
-$\mathbf{I}_4$&  
-$\mathbf{I}_4$&
-$\mathbf{I}_4$\\ 
\hline
${\rm {Vector}} ~[{\bm{\rm W}}_n]_{i}{}^{j}$&  
-$\mathbf{I}_4$&  
-$\mathbf{I}_4$&  
-$\mathbf{I}_4$&
-$\mathbf{I}_4$\\ 
\hline
${\rm {Tensor}} ~[{\bm{\rm W}}_n]_{i}{}^{j}$ &  
-$\mathbf{I}_4$&  
-$\mathbf{I}_4$&  
-$\mathbf{I}_4$&
-$\mathbf{I}_4$\\ 
\hline
${\rm {CLS}} ~[{\bm{\rm W}}_n]_{i}{}^{j}$ &  
-$\mathbf{I}_{12}$&  
-$\mathbf{I}_{12}$&  
-$\mathbf{I}_{12}$&
-$\mathbf{I}_{12}$\\ 
\hline
\end{tabular}
\end{center}
\caption{$\bm{\rm W}$-matrix summary table for the Minimal 4-Color Supermultiplets.}
\label{tabZ}
\end{table}

When we compare bosonic holoraumy matrix and the $\bm{\rm X}$-matrix, based on the definition of ${\bm{\cal B}}_{w_I\,w_J}$ in Eq. (\ref{e:BR1}), Chapter 2:

\begin{eqnarray}
\bm{\rm{X_1}}^{\, ({\cal R})}&={\bm{\cal B}}_{{w_2}\,{w_1}}^{\, ({\cal R})}&=-{\bm{\cal B}}_{{w_1}\,{w_2}}^{\, ({\cal R})}
\nonumber\\
\bm{\rm{X_2}}^{\,({\cal R})}&= 
{\bm{\cal B}}_{{w_3}\,{w_2}}^{\,({\cal R})}&=-{\bm{\cal B}}_{{w_2}\,{w_3}}^{\, ({\cal R})} \nonumber\\
\bm{\rm{X_3}}^{\, ({\cal R})}&= {\bm{\cal B}}_{{w_4}\,{w_3}}^{\,({\cal R})}
&=-{\bm{\cal B}}_{{w_3}\,{w_4}}^{\,({\cal R})}
\nonumber\\
\bm{\rm{X_4}}^{\, ({\cal R})}&=
{\bm{\cal B}}_{{w_1}\,{w_4}}^{\, ({\cal R})}&
\label{eq41}
\end{eqnarray}

The results show that the ${\bm{\cal B}}$-matrix values on Table \ref{t:0braneXZX}, coincide with the $\bm{\rm X}$-matrix values in Tables \ref{tab1}, \ref{tab3a}, and \ref{tab5a}. This confirms that the definition of bosonic holoraumy is correct, and the values are consistent with the raw inverse matrix calculations, $\bm{\rm X}$-matrix.

\subsection{4D $\mathcal{N}$=1 Chiral supermultiplet ${\bm \rL}$-matrix}
In this chapter we will show how to contruct the values of Table \ref{tab1}-\ref{tabZ} and there are regular matrix multiplication rules between the ${\bm \rL}$-matrices. 
We will use the 4D $\mathcal{N}$=1 Chiral multiplet as a toy model.
Firstly, we gave the ${\bm \rL}$-matrices in a
compact form in Table 1 which using our conventions led to
the ${\bm \rL}$-matrices for the chiral multiplet given as, \cite{3}

\begin{equation}
[{\bm \rL}_1]_{i}{}^{\hat k}=
\begin{pmatrix}
1 & 0 & 0 & 0 \\
0 & 0 & 0 & -1 \\
0 & 1 & 0 & 0 \\
0 & 0 & -1 & 0 \\
\end{pmatrix},\quad\quad
[{\bm \rL}_2]_{i}{}^{\hat k}=
\begin{pmatrix}
0 & 1 & 0 & 0 \\
0 & 0 & 1 & 0 \\
-1 & 0 & 0 & 0 \\
0 & 0 & 0 & -1 \\
\end{pmatrix}
\end{equation}

\begin{equation}
[{\bm \rL}_3]_{i}{}^{\hat k}=
\begin{pmatrix}
0 & 0 & 1 & 0 \\
0 & -1 & 0 & 0 \\
0 & 0 & 0 & -1 \\
1 & 0 & 0 & 0 \\
\end{pmatrix}
,\quad\quad
[{\bm \rL}_4]_{i}{}^{\hat k}=
\begin{pmatrix}
0 & 0 & 0 & 1 \\
1 & 0 & 0 & 0 \\
0 & 0 & 1 & 0 \\
0 & 1 & 0 & 0 \\
\end{pmatrix}
\end{equation}

We need to find matrices $[{\bm{\rm X}}_n]_{i}{}^{k}$ which satisfy the
equations
\begin{equation}
[{\bm \rL}_{n+1}]_{i}{}^{\hat k}=[{\bm{\rm X}}_n]_{i}{}^{j}[{\bm \rL}_n]_{j}{}^{\hat k},\quad n\in\{1,2,3,4\}
\label{eq3}
\end{equation}

where here if $n+1=5$ then $n+1$ count as 1.
The similarity of this
equation to the one given in Eq. (\ref{e:BR4}) insures that such matrices exist!  In fact the matrices $[{\bm{\rm X}}_n]_{i}{}^{k}$ must 
correspond to some elements of the bosonic holoraumy matrix.  Explicitly we
find
\begin{equation}
[{\bm{\rm X}}_1]_{i}{}^{k}=
\begin{pmatrix}
0 & 0 & 1 & 0 \\
0 & 0 & 0 & -1 \\
-1 & 0 & 0 & 0 \\
0 & 1 & 0 & 0 \\
\end{pmatrix}
,\quad\quad
[{\bm{\rm X}}_2]_{i}{}^{k}=
\begin{pmatrix}
0 & 1 & 0 & 0 \\
-1 & 0 & 0 & 0 \\
0 & 0 & 0 & 1 \\
0 & 0 & -1 & 0 \\
\end{pmatrix}
\end{equation}

\begin{equation}
[{\bm{\rm X}}_3]_{i}{}^{k}=
\begin{pmatrix}
0 & 0 & -1 & 0 \\
0 & 0 & 0 & 1 \\
1 & 0 & 0 & 0 \\
0 & -1 & 0 & 0 \\
\end{pmatrix}
,\quad\quad
[{\bm{\rm X}}_4]_{i}{}^{k}=
\begin{pmatrix}
0 & 1 & 0 & 0 \\
-1 & 0 & 0 & 0 \\
0 & 0 & 0 & 1 \\
0 & 0 & -1 & 0 \\
\end{pmatrix}
\end{equation}

Now we repeat the same process {\it {but}} starting with the $[{\bm{\rm X}}_n]_{i}{}^{k}$ matrices to obtain matrices $[{\bm{\rm Y}}_n]_{i}{}^{k}$ which satisfy

\begin{equation}
[{\bm{\rm X}}_{n+1}]_{i}{}^{k}=[{\bm{\rm Y}}_n]_{i}{}^{j}[{\bm{\rm X}}_n]_{j}{}^{k},\quad n\in\{1,2,3,4\}
\label{eq6}
\end{equation}

where as before if $n+1=5$ then $n+1$ count as 1.  

We find the solutions are given by
\begin{equation}
[{\bm{\rm Y}}_1]_{i}{}^{k}=
\begin{pmatrix}
0 & 0 & 0 & 1 \\
0 & 0 & 1 & 0 \\
0 & -1 & 0 & 0 \\
-1 & 0 & 0 & 0 \\
\end{pmatrix}
,\quad\quad
[{\bm{\rm Y}}_2]_{i}{}^{k}=
\begin{pmatrix}
0 & 0 & 0 & 1 \\
0 & 0 & 1 & 0 \\
0 & -1 & 0 & 0 \\
-1 & 0 & 0 & 0 \\
\end{pmatrix}
\end{equation}

\begin{equation}
[{\bm{\rm Y}}_3]_{i}{}^{k}=
\begin{pmatrix}
0 & 0 & 0 & -1 \\
0 & 0 & -1 & 0 \\
0 & 1 & 0 & 0 \\
1 & 0 & 0 & 0 \\
\end{pmatrix}
,\quad\quad
[{\bm{\rm Y}}_4]_{i}{}^{k}=
\begin{pmatrix}
0 & 0 & 0 & -1 \\
0 & 0 & -1 & 0 \\
0 & 1 & 0 & 0 \\
1 & 0 & 0 & 0 \\
\end{pmatrix}
\end{equation}

Now we repeat the same process {\it {but}} starting with the
$[{\bm{\rm Y}}_n]_{i}{}^{k}$ matrices to find a set of matrices
$[{\bm{\rm Z}}_n]_{i}{}^{k}$ to satisfy

\begin{equation}
[{\bm{\rm Y}}_{n+1}]_{i}{}^{k}=[{\bm{\rm Z}}_n]_{i}{}^{j}[{\bm{\rm Y}}_n]_{j}{}^{k},\quad n\in\{1,2,3,4\}
\label{eq3.9}
\end{equation}
where once more if $n+1=5$ then $n+1$ count as 1.

Our solutions take the forms of the $[{\bm{\rm Z}}_n]_{i}{}^{k}$ matrices are
\begin{equation}
[{\bm{\rm Z}}_{1,3}]_{i}{}^{k}=[\mathbf{I}_{4}]_{i}{}^{k},\quad [{\bm{\rm Z}}_{2,4}]_{i}{}^{k}=-
[\mathbf{I}_{4}]_{i}{}^{k} 
\end{equation}

Finally we define $\bm{\rm W}$-matrix as, (here if $n+1=5$ then $n+1$ count as 1)
\begin{equation}
[{\bm{\rm Z}}_{n+1}]_{i}{}^{k}=[{\bm{\rm W}}_n]_{i}{}^{j}[{\bm{\rm Z}}_n]_{j}{}^{k},\quad n\in\{1,2,3,4\}
\label{eq4-11}
\end{equation}

Then $[{\bm{\rm W}}_n]_{i}{}^{k}=-[\mathbf{I}_{4}]_{i}{}^{k}$.

Now we have all the necessary basic elements to reconstruct the ${\bm{\rL}}$-matrix. To simplify these relationships, we can start by decomposing the ${\bm{\rm X}}$- and ${\bm{\rm Y}}$- matrices using the Pauli matrix form.
Here we can simplify the $[{\bm{\rm X}}_{n}]_{i}{}^{k}$ and $[{\bm{\rm Y}}_{n}]_{i}{}^{k}$ by pauli matrix

\begin{equation}
[{\bm{\rm X}}_1]_{i}{}^{k}=[i\sigma_2\otimes \sigma_3]_{i}{}^{k}
\end{equation}

\begin{equation}
[{\bm{\rm Y}}_1]_{i}{}^{k}=[i\sigma_2\otimes \sigma_1]_{i}{}^{k}
\end{equation}

Therefore, Eq. (\ref{eq313}) to Eq. (\ref{eq315}) is hold.

\begin{equation}
[{\bm{\rm X}}_1]_{i}{}^{j}[{\bm{\rm Y}}_1]_{j}{}^{k}=-[(\sigma_2\cdot\sigma_2)\otimes(\sigma_3\cdot\sigma_1)]_{i}{}^{k}=-[i\mathbf{I}_2\otimes\sigma_2]_{i}{}^{k}
\label{eq313}
\end{equation}

\begin{equation}
[[{\bm{\rm X}}_1]_{i}{}^{k}]^4=[[{\bm{\rm Y}}_1]_{i}{}^{k}]^4=[\mathbf{I}_4]_{i}{}^{k}
\end{equation}

\begin{equation}
[[{\bm{\rm X}}_1]_{i}{}^{k}]^2=[[{\bm{\rm Y}}_1]_{i}{}^{k}]^2=[[\bm{\rm X_1Y_1}]_{i}{}^{k}]^2=-[\mathbf{I}_4]_{i}{}^{k}
\label{eq315}
\end{equation}

\begin{eqnarray}
[{\bm{\rm X}}_1]_{i}{}^{k}&=&[{\bm{\rm X}}_1]_{i}{}^{k}\label{eq316}\\
\
[{\bm{\rm X}}_2]_{i}{}^{k}&=&[\bm{\rm Y_1X_1}]_{i}{}^{k}\\
\
[{\bm{\rm X}}_3]_{i}{}^{k}&=&[\bm{\rm Y_2X_2}]_{i}{}^{k}
=[\bm{\rm Z_1 Y_1 Y_1 X_1}]_{i}{}^{k}
=[\bm{\rm Y_1^2X_1}]_{i}{}^{k}=-[{\bm{\rm X}}_1]_{i}{}^{k}\\
\
[{\bm{\rm X}}_4]_{i}{}^{k}&=&[\bm{\rm Y_3X_3}]_{i}{}^{k}
=[\bm{\rm Z_2 Y_2}]_{i}{}^{j}[-\bm{\rm X_1}]_{j}{}^{k}
=[-\bm{\rm Y}_1]_{i}{}^{j}[-\bm{\rm X_1}]_{j}{}^{k}
=[\bm{\rm Y_1X_1}]_{i}{}^{k}
\label{eq319}
\end{eqnarray}

Therefore we can use Eq. (\ref{eq316}) to Eq. (\ref{eq319}) on L-matrix, we can obtain 

\begin{eqnarray}
[{\bm \rL}_1]_{i}{}^{\hat k}&=&[{\bm \rL}_1]_{i}{}^{\hat k}\nonumber\\
\
[{\bm \rL}_2]_{i}{}^{\hat k}&=&[{\bm{\rm X}}_1]_{i}{}^{\hat j}[{\bm \rL}_1]_{j}{}^{\hat k}\nonumber\\
\
[{\bm \rL}_3]_{i}{}^{\hat k}&=&[{\bm{\rm X}}_2]_{i}{}^{\hat j}[{\bm \rL}_2]_{j}{}^{\hat k}=-[\bm{\rm Y_1L_1}]_{j}{}^{\hat k}\nonumber\\
\
[{\bm \rL}_4]_{i}{}^{\hat k}&=&[{\bm{\rm X}}_3]_{i}{}^{\hat j}[{\bm \rL}_3]_{j}{}^{\hat k}=[\bm{\rm X_1Y_1L_1}]_{j}{}^{\hat k}
\label{eq8.24}
\end{eqnarray}

\subsection{4D $\mathcal{N}$=1 Chiral supermultiplet $\bm\rR$-matrix}

We can follow a similar process on the $\bm\rR$-matrix to demonstrate the regular matrix multiplication rules. The $\bm\rR$-matrix in chiral multiplet is given as, \cite{3}

\begin{equation}
[{\bm \rR}_1]_{\hat i}{}^{k}=
\begin{pmatrix}
1 & 0 & 0 & 0 \\
0 & 0 & 1 & 0 \\
0 & 0 & 0 & -1 \\
0 & -1 & 0 & 0 \\
\end{pmatrix}
,\quad\quad
[{\bm \rR}_2]_{\hat i}{}^{k}=
\begin{pmatrix}
0 & 0 & -1 & 0 \\
1 & 0 & 0 & 0 \\
0 & 1 & 0 & 0 \\
0 & 0 & 0 & -1 \\
\end{pmatrix}
\end{equation}

\begin{equation}
[{\bm{\rR}}_3]_{\hat i}{}^{k}=
\begin{pmatrix}
0 & 0 & 0 & 1 \\
0 & -1 & 0 & 0 \\
1 & 0 & 0 & 0 \\
0 & 0 & -1 & 0 \\
\end{pmatrix}
,\quad\quad
[{\bm{\rR}}_4]_{\hat i}{}^{k}=
\begin{pmatrix}
0 & 1 & 0 & 0 \\
0 & 0 & 0 & 1 \\
0 & 0 & 1 & 0 \\
1 & 0 & 0 & 0 \\
\end{pmatrix}
\end{equation}

And we define the matrix $[{\bm{\rm X'}}_n]_{\hat i}{}^{\hat k}$ which satisfy

\begin{equation}
[{\bm \rR}_{n+1}]_{\hat i}{}^{k}=[\bm{\rm X'}_n]_{\hat i}{}^{\hat j}[\bm{\rm R}_n]_{\hat i}{}^{k},\quad n\in\{1,2,3,4\}
\label{eq23}
\end{equation}
\begin{equation}
[{\bm{\rm X}}'_1]_{\hat i}{}^{\hat k}=
\begin{pmatrix}
0 & -1 & 0 & 0 \\
1 & 0 & 0 & 0 \\
0 & 0 & 0 & -1 \\
0 & 0 & 1 & 0 \\
\end{pmatrix}
,\quad\quad
[{\bm{\rm X}}'_2]_{\hat i}{}^{\hat k}=
\begin{pmatrix}
0 & 0 & 0 & -1 \\
0 & 0 & -1 & 0 \\
0 & 1 & 0 & 0 \\
1 & 0 & 0 & 0 \\
\end{pmatrix}
\end{equation}

\begin{equation}
[{\bm{\rm X}}'_3]_{\hat i}{}^{\hat k}=
\begin{pmatrix}
0 & -1 & 0 & 0 \\
1 & 0 & 0 & 0 \\
0 & 0 & 0 & -1 \\
0 & 0 & 1 & 0 \\
\end{pmatrix}
,\quad\quad
[{\bm{\rm X}}'_4]_{\hat i}{}^{\hat k}=
\begin{pmatrix}
0 & 0 & 0 & 1 \\
0 & 0 & 1 & 0 \\
0 & -1 & 0 & 0 \\
-1 & 0 & 0 & 0 \\
\end{pmatrix}
\end{equation}

And by using the same logic we can repeat same process for $[{\bm{\rm X'}}_n]_{\hat i}{}^{\hat k}$ and can obtain the matrix $[{\bm{\rm Y'}}_n]_{\hat i}{}^{\hat k}$ which satisfy

\begin{equation}
[{\bm{\rm X}}'_{n+1}]_{\hat i}{}^{\hat k}=[{\bm{\rm Y}}'_n]_{\hat i}{}^{\hat j}[{\bm{\rm X}}'_n]_{\hat j}{}^{\hat k},\quad n\in\{1,2,3,4\}
\label{eq26}
\end{equation}

\begin{equation}
[{\bm{\rm Y}}'_1]_{\hat i}{}^{\hat k}=
\begin{pmatrix}
0 & 0 & 1 & 0 \\
0 & 0 & 0 & -1 \\
-1 & 0 & 0 & 0 \\
0 & 1 & 0 & 0 \\
\end{pmatrix}
,\quad\quad
[{\bm{\rm Y}}'_2]_{\hat i}{}^{\hat k}=
\begin{pmatrix}
0 & 0 & -1 & 0 \\
0 & 0 & 0 & 1 \\
1 & 0 & 0 & 0 \\
0 & -1 & 0 & 0 \\
\end{pmatrix}
\end{equation}

\begin{equation}
[{\bm{\rm Y}}'_3]_{\hat i}{}^{\hat k}=
\begin{pmatrix}
0 & 0 & -1 & 0 \\
0 & 0 & 0 & 1 \\
1 & 0 & 0 & 0 \\
0 & -1 & 0 & 0 \\
\end{pmatrix}
,\quad\quad
[{\bm{\rm Y}}'_4]_{\hat i}{}^{\hat k}=
\begin{pmatrix}
0 & 0 & 1 & 0 \\
0 & 0 & 0 & -1 \\
-1 & 0 & 0 & 0 \\
0 & 1 & 0 & 0 \\
\end{pmatrix}
\end{equation}

Finally when we see the relation between $[{\bm{\rm Y'}}_n]_{\hat i}{}^{\hat k}$ then we can obtain the matrix $[{\bm{\rm Z'}}_n]_{\hat i}{}^{\hat k}$ which satisfy

\begin{equation}
[{\bm{\rm Y}}'_{n+1}]_{\hat i}{}^{\hat k}=[{\bm{\rm Z}}'_n]_{\hat i}{}^{\hat j}[{\bm{\rm Y}}'_n]_{\hat j}{}^{\hat k},\quad n\in\{1,2,3,4\}
\label{eq29}
\end{equation}

\begin{equation}
[{\bm{\rm Z}}'_{1,3}]_{\hat i}{}^{\hat k}=-[\mathbf{I}_4]_{\hat i}{}^{\hat k},\quad [{\bm{\rm Z}}'_{2,4}]_{\hat i}{}^{\hat k}=[\mathbf{I}_4]_{\hat i}{}^{\hat k}
\end{equation}

Finally we define $\bm{\rm W}'$-matrix as, (here if $n+1=5$ then $n+1$ count as 1)
\begin{equation}
[{\bm{\rm Z}}'_{n+1}]_{i}{}^{k}=[{\bm{\rm W}}'_n]_{i}{}^{j}[{\bm{\rm Z}}'_n]_{j}{}^{k},\quad n\in\{1,2,3,4\}
\label{eq4-32}
\end{equation}

Then $[{\bm{\rm W}}'_n]_{i}{}^{k}=-[\mathbf{I}_{4}]_{i}{}^{k}$.

In the same way in $\bm\rL$-matrix case, here we can simplify the $[{\bm{\rm X}}'_{1}]_{\hat i}{}^{\hat k}$ and $[{\bm{\rm Y}}'_{1}]_{\hat i}{}^{\hat k}$ matrix with pauli matrix.

\begin{equation}
[{\bm{\rm X}}'_{1}]_{\hat i}{}^{\hat k}=-[i\mathbf{I}_2\otimes \sigma_2]_{\hat i}{}^{\hat k}=[{\bm{\rm X_1Y_1}}]_{\hat i}{}^{\hat k}
\end{equation}

\begin{equation}
[{\bm{\rm Y}}'_{1}]_{\hat i}{}^{\hat k}=[i\sigma_2\otimes \sigma_3]_{\hat i}{}^{\hat k}=[{\bm{\rm X_1}}]_{\hat i}{}^{\hat k}
\end{equation}

Therefore, Eq. (\ref{eq33}) to Eq. (\ref{eq35}) is hold

\begin{equation}
[{\bm{\rm X'_{1}Y'_1}}]_{\hat i}{}^{\hat k}=[(\mathbf{I}_2\cdot\sigma_2)\otimes(\sigma_2\cdot\sigma_3)]_{\hat i}{}^{\hat k}=[i\sigma_2\otimes\sigma_1]_{\hat i}{}^{\hat k}=[{\bm{\rm Y}}_1]_{\hat i}{}^{\hat k}
\label{eq33}
\end{equation}

\begin{equation}
[[{\bm{\rm X}}'_{1}]_{\hat i}{}^{\hat k}]^4=[[{\bm{\rm Y}}'_1]_{\hat i}{}^{\hat k}]^4=[\mathbf{I}_4]_{\hat i}{}^{\hat k}
\end{equation}

\begin{equation}
[[{\bm{\rm X}}'_{1}]_{i}{}^{k}]^2=[[{\bm{\rm Y}}'_1]_{\hat i}{}^{\hat k}]^2=[[ {\bm{\rm X'_1Y'_1}}]_{i}{}^{k}]^2=-[\mathbf{I}_4]_{\hat i}{}^{\hat k}
\label{eq35}
\end{equation}

\begin{eqnarray}
[{\bm{\rm X}}'_1]_{\hat i}{}^{\hat k}&=&[{\bm{\rm X}}'_1]_{\hat i}{}^{\hat k}
\label{eq36}\\
\
[{\bm{\rm X}}'_2]_{\hat i}{}^{\hat k}&=&[{\bm{\rm Y'_1X'_1}}]_{\hat i}{}^{\hat k}\\
\
[{\bm{\rm X}}'_3]_{\hat i}{}^{\hat k}&=&[{\bm{\rm Y'_2X'_2}}]_{\hat i}{}^{\hat k}
=[{\bm{\rm Z'_1Y'_1Y'_1X'_1}}]_{\hat i}{}^{\hat k}
=[{\bm{\rm X}}'_1]_{\hat i}{}^{\hat k}\\
\
[{\bm{\rm X}}'_4]_{\hat i}{}^{\hat k}&=&[{\bm{\rm Y'_3X'_3}}]_{\hat i}{}^{\hat k}
=-[{\bm{\rm Z'_2Y'_2Y'_2X'_2}}]_{\hat i}{}^{\hat k}
=-[{\bm{\rm Y'_1X'_1}}]_{\hat i}{}^{\hat k}
\label{eq39}
\end{eqnarray}

Therefore we can use Eq. (\ref{eq36}) to Eq. (\ref{eq39}) on $\bm\rR$-matrix, we can obtain

\begin{eqnarray}
[{\bm \rR}_1]_{\hat i}{}^{ k}&=&[{\bm \rR}_1]_{\hat i}{}^{k}\nonumber\\
\
[{\bm \rR}_2]_{\hat i}{}^{k}&=&[{\bm{\rm X'_1R_1}}]_{\hat i}{}^{k}=[{\bm{\rm X_1Y_1R_1}}]_{\hat i}{}^{k}\nonumber\\
\
[{\bm{\rR}}_3]_{\hat i}{}^{k}&=&[{\bm{\rm X'_2R_2}}]_{\hat i}{}^{k}
=[{\bm{\rm Y'_1X'_1X'_1R_1}}]_{\hat i}{}^{k}
=-[{\bm{\rm Y'_1R_1}}]_{\hat i}{}^{k}
=-[{\bm{\rm X_1R_1}}]_{\hat i}{}^{k}\nonumber\\
\
[{\bm{\rR}}_4]_{\hat i}{}^{k}&=&[{\bm{\rm X'_3R_3}}]_{\hat i}{}^{k}
=[{\bm{\rm X'_1X'_2R_2}}]_{\hat i}{}^{k}
=[{\bm{\rm X'_1Y'_1X'_1X'_1R_1}}]_{\hat i}{}^{k}
=-[{\bm{\rm X'_1Y'_1R_1}}]_{\hat i}{}^{k}
=-[{\bm{\rm Y_1R_1}}]_{\hat i}{}^{k}\nonumber\\
\
\label{eq8.47}
\end{eqnarray}

\subsection{Hopping between $\bm\rL$ and $\bm\rR$-matrix for 4D $\mathcal{N}=1$ Chiral supermultiplet}
If we summarize all the results from Eq. (\ref{eq8.24}) and Eq. (\ref{eq8.47}), we can obtain the following equation:

\begin{eqnarray}
[{\bm \rL}_2]_{i}{}^{\hat k}&=&[{\bm{\rm X_1L_1}}]_{i}{}^{\hat k}\nonumber\\
\
[{\bm \rL}_3]_{i}{}^{\hat k}&=&-[{\bm{\rm Y_1L_1}}]_{i}{}^{\hat k}\nonumber\\
\
[{\bm \rL}_4]_{i}{}^{\hat k}&=&[{\bm{\rm X_1Y_1}}]_{i}{}^{j}[{\bm{\rm L_1}}]_{j}{}^{\hat k}=-[{\bm{\rm X_1L_3}}]_{i}{}^{\hat k}\nonumber\\
\
[{\bm \rL}_1]_{i}{}^{\hat k}&=&[{\bm{\rm Y_1L_3}}]_{i}{}^{\hat k}
\end{eqnarray}

For $\bm{\rR}$-matrix we can derive the similar relation with $\bm{\rL}$-matrix relation:

\begin{eqnarray}
[{\bm \rR}_2]_{\hat i}{}^{k}&=&[{\bm{\rm X_1Y_1}}]_{\hat i}{}^{\hat j}[{\bm \rR}_1]_{\hat j}{}^{k}=[{\bm{\rm X_1R_4}}]_{\hat i}{}^{k}\nonumber\\
\
[{\bm{\rR}}_3]_{\hat i}{}^{k}&=&-[{\bm{\rm X_1R_1}}]_{\hat i}{}^{k}\nonumber\\
\
[{\bm{\rR}}_4]_{\hat i}{}^{k}&=&-[{\bm{\rm Y_1R_1}}]_{\hat i}{}^{k}\nonumber\\
\
[{\bm \rR}_1]_{\hat i}{}^{k}&=&[{\bm{\rm Y_1R_4}}]_{\hat i}{}^{k}
\end{eqnarray}

\newpage
\section{4D $\mathcal{N}$=1 Complex linear supermultiplet}
As a final step, we will now use the same logic that we used for the minimal supermultiplet to calculate the hopping matrix, for the CLS supermultiplets case.
The complex linear supermultiplet (CLS) contains scalar $K$, pseudoscalar $L$, Majorana spinor $\zeta_a$ and auxiliary scalar $M$, auxiliary pseudoscalar $N$, auxiliary vector $V_\mu$, auxiliary axial-vector $U_\mu$, and auxiliary Majorana spinors $\rho_a$ and $\beta_a$.
According to \cite{5} Table. 4, we can draw the Adinkra between bosonic and fermionic fields. \footnote{The more detail field definition is referenced in Eq. (3.16) on \cite{5}}

\begin{figure}[H]
    \centering
    \includegraphics[scale=0.5]{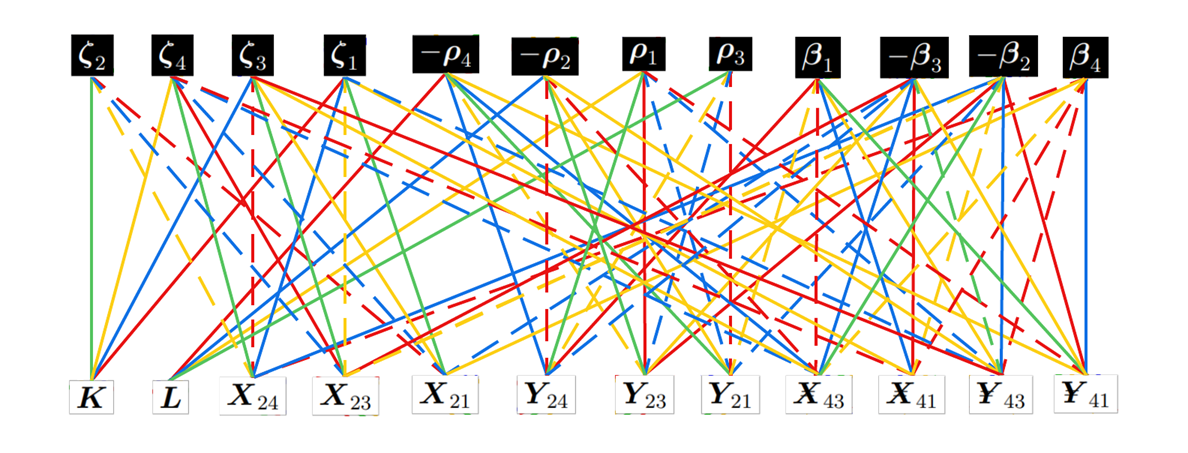}
    \caption{Valise Adinkra for CLS}
    \label{fig3}
\end{figure}
We can draw the three separate adinkras for the CLS based on calculated $\bm\rL$- and $\bm\rR$-matrices.
Here, red, green blue, yellow indicate $\rm D_1, D_2, D_3, D_4$ respectively. A dotted line represents a negative relationship and a solid line represents a positive relationship between the two fields. 
(Appendix C there are standard CLS Adinkra block diagronal forms of $\bm\rL$- and $\bm\rR$-matrix.\cite{2,38})

\subsection{4D $\mathcal{N}$=1 CLS $\bm\rL$-matrix}

Based on Fig. \ref{fig3} we can construct the $\bm\rL$-matrix for CLS.

\[
[{\bm{\rm L}_1}]_{i}{}^{\hat k}  = \left[\begin{array}{cccccccc|cccc}
    0 & 0 & 0 & 1 & 0 & 0 & 0 & 0 & 0 & 0 & 0 & 0 \\ 
    0 & 0 & 0 & 0 & 0 & 0 & 0 & 1 & 0 & 0 & 0 & 0 \\
    \hline
    0 & 0 & -1 & 0 & 0 & 0 & 0 & 0 & 0 & 0 & -1 & 0 \\ 
    0 & 1 & 0 & 0 & 0 & 0 & 0 & 0 & 0 & 1 & 0 & 0 \\ 
    -1 & 0 & 0 & 0 & 0 & 0 & 0 & 0 & 0 & 0 & 0 & 0 \\ 
    0 & 0 & 0 & 0 & 0 & 0 & -1 & 0 & 1 & 0 & 0 & 0 \\ 
    0 & 0 & 0 & 0 & 0 & 1 & 0 & 0 & 0 & 0 & 0 & 1 \\ 
    0 & 0 & 0 & 0 & -1 & 0 & 0 & 0 & 0 & 0 & 0 & 0 \\ 
    \hline
    0 & 0 & 0 & 0 & 0 & 0 & 0 & 0 & -1 & 0 & 0 & 0 \\ 
    0 & 0 & 0 & 0 & 0 & 0 & 0 & 0 & 0 & 1 & 0 & 0 \\ 
    0 & 0 & 0 & 0 & 0 & 0 & 0 & 0 & 0 & 0 & -1 & 0 \\ 
    0 & 0 & 0 & 0 & 0 & 0 & 0 & 0 & 0 & 0 & 0 & 1
\numberthis\end{array}\right] 
\label{eq6.1}\]

\[
[{\bm{\rm L}_2}]_{i}{}^{\hat k}  = \left[\begin{array}{cccccccc|cccc}
    1 & 0 & 0 & 0 & 0 & 0 & 0 & 0 & 0 & 0 & 0 & 0 \\ 
    0 & 0 & 0 & 0 & 1 & 0 & 0 & 0 & 0 & 0 & 0 & 0 \\ 
    \hline
    0 & 1 & 0 & 0 & 0 & 0 & 0 & 0 & 0 & 0 & 0 & 0 \\ 
    0 & 0 & 1 & 0 & 0 & 0 & 0 & 0 & 0 & 0 & 0 & 0 \\ 
    0 & 0 & 0 & 1 & 0 & 0 & 0 & 0 & 0 & 0 & 0 & 0 \\ 
    0 & 0 & 0 & 0 & 0 & 1 & 0 & 0 & 0 & 0 & 0 & 0 \\ 
    0 & 0 & 0 & 0 & 0 & 0 & 1 & 0 & 0 & 0 & 0 & 0 \\ 
    0 & 0 & 0 & 0 & 0 & 0 & 0 & 1 & 0 & 0 & 0 & 0 \\ 
    \hline
    0 & 0 & 0 & 0 & 0 & 0 & 0 & 0 & 0 & 0 & 0 & 1 \\ 
    0 & 0 & 0 & 0 & 0 & 0 & 0 & 0 & 0 & 0 & -1 & 0 \\ 
    0 & 0 & 0 & 0 & 0 & 0 & 0 & 0 & 0 & -1 & 0 & 0 \\ 
    0 & 0 & 0 & 0 & 0 & 0 & 0 & 0 & 1 & 0 & 0 & 0
\numberthis\end{array}\right] \]

\[
[{\bm{\rm L}_3}]_{i}{}^{\hat k}  = \left[\begin{array}{cccccccc|cccc}
    0 & 0 & 1 & 0 & 0 & 0 & 0 & 0 & 0 & 0 & 0 & 0 \\ 
    0 & 0 & 0 & 0 & 0 & 0 & 1 & 0 & 0 & 0 & 0 & 0 \\
    \hline
    0 & 0 & 0 & 1 & 0 & 0 & 0 & 0 & 0 & 0 & 0 & 1 \\ 
    -1 & 0 & 0 & 0 & 0 & 0 & 0 & 0 & 0 & 0 & 0 & 0 \\ 
    0 & -1 & 0 & 0 & 0 & 0 & 0 & 0 & 0 & -1 & 0 & 0 \\ 
    0 & 0 & 0 & 0 & 0 & 0 & 0 & -1 & 0 & -1 & 0 & 0 \\ 
    0 & 0 & 0 & 0 & -1 & 0 & 0 & 0 & 0 & 0 & 0 & 0 \\ 
    0 & 0 & 0 & 0 & 0 & -1 & 0 & 0 & 0 & 0 & 0 & -1 \\ 
    \hline
    0 & 0 & 0 & 0 & 0 & 0 & 0 & 0 & 0 & 1 & 0 & 0 \\ 
    0 & 0 & 0 & 0 & 0 & 0 & 0 & 0 & 1 & 0 & 0 & 0 \\ 
    0 & 0 & 0 & 0 & 0 & 0 & 0 & 0 & 0 & 0 & 0 & 1 \\ 
    0 & 0 & 0 & 0 & 0 & 0 & 0 & 0 & 0 & 0 & 1 & 0
\numberthis\end{array}\right] \]

\[
[{\bm{\rm L}_4}]_{i}{}^{\hat k}  = \left[\begin{array}{cccccccc|cccc}
    0 & 1 & 0 & 0 & 0 & 0 & 0 & 0 & 0 & 0 & 0 & 0 \\ 
    0 & 0 & 0 & 0 & 0 & 1 & 0 & 0 & 0 & 0 & 0 & 0 \\
    \hline
    -1 & 0 & 0 & 0 & 0 & 0 & 0 & 0 & 0 & 0 & 0 & 0 \\ 
    0 & 0 & 0 & -1 & 0 & 0 & 0 & 0 & 0 & 0 & 0 & -1 \\ 
    0 & 0 & 1 & 0 & 0 & 0 & 0 & 0 & 0 & 0 & 1 & 0 \\ 
    0 & 0 & 0 & 0 & -1 & 0 & 0 & 0 & 0 & 0 & 0 & 0 \\ 
    0 & 0 & 0 & 0 & 0 & 0 & 0 & -1 & 0 & 1 & 0 & 0 \\ 
    0 & 0 & 0 & 0 & 0 & 0 & 1 & 0 & -1 & 0 & 0 & 0 \\ 
    \hline
    0 & 0 & 0 & 0 & 0 & 0 & 0 & 0 & 0 & 0 & -1 & 0 \\ 
    0 & 0 & 0 & 0 & 0 & 0 & 0 & 0 & 0 & 0 & 0 & -1 \\ 
    0 & 0 & 0 & 0 & 0 & 0 & 0 & 0 & 1 & 0 & 0 & 0 \\ 
    0 & 0 & 0 & 0 & 0 & 0 & 0 & 0 & 0 & 1 & 0 & 0
\numberthis\end{array}\right] 
\label{eq6.4}\]

\subsubsection{$\bm{\rm X}$-matrix}

Based on Eq. (\ref{eq3}) we can construct the $\bm{\rm{X}}$-matrix for CLS.

\[[{\bm{\rm X}_1}]_{i}{}^{k}= 
\left[\begin{array}{cccccccc|cccc}
 0 & 0 & 0 & 0 & -1 & 0 & 0 & 0 & 0 & 0 & 0 & 0 \\
 0 & 0 & 0 & 0 & 0 & 0 & 0 & -1 & 0 & 0 & 0 & 0 \\
  \hline
 0 & 0 & 0 & 1 & 0 & 0 & 0 & 0 & 0 & -1 & 0 & 0 \\
 0 & 0 & -1 & 0 & 0 & 0 & 0 & 0 & 0 & 0 & 1 & 0 \\
 1 & 0 & 0 & 0 & 0 & 0 & 0 & 0 & 0 & 0 & 0 & 0 \\
 0 & 0 & 0 & 0 & 0 & 0 & 1 & 0 & 0 & 0 & 0 & -1 \\
 0 & 0 & 0 & 0 & 0 & -1 & 0 & 0 & -1 & 0 & 0 & 0 \\
 0 & 1 & 0 & 0 & 0 & 0 & 0 & 0 & 0 & 0 & 0 & 0 \\
  \hline
 0 & 0 & 0 & 0 & 0 & 0 & 0 & 0 & 0 & 0 & 0 & 1 \\
 0 & 0 & 0 & 0 & 0 & 0 & 0 & 0 & 0 & 0 & 1 & 0 \\
 0 & 0 & 0 & 0 & 0 & 0 & 0 & 0 & 0 & -1 & 0 & 0 \\
 0 & 0 & 0 & 0 & 0 & 0 & 0 & 0 & -1 & 0 & 0 & 0
\numberthis\end{array}\right] \label{eq55}\]

\[[{\bm{\rm X}_2}]_{i}{}^{k}= 
\left[\begin{array}{cccccccc|cccc}
 0 & 0 & 0 & 1 & 0 & 0 & 0 & 0 & 0 & 0 & 0 & 0 \\
 0 & 0 & 0 & 0 & 0 & 0 & 1 & 0 & 0 & 0 & 0 & 0 \\
  \hline
 0 & 0 & 0 & 0 & 1 & 0 & 0 & 0 & 1 & 0 & 0 & 0 \\
 -1 & 0 & 0 & 0 & 0 & 0 & 0 & 0 & 0 & 0 & 0 & 0 \\
 0 & 0 & -1 & 0 & 0 & 0 & 0 & 0 & 0 & 0 & 1 & 0 \\
 0 & 0 & 0 & 0 & 0 & 0 & 0 & 1 & 0 & 0 & 1 & 0 \\
 0 & -1 & 0 & 0 & 0 & 0 & 0 & 0 & 0 & 0 & 0 & 0 \\
 0 & 0 & 0 & 0 & 0 & -1 & 0 & 0 & -1 & 0 & 0 & 0 \\
  \hline
 0 & 0 & 0 & 0 & 0 & 0 & 0 & 0 & 0 & 0 & -1 & 0 \\
 0 & 0 & 0 & 0 & 0 & 0 & 0 & 0 & 0 & 0 & 0 & 1 \\
 0 & 0 & 0 & 0 & 0 & 0 & 0 & 0 & 1 & 0 & 0 & 0 \\
 0 & 0 & 0 & 0 & 0 & 0 & 0 & 0 & 0 & -1 & 0 & 0
\numberthis\end{array}\right] \]

\[[{\bm{\rm X}_3}]_{i}{}^{k}= 
\left[\begin{array}{cccccccc|cccc}
 0 & 0 & 0 & 0 & -1 & 0 & 0 & 0 & -1 & 0 & 0 & 0 \\
 0 & 0 & 0 & 0 & 0 & 0 & 0 & -1 & 0 & 0 & -1 & 0 \\
  \hline
 0 & 0 & 0 & 1 & 0 & 0 & 0 & 0 & 0 & 0 & 0 & 0 \\
 0 & 0 & -1 & 0 & 0 & 0 & 0 & 0 & 0 & 0 & 0 & 0 \\
 1 & 0 & 0 & 0 & 0 & 0 & 0 & 0 & 0 & 0 & 0 & 1 \\
 0 & 0 & 0 & 0 & 0 & 0 & 1 & 0 & 0 & 0 & 0 & 0 \\
 0 & 0 & 0 & 0 & 0 & -1 & 0 & 0 & 0 & 0 & 0 & 0 \\
 0 & 1 & 0 & 0 & 0 & 0 & 0 & 0 & 0 & -1 & 0 & 0 \\
  \hline
 0 & 0 & 0 & 0 & 0 & 0 & 0 & 0 & 0 & 0 & 0 & -1 \\
 0 & 0 & 0 & 0 & 0 & 0 & 0 & 0 & 0 & 0 & -1 & 0 \\
 0 & 0 & 0 & 0 & 0 & 0 & 0 & 0 & 0 & 1 & 0 & 0 \\
 0 & 0 & 0 & 0 & 0 & 0 & 0 & 0 & 1 & 0 & 0 & 0
\numberthis\end{array}\right] \]

\[[{\bm{\rm X}_4}]_{i}{}^{k}= 
\left[\begin{array}{cccccccc|cccc}
 0 & 0 & 0 & -1 & 0 & 0 & 0 & 0 & 0 & -1 & 0 & 0 \\
 0 & 0 & 0 & 0 & 0 & 0 & -1 & 0 & 0 & 0 & 0 & 1 \\
  \hline
 0 & 0 & 0 & 0 & -1 & 0 & 0 & 0 & 0 & 0 & 0 & 0 \\
 1 & 0 & 0 & 0 & 0 & 0 & 0 & 0 & 0 & 0 & 0 & 1 \\
 0 & 0 & 1 & 0 & 0 & 0 & 0 & 0 & 0 & 0 & 0 & 0 \\
 0 & 0 & 0 & 0 & 0 & 0 & 0 & -1 & 0 & 0 & 0 & 0 \\
 0 & 1 & 0 & 0 & 0 & 0 & 0 & 0 & 0 & -1 & 0 & 0 \\
 0 & 0 & 0 & 0 & 0 & 1 & 0 & 0 & 0 & 0 & 0 & 0 \\
  \hline
 0 & 0 & 0 & 0 & 0 & 0 & 0 & 0 & 0 & 0 & -1 & 0 \\
 0 & 0 & 0 & 0 & 0 & 0 & 0 & 0 & 0 & 0 & 0 & 1 \\
 0 & 0 & 0 & 0 & 0 & 0 & 0 & 0 & 1 & 0 & 0 & 0 \\
 0 & 0 & 0 & 0 & 0 & 0 & 0 & 0 & 0 & -1 & 0 & 0
\numberthis\end{array}\right] \]

\subsubsection{$\bm{\rm Y}$-matrix}

Based on Eq. (\ref{eq6}) we can construct the $\bm{\rm{Y}}$-matrix for CLS.

\[[{\bm{\rm Y}_1}]_{i}{}^{k}= 
\left[\begin{array}{cccccccc|cccc}
 0 & 0 & 1 & 0 & 0 & 0 & 0 & 0 & 0 & 0 & -1 & 0 \\
 0 & 0 & 0 & 0 & 0 & 1 & 0 & 0 & 1 & 0 & 0 & 0 \\
  \hline
 -1 & 0 & 0 & 0 & 0 & 0 & 0 & 0 & 0 & 0 & 0 & -1 \\
 0 & 0 & 0 & 0 & -1 & 0 & 0 & 0 & 0 & 0 & 0 & 0 \\
 0 & 0 & 0 & 1 & 0 & 0 & 0 & 0 & 0 & 0 & 0 & 0 \\
 0 & -1 & 0 & 0 & 0 & 0 & 0 & 0 & 0 & 1 & 0 & 0 \\
 0 & 0 & 0 & 0 & 0 & 0 & 0 & -1 & 0 & 0 & 0 & 0 \\
 0 & 0 & 0 & 0 & 0 & 0 & 1 & 0 & 0 & 0 & 0 & 0 \\
  \hline
 0 & 0 & 0 & 0 & 0 & 0 & 0 & 0 & 0 & -1 & 0 & 0 \\
 0 & 0 & 0 & 0 & 0 & 0 & 0 & 0 & 1 & 0 & 0 & 0 \\
 0 & 0 & 0 & 0 & 0 & 0 & 0 & 0 & 0 & 0 & 0 & -1 \\
 0 & 0 & 0 & 0 & 0 & 0 & 0 & 0 & 0 & 0 & 1 & 0 
\numberthis\end{array}\right] \]

\[[{\bm{\rm Y}_2}]_{i}{}^{k}= 
\left[\begin{array}{cccccccc|cccc}
 0 & 0 & -1 & 0 & 0 & 0 & 0 & 0 & 0 & 0 & 0 & 0 \\
 0 & 0 & 0 & 0 & 0 & -1 & 0 & 0 & 0 & 0 & 0 & 0 \\
  \hline
  1 & 0 & 0 & 0 & 0 & 0 & 0 & 0 & 0 & 0 & 0 & 0 \\
 0 & 0 & 0 & 0 & 1 & 0 & 0 & 0 & 1 & 0 & 0 & 0 \\
 0 & 0 & 0 & -1 & 0 & 0 & 0 & 0 & 0 & 1 & 0 & 0 \\
 0 & 1 & 0 & 0 & 0 & 0 & 0 & 0 & 0 & 0 & 0 & 0 \\
 0 & 0 & 0 & 0 & 0 & 0 & 0 & 1 & 0 & 0 & 1 & 0 \\
 0 & 0 & 0 & 0 & 0 & 0 & -1 & 0 & 0 & 0 & 0 & 1 \\
  \hline
 0 & 0 & 0 & 0 & 0 & 0 & 0 & 0 & 0 & -1 & 0 & 0 \\
 0 & 0 & 0 & 0 & 0 & 0 & 0 & 0 & 1 & 0 & 0 & 0 \\
 0 & 0 & 0 & 0 & 0 & 0 & 0 & 0 & 0 & 0 & 0 & -1 \\
 0 & 0 & 0 & 0 & 0 & 0 & 0 & 0 & 0 & 0 & 1 & 0
\numberthis\end{array}\right] \]

\[[{\bm{\rm Y}_3}]_{i}{}^{k}= 
\left[\begin{array}{cccccccc|cccc}
 0 & 0 & -1 & 0 & 0 & 0 & 0 & 0 & 0 & 0 & 1 & 0 \\
 0 & 0 & 0 & 0 & 0 & -1 & 0 & 0 & -1 & 0 & 0 & 0 \\
  \hline
 1 & 0 & 0 & 0 & 0 & 0 & 0 & 0 & 0 & 0 & 0 & 1 \\
 0 & 0 & 0 & 0 & 1 & 0 & 0 & 0 & 0 & 0 & 0 & 0 \\
 0 & 0 & 0 & -1 & 0 & 0 & 0 & 0 & 0 & 0 & 0 & 0 \\
 0 & 1 & 0 & 0 & 0 & 0 & 0 & 0 & 0 & -1 & 0 & 0 \\
 0 & 0 & 0 & 0 & 0 & 0 & 0 & 1 & 0 & 0 & 0 & 0 \\
 0 & 0 & 0 & 0 & 0 & 0 & -1 & 0 & 0 & 0 & 0 & 0 \\
  \hline
 0 & 0 & 0 & 0 & 0 & 0 & 0 & 0 & 0 & 1 & 0 & 0 \\
 0 & 0 & 0 & 0 & 0 & 0 & 0 & 0 & -1 & 0 & 0 & 0 \\
 0 & 0 & 0 & 0 & 0 & 0 & 0 & 0 & 0 & 0 & 0 & 1 \\
 0 & 0 & 0 & 0 & 0 & 0 & 0 & 0 & 0 & 0 & -1 & 0
\numberthis\end{array}\right] \]

\[[{\bm{\rm Y}_4}]_{i}{}^{k}= 
\left[\begin{array}{cccccccc|cccc}
 0 & 0 & 1 & 0 & 0 & 0 & 0 & 0 & 0 & 0 & 0 & 0 \\
 0 & 0 & 0 & 0 & 0 & 1 & 0 & 0 & 0 & 0 & 0 & 0 \\
 \hline
 -1 & 0 & 0 & 0 & 0 & 0 & 0 & 0 & 0 & 0 & 0 & 0 \\
 0 & 0 & 0 & 0 & -1 & 0 & 0 & 0 & -1 & 0 & 0 & 0 \\
 0 & 0 & 0 & 1 & 0 & 0 & 0 & 0 & 0 & -1 & 0 & 0 \\
 0 & -1 & 0 & 0 & 0 & 0 & 0 & 0 & 0 & 0 & 0 & 0 \\
 0 & 0 & 0 & 0 & 0 & 0 & 0 & -1 & 0 & 0 & -1 & 0 \\
 0 & 0 & 0 & 0 & 0 & 0 & 1 & 0 & 0 & 0 & 0 & -1 \\
  \hline
 0 & 0 & 0 & 0 & 0 & 0 & 0 & 0 & 0 & 1 & 0 & 0 \\
 0 & 0 & 0 & 0 & 0 & 0 & 0 & 0 & -1 & 0 & 0 & 0 \\
 0 & 0 & 0 & 0 & 0 & 0 & 0 & 0 & 0 & 0 & 0 & 1 \\
 0 & 0 & 0 & 0 & 0 & 0 & 0 & 0 & 0 & 0 & -1 & 0 
\numberthis\end{array}\right] \label{eq512}\]

\subsubsection{$\bm{\rm Z}$-matrix}
Based on Eq. (\ref{eq3.9}) we can construct the $\bm{\rm{Z}}$-matrix for CLS.

\[[{\bm{\rm Z}_1}]_{i}{}^{k}= 
\left[\begin{array}{cccccccc|cccc}
 -1 & 0 & 0 & 0 & 0 & 0 & 0 & 0 & 0 & 0 & 0 & -1 \\
 0 & -1 & 0 & 0 & 0 & 0 & 0 & 0 & 0 & 1 & 0 & 0 \\
 \hline
 0 & 0 & -1 & 0 & 0 & 0 & 0 & 0 & 0 & 0 & 1 & 0 \\
 0 & 0 & 0 & -1 & 0 & 0 & 0 & 0 & 0 & 1 & 0 & 0 \\
 0 & 0 & 0 & 0 & -1 & 0 & 0 & 0 & -1 & 0 & 0 & 0 \\
 0 & 0 & 0 & 0 & 0 & -1 & 0 & 0 & -1 & 0 & 0 & 0 \\
 0 & 0 & 0 & 0 & 0 & 0 & -1 & 0 & 0 & 0 & 0 & 1 \\
 0 & 0 & 0 & 0 & 0 & 0 & 0 & -1 & 0 & 0 & -1 & 0 \\
 \hline
 0 & 0 & 0 & 0 & 0 & 0 & 0 & 0 & 1 & 0 & 0 & 0 \\
 0 & 0 & 0 & 0 & 0 & 0 & 0 & 0 & 0 & 1 & 0 & 0 \\
 0 & 0 & 0 & 0 & 0 & 0 & 0 & 0 & 0 & 0 & 1 & 0 \\
 0 & 0 & 0 & 0 & 0 & 0 & 0 & 0 & 0 & 0 & 0 & 1 
\numberthis\end{array}\right] \]

\[[{\bm{\rm Z}_2}]_{i}{}^{k}= 
\left[\begin{array}{cccccccc|cccc}
 1 & 0 & 0 & 0 & 0 & 0 & 0 & 0 & 0 & 0 & 0 & 1 \\
 0 & 1 & 0 & 0 & 0 & 0 & 0 & 0 & 0 & -1 & 0 & 0 \\
 \hline
 0 & 0 & 1 & 0 & 0 & 0 & 0 & 0 & 0 & 0 & -1 & 0 \\
 0 & 0 & 0 & 1 & 0 & 0 & 0 & 0 & 0 & -1 & 0 & 0 \\
 0 & 0 & 0 & 0 & 1 & 0 & 0 & 0 & 1 & 0 & 0 & 0 \\
 0 & 0 & 0 & 0 & 0 & 1 & 0 & 0 & 1 & 0 & 0 & 0 \\
 0 & 0 & 0 & 0 & 0 & 0 & 1 & 0 & 0 & 0 & 0 & -1 \\
 0 & 0 & 0 & 0 & 0 & 0 & 0 & 1 & 0 & 0 & 1 & 0 \\
 \hline
 0 & 0 & 0 & 0 & 0 & 0 & 0 & 0 & -1 & 0 & 0 & 0 \\
 0 & 0 & 0 & 0 & 0 & 0 & 0 & 0 & 0 & -1 & 0 & 0 \\
 0 & 0 & 0 & 0 & 0 & 0 & 0 & 0 & 0 & 0 & -1 & 0 \\
 0 & 0 & 0 & 0 & 0 & 0 & 0 & 0 & 0 & 0 & 0 & -1
\numberthis\end{array}\right] \]

\[[{\bm{\rm Z}_3}]_{i}{}^{k}= 
\left[\begin{array}{cccccccc|cccc}
 -1 & 0 & 0 & 0 & 0 & 0 & 0 & 0 & 0 & 0 & 0 & -1 \\
 0 & -1 & 0 & 0 & 0 & 0 & 0 & 0 & 0 & 1 & 0 & 0 \\
 \hline
 0 & 0 & -1 & 0 & 0 & 0 & 0 & 0 & 0 & 0 & 1 & 0 \\
 0 & 0 & 0 & -1 & 0 & 0 & 0 & 0 & 0 & 1 & 0 & 0 \\
 0 & 0 & 0 & 0 & -1 & 0 & 0 & 0 & -1 & 0 & 0 & 0 \\
 0 & 0 & 0 & 0 & 0 & -1 & 0 & 0 & -1 & 0 & 0 & 0 \\
 0 & 0 & 0 & 0 & 0 & 0 & -1 & 0 & 0 & 0 & 0 & 1 \\
 0 & 0 & 0 & 0 & 0 & 0 & 0 & -1 & 0 & 0 & -1 & 0 \\
\hline
 0 & 0 & 0 & 0 & 0 & 0 & 0 & 0 & 1 & 0 & 0 & 0 \\
 0 & 0 & 0 & 0 & 0 & 0 & 0 & 0 & 0 & 1 & 0 & 0 \\
 0 & 0 & 0 & 0 & 0 & 0 & 0 & 0 & 0 & 0 & 1 & 0 \\
 0 & 0 & 0 & 0 & 0 & 0 & 0 & 0 & 0 & 0 & 0 & 1 
\numberthis\end{array}\right] \]

\[[{\bm{\rm Z}_4}]_{i}{}^{k}= 
\left[\begin{array}{cccccccc|cccc}
 1 & 0 & 0 & 0 & 0 & 0 & 0 & 0 & 0 & 0 & 0 & 1 \\
 0 & 1 & 0 & 0 & 0 & 0 & 0 & 0 & 0 & -1 & 0 & 0 \\
 \hline
 0 & 0 & 1 & 0 & 0 & 0 & 0 & 0 & 0 & 0 & -1 & 0 \\
 0 & 0 & 0 & 1 & 0 & 0 & 0 & 0 & 0 & -1 & 0 & 0 \\
 0 & 0 & 0 & 0 & 1 & 0 & 0 & 0 & 1 & 0 & 0 & 0 \\
 0 & 0 & 0 & 0 & 0 & 1 & 0 & 0 & 1 & 0 & 0 & 0 \\
 0 & 0 & 0 & 0 & 0 & 0 & 1 & 0 & 0 & 0 & 0 & -1 \\
 0 & 0 & 0 & 0 & 0 & 0 & 0 & 1 & 0 & 0 & 1 & 0 \\
\hline
 0 & 0 & 0 & 0 & 0 & 0 & 0 & 0 & -1 & 0 & 0 & 0 \\
 0 & 0 & 0 & 0 & 0 & 0 & 0 & 0 & 0 & -1 & 0 & 0 \\
 0 & 0 & 0 & 0 & 0 & 0 & 0 & 0 & 0 & 0 & -1 & 0 \\
 0 & 0 & 0 & 0 & 0 & 0 & 0 & 0 & 0 & 0 & 0 & -1
\numberthis\end{array}\right] \]

\subsubsection{$\bm{\rm W}$-matrix}
Based on Eq. (\ref{eq4-11}) we can construct the $\bm{\rm{W}}$-matrix for CLS.

\begin{equation}
[{\bm{\rm W}}_n]_{i}{}^{k}=-[\mathbf{I}_{12}]_{i}{}^{k}\quad
n\in\{1,2,3,4\}
\end{equation}

\subsection{4D $\mathcal{N}$=1 CLS $\bm\rR$-matrix}

We can transpose the $\bm{\rL}$-matrix to obtain $\bm{\rR}$-matrix from Eq. (\ref{eq6.1}) to (\ref{eq6.4})

\[
[{\bm{\rm R}_1}]_{\hat i}{}^{k} = \left[\begin{array}{cc|cccccc|cccc}
 0 & 0 & 0 & 0 & -1 & 0 & 0 & 0 & 0 & 0 & 0 & 0 \\
 0 & 0 & 0 & 1 & 0 & 0 & 0 & 0 & 0 & -1 & 0 & 0 \\
 0 & 0 & -1 & 0 & 0 & 0 & 0 & 0 & 0 & 0 & 1 & 0 \\
 1 & 0 & 0 & 0 & 0 & 0 & 0 & 0 & 0 & 0 & 0 & 0 \\
 0 & 0 & 0 & 0 & 0 & 0 & 0 & -1 & 0 & 0 & 0 & 0 \\
 0 & 0 & 0 & 0 & 0 & 0 & 1 & 0 & 0 & 0 & 0 & -1 \\
 0 & 0 & 0 & 0 & 0 & -1 & 0 & 0 & -1 & 0 & 0 & 0 \\
 0 & 1 & 0 & 0 & 0 & 0 & 0 & 0 & 0 & 0 & 0 & 0 \\
\hline
 0 & 0 & 0 & 0 & 0 & 0 & 0 & 0 & -1 & 0 & 0 & 0 \\
 0 & 0 & 0 & 0 & 0 & 0 & 0 & 0 & 0 & 1 & 0 & 0 \\
 0 & 0 & 0 & 0 & 0 & 0 & 0 & 0 & 0 & 0 & -1 & 0 \\
 0 & 0 & 0 & 0 & 0 & 0 & 0 & 0 & 0 & 0 & 0 & 1  
\numberthis\end{array}\right] \]

\[
[{\bm{\rm R}_2}]_{\hat i}{}^{k} = \left[\begin{array}{cc|cccccc|cccc}
 1 & 0 & 0 & 0 & 0 & 0 & 0 & 0 & 0 & 0 & 0 & 0 \\
 0 & 0 & 1 & 0 & 0 & 0 & 0 & 0 & 0 & 0 & 0 & 0 \\
 0 & 0 & 0 & 1 & 0 & 0 & 0 & 0 & 0 & 0 & 0 & 0 \\
 0 & 0 & 0 & 0 & 1 & 0 & 0 & 0 & 0 & 0 & 0 & 0 \\
 0 & 1 & 0 & 0 & 0 & 0 & 0 & 0 & 0 & 0 & 0 & 0 \\
 0 & 0 & 0 & 0 & 0 & 1 & 0 & 0 & 0 & 0 & 0 & 0 \\
 0 & 0 & 0 & 0 & 0 & 0 & 1 & 0 & 0 & 0 & 0 & 0 \\
 0 & 0 & 0 & 0 & 0 & 0 & 0 & 1 & 0 & 0 & 0 & 0 \\
\hline
 0 & 0 & 0 & 0 & 0 & 0 & 0 & 0 & 0 & 0 & 0 & 1 \\
 0 & 0 & 0 & 0 & 0 & 0 & 0 & 0 & 0 & 0 & -1 & 0 \\
 0 & 0 & 0 & 0 & 0 & 0 & 0 & 0 & 0 & -1 & 0 & 0 \\
 0 & 0 & 0 & 0 & 0 & 0 & 0 & 0 & 1 & 0 & 0 & 0
\numberthis\end{array}\right] \]

\[[{\bm{\rm R}_3}]_{\hat i}{}^{k} = \left[\begin{array}{cc|cccccc|cccc}
 0 & 0 & 0 & -1 & 0 & 0 & 0 & 0 & 0 & 0 & 0 & 0 \\
 0 & 0 & 0 & 0 & -1 & 0 & 0 & 0 & -1 & 0 & 0 & 0 \\
 1 & 0 & 0 & 0 & 0 & 0 & 0 & 0 & 0 & 0 & 0 & 0 \\
 0 & 0 & 1 & 0 & 0 & 0 & 0 & 0 & 0 & 0 & -1 & 0 \\
 0 & 0 & 0 & 0 & 0 & 0 & -1 & 0 & 0 & 0 & 0 & 0 \\
 0 & 0 & 0 & 0 & 0 & 0 & 0 & -1 & 0 & 0 & -1 & 0 \\
 0 & 1 & 0 & 0 & 0 & 0 & 0 & 0 & 0 & 0 & 0 & 0 \\
 0 & 0 & 0 & 0 & 0 & 1 & 0 & 0 & 1 & 0 & 0 & 0 \\
 \hline
 0 & 0 & 0 & 0 & 0 & 0 & 0 & 0 & 0 & 1 & 0 & 0 \\
 0 & 0 & 0 & 0 & 0 & 0 & 0 & 0 & 1 & 0 & 0 & 0 \\
 0 & 0 & 0 & 0 & 0 & 0 & 0 & 0 & 0 & 0 & 0 & 1 \\
 0 & 0 & 0 & 0 & 0 & 0 & 0 & 0 & 0 & 0 & 1 & 0
\numberthis\end{array}\right] \]

\[[{\bm{\rm R}_4}]_{\hat i}{}^{k} = \left[\begin{array}{cc|cccccc|cccc}
 0 & 0 & -1 & 0 & 0 & 0 & 0 & 0 & 0 & 0 & 0 & 0 \\
 1 & 0 & 0 & 0 & 0 & 0 & 0 & 0 & 0 & 0 & 0 & 0 \\
 0 & 0 & 0 & 0 & 1 & 0 & 0 & 0 & 1 & 0 & 0 & 0 \\
 0 & 0 & 0 & -1 & 0 & 0 & 0 & 0 & 0 & 1 & 0 & 0 \\
 0 & 0 & 0 & 0 & 0 & -1 & 0 & 0 & 0 & 0 & 0 & 0 \\
 0 & 1 & 0 & 0 & 0 & 0 & 0 & 0 & 0 & 0 & 0 & 0 \\
 0 & 0 & 0 & 0 & 0 & 0 & 0 & 1 & 0 & 0 & 1 & 0 \\
 0 & 0 & 0 & 0 & 0 & 0 & -1 & 0 & 0 & 0 & 0 & 1 \\
 \hline
 0 & 0 & 0 & 0 & 0 & 0 & 0 & 0 & 0 & 0 & 1 & 0 \\
 0 & 0 & 0 & 0 & 0 & 0 & 0 & 0 & 0 & 0 & 0 & 1 \\
 0 & 0 & 0 & 0 & 0 & 0 & 0 & 0 & -1 & 0 & 0 & 0 \\
 0 & 0 & 0 & 0 & 0 & 0 & 0 & 0 & 0 & -1 & 0 & 0 
\numberthis\end{array}\right] \]

\subsubsection{$\bm{\rm X'}$-matrix}
Based on Eq. (\ref{eq23}) we can construct the $\bm{\rm{X'}}$-matrix for CLS.

\[[{\bm{\rm X'}_1}]_{\hat i}{}^{\hat k}= 
\left[\begin{array}{cc|cccccc|cccc}
 0 & 0 & 0 & 1 & 0 & 0 & 0 & 0 & 0 & 0 & 0 & 0 \\
 0 & 0 & -1 & 0 & 0 & 0 & 0 & 0 & 0 & 0 & -1 & 0 \\
 0 & 1 & 0 & 0 & 0 & 0 & 0 & 0 & 0 & 1 & 0 & 0 \\
 -1 & 0 & 0 & 0 & 0 & 0 & 0 & 0 & 0 & 0 & 0 & 0 \\
 0 & 0 & 0 & 0 & 0 & 0 & 0 & 1 & 0 & 0 & 0 & 0 \\
 0 & 0 & 0 & 0 & 0 & 0 & -1 & 0 & 1 & 0 & 0 & 0 \\
 0 & 0 & 0 & 0 & 0 & 1 & 0 & 0 & 0 & 0 & 0 & 1 \\
 0 & 0 & 0 & 0 & -1 & 0 & 0 & 0 & 0 & 0 & 0 & 0 \\
 \hline
 0 & 0 & 0 & 0 & 0 & 0 & 0 & 0 & 0 & 0 & 0 & 1 \\
 0 & 0 & 0 & 0 & 0 & 0 & 0 & 0 & 0 & 0 & 1 & 0 \\
 0 & 0 & 0 & 0 & 0 & 0 & 0 & 0 & 0 & -1 & 0 & 0 \\
 0 & 0 & 0 & 0 & 0 & 0 & 0 & 0 & -1 & 0 & 0 & 0 
\numberthis\end{array}\right] \]

\[[{\bm{\rm X'}_2}]_{\hat i}{}^{\hat k}= 
\left[\begin{array}{cc|cccccc|cccc}
 0 & 0 & -1 & 0 & 0 & 0 & 0 & 0 & 0 & 0 & 0 & 0 \\
 0 & 0 & 0 & -1 & 0 & 0 & 0 & 0 & 0 & 0 & 0 & -1 \\
 1 & 0 & 0 & 0 & 0 & 0 & 0 & 0 & 0 & 0 & 0 & 0 \\
 0 & 1 & 0 & 0 & 0 & 0 & 0 & 0 & 0 & 1 & 0 & 0 \\
 0 & 0 & 0 & 0 & 0 & 0 & -1 & 0 & 0 & 0 & 0 & 0 \\
 0 & 0 & 0 & 0 & 0 & 0 & 0 & -1 & 0 & 1 & 0 & 0 \\
 0 & 0 & 0 & 0 & 1 & 0 & 0 & 0 & 0 & 0 & 0 & 0 \\
 0 & 0 & 0 & 0 & 0 & 1 & 0 & 0 & 0 & 0 & 0 & 1 \\
\hline
 0 & 0 & 0 & 0 & 0 & 0 & 0 & 0 & 0 & 0 & -1 & 0 \\
 0 & 0 & 0 & 0 & 0 & 0 & 0 & 0 & 0 & 0 & 0 & 1 \\
 0 & 0 & 0 & 0 & 0 & 0 & 0 & 0 & 1 & 0 & 0 & 0 \\
 0 & 0 & 0 & 0 & 0 & 0 & 0 & 0 & 0 & -1 & 0 & 0 
\numberthis\end{array}\right] \]

\[[{\bm{\rm X'}_3}]_{\hat i}{}^{\hat k}= 
\left[\begin{array}{cc|cccccc|cccc}
 0 & 0 & 0 & -1 & 0 & 0 & 0 & 0 & 0 & 0 & 0 & -1 \\
 0 & 0 & 1 & 0 & 0 & 0 & 0 & 0 & 0 & 0 & 0 & 0 \\
 0 & -1 & 0 & 0 & 0 & 0 & 0 & 0 & 0 & 0 & 0 & 0 \\
 1 & 0 & 0 & 0 & 0 & 0 & 0 & 0 & 1 & 0 & 0 & 0 \\
 0 & 0 & 0 & 0 & 0 & 0 & 0 & -1 & 0 & 1 & 0 & 0 \\
 0 & 0 & 0 & 0 & 0 & 0 & 1 & 0 & 0 & 0 & 0 & 0 \\
 0 & 0 & 0 & 0 & 0 & -1 & 0 & 0 & 0 & 0 & 0 & 0 \\
 0 & 0 & 0 & 0 & 1 & 0 & 0 & 0 & 0 & 0 & 1 & 0 \\
 \hline
 0 & 0 & 0 & 0 & 0 & 0 & 0 & 0 & 0 & 0 & 0 & 1 \\
 0 & 0 & 0 & 0 & 0 & 0 & 0 & 0 & 0 & 0 & 1 & 0 \\
 0 & 0 & 0 & 0 & 0 & 0 & 0 & 0 & 0 & -1 & 0 & 0 \\
 0 & 0 & 0 & 0 & 0 & 0 & 0 & 0 & -1 & 0 & 0 & 0 
\numberthis\end{array}\right] \]

\[[{\bm{\rm X'}_4}]_{\hat i}{}^{\hat k}= 
\left[\begin{array}{cc|cccccc|cccc}
 0 & 0 & -1 & 0 & 0 & 0 & 0 & 0 & 0 & 0 & -1 & 0 \\
 0 & 0 & 0 & -1 & 0 & 0 & 0 & 0 & 0 & 0 & 0 & 0 \\
 1 & 0 & 0 & 0 & 0 & 0 & 0 & 0 & 1 & 0 & 0 & 0 \\
 0 & 1 & 0 & 0 & 0 & 0 & 0 & 0 & 0 & 0 & 0 & 0 \\
 0 & 0 & 0 & 0 & 0 & 0 & -1 & 0 & 1 & 0 & 0 & 0 \\
 0 & 0 & 0 & 0 & 0 & 0 & 0 & -1 & 0 & 0 & 0 & 0 \\
 0 & 0 & 0 & 0 & 1 & 0 & 0 & 0 & 0 & 0 & 1 & 0 \\
 0 & 0 & 0 & 0 & 0 & 1 & 0 & 0 & 0 & 0 & 0 & 0 \\
 \hline
 0 & 0 & 0 & 0 & 0 & 0 & 0 & 0 & 0 & 0 & 1 & 0 \\
 0 & 0 & 0 & 0 & 0 & 0 & 0 & 0 & 0 & 0 & 0 & -1 \\
 0 & 0 & 0 & 0 & 0 & 0 & 0 & 0 & -1 & 0 & 0 & 0 \\
 0 & 0 & 0 & 0 & 0 & 0 & 0 & 0 & 0 & 1 & 0 & 0 
\numberthis\end{array}\right] \]

\subsubsection{$\bm{\rm Y'}$-matrix}
Based on Eq. (\ref{eq26}) we can construct the $\bm{\rm{Y'}}$-matrix for CLS.

\[[{\bm{\rm Y'}_1}]_{\hat i}{}^{\hat k}= 
\left[\begin{array}{cc|cccccc|cccc}
 0 & 1 & 0 & 0 & 0 & 0 & 0 & 0 & 0 & 1 & 0 & 0 \\
 -1 & 0 & 0 & 0 & 0 & 0 & 0 & 0 & -1 & 0 & 0 & 0 \\
 0 & 0 & 0 & -1 & 0 & 0 & 0 & 0 & 0 & 0 & 0 & 0 \\
 0 & 0 & 1 & 0 & 0 & 0 & 0 & 0 & 0 & 0 & 0 & 0 \\
 0 & 0 & 0 & 0 & 0 & 1 & 0 & 0 & 0 & 0 & 0 & 1 \\
 0 & 0 & 0 & 0 & -1 & 0 & 0 & 0 & 0 & 0 & -1 & 0 \\
 0 & 0 & 0 & 0 & 0 & 0 & 0 & -1 & 0 & 0 & 0 & 0 \\
 0 & 0 & 0 & 0 & 0 & 0 & 1 & 0 & 0 & 0 & 0 & 0 \\
 \hline
 0 & 0 & 0 & 0 & 0 & 0 & 0 & 0 & 0 & -1 & 0 & 0 \\
 0 & 0 & 0 & 0 & 0 & 0 & 0 & 0 & 1 & 0 & 0 & 0 \\
 0 & 0 & 0 & 0 & 0 & 0 & 0 & 0 & 0 & 0 & 0 & -1 \\
 0 & 0 & 0 & 0 & 0 & 0 & 0 & 0 & 0 & 0 & 1 & 0 
\numberthis\end{array}\right] \]

\[[{\bm{\rm Y'}_2}]_{\hat i}{}^{\hat k}= 
\left[\begin{array}{cc|cccccc|cccc}
 0 & 1 & 0 & 0 & 0 & 0 & 0 & 0 & 0 & 0 & 0 & 0 \\
 -1 & 0 & 0 & 0 & 0 & 0 & 0 & 0 & 0 & 0 & 0 & 0 \\
 0 & 0 & 0 & -1 & 0 & 0 & 0 & 0 & 0 & 0 & 0 & -1 \\
 0 & 0 & 1 & 0 & 0 & 0 & 0 & 0 & 0 & 0 & 1 & 0 \\
 0 & 0 & 0 & 0 & 0 & 1 & 0 & 0 & 0 & 0 & 0 & 0 \\
 0 & 0 & 0 & 0 & -1 & 0 & 0 & 0 & 0 & 0 & 0 & 0 \\
 0 & 0 & 0 & 0 & 0 & 0 & 0 & -1 & 0 & 1 & 0 & 0 \\
 0 & 0 & 0 & 0 & 0 & 0 & 1 & 0 & -1 & 0 & 0 & 0 \\
 \hline
 0 & 0 & 0 & 0 & 0 & 0 & 0 & 0 & 0 & 1 & 0 & 0 \\
 0 & 0 & 0 & 0 & 0 & 0 & 0 & 0 & -1 & 0 & 0 & 0 \\
 0 & 0 & 0 & 0 & 0 & 0 & 0 & 0 & 0 & 0 & 0 & 1 \\
 0 & 0 & 0 & 0 & 0 & 0 & 0 & 0 & 0 & 0 & -1 & 0 
\numberthis\end{array}\right] \]

\[[{\bm{\rm Y'}_3}]_{\hat i}{}^{\hat k}= 
\left[\begin{array}{cc|cccccc|cccc}
 0 & -1 & 0 & 0 & 0 & 0 & 0 & 0 & 0 & -1 & 0 & 0 \\
 1 & 0 & 0 & 0 & 0 & 0 & 0 & 0 & 1 & 0 & 0 & 0 \\
 0 & 0 & 0 & 1 & 0 & 0 & 0 & 0 & 0 & 0 & 0 & 0 \\
 0 & 0 & -1 & 0 & 0 & 0 & 0 & 0 & 0 & 0 & 0 & 0 \\
 0 & 0 & 0 & 0 & 0 & -1 & 0 & 0 & 0 & 0 & 0 & -1 \\
 0 & 0 & 0 & 0 & 1 & 0 & 0 & 0 & 0 & 0 & 1 & 0 \\
 0 & 0 & 0 & 0 & 0 & 0 & 0 & 1 & 0 & 0 & 0 & 0 \\
 0 & 0 & 0 & 0 & 0 & 0 & -1 & 0 & 0 & 0 & 0 & 0 \\
 \hline
 0 & 0 & 0 & 0 & 0 & 0 & 0 & 0 & 0 & 1 & 0 & 0 \\
 0 & 0 & 0 & 0 & 0 & 0 & 0 & 0 & -1 & 0 & 0 & 0 \\
 0 & 0 & 0 & 0 & 0 & 0 & 0 & 0 & 0 & 0 & 0 & 1 \\
 0 & 0 & 0 & 0 & 0 & 0 & 0 & 0 & 0 & 0 & -1 & 0 
\numberthis\end{array}\right] \]

\[[{\bm{\rm Y'}_4}]_{\hat i}{}^{\hat k}= 
\left[\begin{array}{cc|cccccc|cccc}
 0 & -1 & 0 & 0 & 0 & 0 & 0 & 0 & 0 & 0 & 0 & 0 \\
 1 & 0 & 0 & 0 & 0 & 0 & 0 & 0 & 0 & 0 & 0 & 0 \\
 0 & 0 & 0 & 1 & 0 & 0 & 0 & 0 & 0 & 0 & 0 & 1 \\
 0 & 0 & -1 & 0 & 0 & 0 & 0 & 0 & 0 & 0 & -1 & 0 \\
 0 & 0 & 0 & 0 & 0 & -1 & 0 & 0 & 0 & 0 & 0 & 0 \\
 0 & 0 & 0 & 0 & 1 & 0 & 0 & 0 & 0 & 0 & 0 & 0 \\
 0 & 0 & 0 & 0 & 0 & 0 & 0 & 1 & 0 & -1 & 0 & 0 \\
 0 & 0 & 0 & 0 & 0 & 0 & -1 & 0 & 1 & 0 & 0 & 0 \\
 \hline
 0 & 0 & 0 & 0 & 0 & 0 & 0 & 0 & 0 & -1 & 0 & 0 \\
 0 & 0 & 0 & 0 & 0 & 0 & 0 & 0 & 1 & 0 & 0 & 0 \\
 0 & 0 & 0 & 0 & 0 & 0 & 0 & 0 & 0 & 0 & 0 & -1 \\
 0 & 0 & 0 & 0 & 0 & 0 & 0 & 0 & 0 & 0 & 1 & 0
\numberthis\end{array}\right] \]

\subsubsection{$\bm{\rm Z'}$-matrix}
Based on Eq. (\ref{eq29}) we can construct the $\bm{\rm{Z'}}$-matrix for CLS.
\[[{\bm{\rm Z'}_1}]_{\hat i}{}^{\hat k}= 
\left[\begin{array}{cc|cccccc|cccc}
 1 & 0 & 0 & 0 & 0 & 0 & 0 & 0 & 1 & 0 & 0 & 0 \\
 0 & 1 & 0 & 0 & 0 & 0 & 0 & 0 & 0 & 1 & 0 & 0 \\
 0 & 0 & 1 & 0 & 0 & 0 & 0 & 0 & 0 & 0 & 1 & 0 \\
 0 & 0 & 0 & 1 & 0 & 0 & 0 & 0 & 0 & 0 & 0 & 1 \\
 0 & 0 & 0 & 0 & 1 & 0 & 0 & 0 & 0 & 0 & 1 & 0 \\
 0 & 0 & 0 & 0 & 0 & 1 & 0 & 0 & 0 & 0 & 0 & 1 \\
 0 & 0 & 0 & 0 & 0 & 0 & 1 & 0 & -1 & 0 & 0 & 0 \\
 0 & 0 & 0 & 0 & 0 & 0 & 0 & 1 & 0 & -1 & 0 & 0 \\
 \hline
  0 & 0 & 0 & 0 & 0 & 0 & 0 & 0 & -1 & 0 & 0 & 0 \\
 0 & 0 & 0 & 0 & 0 & 0 & 0 & 0 & 0 & -1 & 0 & 0 \\
 0 & 0 & 0 & 0 & 0 & 0 & 0 & 0 & 0 & 0 & -1 & 0 \\
 0 & 0 & 0 & 0 & 0 & 0 & 0 & 0 & 0 & 0 & 0 & -1
\numberthis\end{array}\right] \]

\[[{\bm{\rm Z'}_2}]_{\hat i}{}^{\hat k}= 
\left[\begin{array}{cc|cccccc|cccc}
-1 & 0 & 0 & 0 & 0 & 0 & 0 & 0 & -1 & 0 & 0 & 0 \\
 0 & -1 & 0 & 0 & 0 & 0 & 0 & 0 & 0 & -1 & 0 & 0 \\
 0 & 0 & -1 & 0 & 0 & 0 & 0 & 0 & 0 & 0 & -1 & 0 \\
 0 & 0 & 0 & -1 & 0 & 0 & 0 & 0 & 0 & 0 & 0 & -1 \\
 0 & 0 & 0 & 0 & -1 & 0 & 0 & 0 & 0 & 0 & -1 & 0 \\
 0 & 0 & 0 & 0 & 0 & -1 & 0 & 0 & 0 & 0 & 0 & -1 \\
 0 & 0 & 0 & 0 & 0 & 0 & -1 & 0 & 1 & 0 & 0 & 0 \\
 0 & 0 & 0 & 0 & 0 & 0 & 0 & -1 & 0 & 1 & 0 & 0 \\
 \hline
0 & 0 & 0 & 0 & 0 & 0 & 0 & 0 & 1 & 0 & 0 & 0 \\
 0 & 0 & 0 & 0 & 0 & 0 & 0 & 0 & 0 & 1 & 0 & 0 \\
 0 & 0 & 0 & 0 & 0 & 0 & 0 & 0 & 0 & 0 & 1 & 0 \\
 0 & 0 & 0 & 0 & 0 & 0 & 0 & 0 & 0 & 0 & 0 & 1
\numberthis\end{array}\right] \]
\[[{\bm{\rm Z'}_3}]_{\hat i}{}^{\hat k}= 
\left[\begin{array}{cc|cccccc|cccc}
1 & 0 & 0 & 0 & 0 & 0 & 0 & 0 & 1 & 0 & 0 & 0 \\
 0 & 1 & 0 & 0 & 0 & 0 & 0 & 0 & 0 & 1 & 0 & 0 \\
 0 & 0 & 1 & 0 & 0 & 0 & 0 & 0 & 0 & 0 & 1 & 0 \\
 0 & 0 & 0 & 1 & 0 & 0 & 0 & 0 & 0 & 0 & 0 & 1 \\
 0 & 0 & 0 & 0 & 1 & 0 & 0 & 0 & 0 & 0 & 1 & 0 \\
 0 & 0 & 0 & 0 & 0 & 1 & 0 & 0 & 0 & 0 & 0 & 1 \\
 0 & 0 & 0 & 0 & 0 & 0 & 1 & 0 & -1 & 0 & 0 & 0 \\
 0 & 0 & 0 & 0 & 0 & 0 & 0 & 1 & 0 & -1 & 0 & 0 \\
 \hline
 0 & 0 & 0 & 0 & 0 & 0 & 0 & 0 & -1 & 0 & 0 & 0 \\
 0 & 0 & 0 & 0 & 0 & 0 & 0 & 0 & 0 & -1 & 0 & 0 \\
 0 & 0 & 0 & 0 & 0 & 0 & 0 & 0 & 0 & 0 & -1 & 0 \\
 0 & 0 & 0 & 0 & 0 & 0 & 0 & 0 & 0 & 0 & 0 & -1
\numberthis\end{array}\right] \]

\[[{\bm{\rm Z'}_4}]_{\hat i}{}^{\hat k}= 
\left[\begin{array}{cc|cccccc|cccc}
 -1 & 0 & 0 & 0 & 0 & 0 & 0 & 0 & -1 & 0 & 0 & 0 \\
 0 & -1 & 0 & 0 & 0 & 0 & 0 & 0 & 0 & -1 & 0 & 0 \\
 0 & 0 & -1 & 0 & 0 & 0 & 0 & 0 & 0 & 0 & -1 & 0 \\
 0 & 0 & 0 & -1 & 0 & 0 & 0 & 0 & 0 & 0 & 0 & -1 \\
 0 & 0 & 0 & 0 & -1 & 0 & 0 & 0 & 0 & 0 & -1 & 0 \\
 0 & 0 & 0 & 0 & 0 & -1 & 0 & 0 & 0 & 0 & 0 & -1 \\
 0 & 0 & 0 & 0 & 0 & 0 & -1 & 0 & 1 & 0 & 0 & 0 \\
 0 & 0 & 0 & 0 & 0 & 0 & 0 & -1 & 0 & 1 & 0 & 0 \\
 \hline
 0 & 0 & 0 & 0 & 0 & 0 & 0 & 0 & 1 & 0 & 0 & 0 \\
 0 & 0 & 0 & 0 & 0 & 0 & 0 & 0 & 0 & 1 & 0 & 0 \\
 0 & 0 & 0 & 0 & 0 & 0 & 0 & 0 & 0 & 0 & 1 & 0 \\
 0 & 0 & 0 & 0 & 0 & 0 & 0 & 0 & 0 & 0 & 0 & 1 
\numberthis\end{array}\right] \]

\subsubsection{$\bm{\rm W'}$-matrix}
Based on Eq. (\ref{eq4-32}) we can construct the $\bm{\rm{W'}}$-matrix for CLS.

\begin{equation}
[{\bm{\rm W'}}_n]_{\hat i}{}^{\hat k}=-[\mathbf{I}_{12}]_{\hat i}{}^{\hat k},\quad
n\in\{1,2,3,4\}
\end{equation}

\subsection{Comparison with Bosonic Holoraumy}
\begin{table}[H]
\setlength\extrarowheight{2pt}
$$
\begin{array}{|c|c|c|c|c|c|c|c|c|} 
\hline
(\cal R)
 & {\bm{\cal B}}^{({\cal R})}_{{w_1} \, {w_2} } & {\bm{\cal B}}^{({\cal R})}_{{w_1} \, {w_3} }  & {\bm{\cal B}}^{({\cal R})}_{{w_1} \, {w_4} } 
 & {\bm{\cal B}}^{({\cal R})}_{{w_2} \, {w_3} }  & {\bm{\cal B}}^{({\cal R})}_{{w_2} \, {w_4} } & {\bm{\cal B}}^{({\cal R})}_{{w_3} \, {w_4} } \\ \hline  
{\rm CLS} & -\, \bm{\rm X_1}  &  \bm{\rm Y_1} &   \bm{\rm X_4} & -\bm{\rm X_2}  & \bm{\rm Y_2} &  -\bm{\rm X_{3}}  \\ \hline
\end{array}
$$    
\caption{Weighted Ordered Holoraumy of CLS Reps}
\label{tb014}
\end{table} 

Based on Eq. (\ref{eq41}), the results show that the ${\bm{\cal B}}$-matrix values on Table \ref{tb014}, coincide with the $\bm{\rm X}$-matrix values in Eq. (\ref{eq55})-(\ref{eq512}). 
This confirms that the definition of bosonic holoraumy is also correct in CLS, and the values are consistent with the raw inverse matrix calculations, $\bm{\rm X}$-matrix.

\newpage
\section{Mathematica Coding}
First let us brute-force to make Chiral supermultiplet adinkra structure by using $[{\bm \rL}_1]_{i}{}^{\hat j}$, $[{\bm \rL}_2]_{i}{}^{\hat j}$, $[{\bm \rL}_3]_{i}{}^{\hat j}$, $[{\bm \rL}_4]_{i}{}^{\hat j}$ as an input for rewriting rules.

\begin{lcverbatim}
ResourceFunction["WolframModel"][
 {
  {{1, x}, {2, y}, {3, z}, {4, w}} -> 
   {
    {1, x}, {2, w}, {3, y}, {4, z},
    {1, y}, {2, z}, {3, x}, {4, w},
    {1, z}, {2, y}, {3, w}, {4, x},
    {1, w}, {2, x}, {3, z}, {4, y}
    }
  }, 
 {{1, x}, {2, y}, {3, z}, {4, w}}, 2, "StatesPlotsList"]
\end{lcverbatim}

\begin{figure}[H]
    \centering
    \includegraphics[scale=0.8]{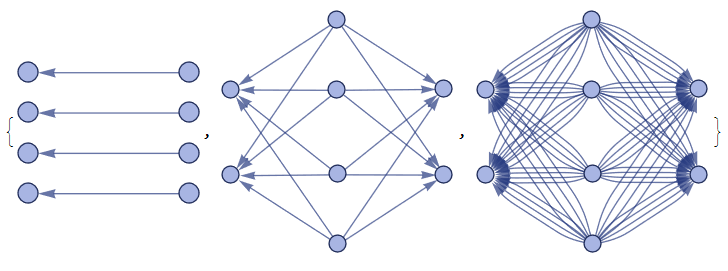}
    \caption{Graphical Representation of Specific Rewriting Rule for Adinkra Generation}
    \label{fig49}
\end{figure}

\begin{figure}[H]
    \centering
    \includegraphics[scale=0.65]{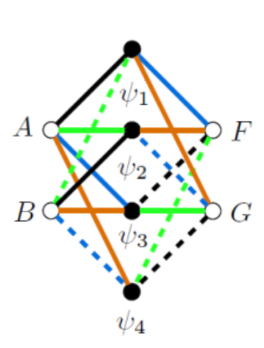}
    \caption{Chiral Supermultiplet Adinkra \protect\cite{3}}
    \label{fig8.4}
\end{figure}

We can observe that the second graph in Fig. \ref{fig49} has the same topological structure as the 4D $\mathcal{N}=1$ Chiral Adinkra, Fig. \ref{fig8.4}.

\begin{figure}[H]
    \centering
    \includegraphics[scale=0.5]{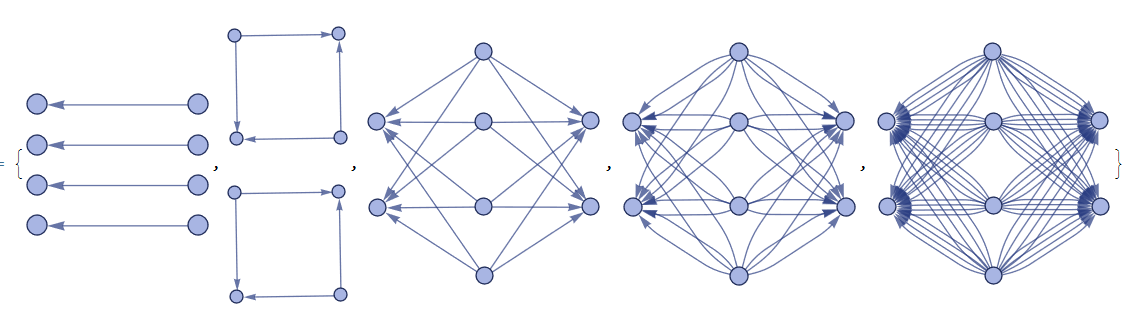}
    \caption{Recursive Chiral Supermultiplet Adinkra by uisng $[{\bm \rL}_1]_{i}{}^{\hat j}$, $[{\bm{\rm Y}}_1]_{i}{}^{j}$ and $[{\bm{\rm Z}}_1]_{i}{}^{j}$. \protect\cite{37}}
    \label{fig8.5}
\end{figure}

On the other hand, empirically we can use the $[{\bm \rL}_1]_{i}{}^{\hat j}$, $[{\bm{\rm Y}}_1]_{i}{}^{j}$, and $[{\bm{\rm Z}}_1]_{i}{}^{j}$ matrix as inputs for the rewriting rules that recursively generate Chiral Supermultiplet Adinkra. Then we can make the observation that the third graph in Fig. \ref{fig8.5} has the same topological structure as the 4D $\mathcal{N}=1$ Chiral Adinkra shown in Fig. \ref{fig8.4}.

\begin{lcverbatim}
ResourceFunction["WolframModel"][
 {
  {{1, x}, {2, w}, {3, y}, {4, z}} ->
   {
    {1, w}, {2, z}, {3, y}, {4, x},
    {1, x}, {2, y}, {3, z}, {4, w}
    }
  },
 {{1, x}, {2, w}, {3, y}, {4, z}}, 4, "StatesPlotsList"]
\end{lcverbatim}

\newpage
\section{Rewriting rules for $\bm\rL$-/$\bm\rR$-matrix}
And to find the proper rewriting rules for generating the same topological structure for adinkra, we can first consider is there any matrix that can transforms

\subsection{4D $\mathcal{N}=1$ Chiral supermultiplets}

First let us consider the summation of the all ${\bm{\rm L}}$-matrix of chiral supermultiplets, ${\bm{\rm L}}$. And let us assume there is a matrix ``${\bm{\rm G}}$" which satisfy Eq. (\ref{eq8.1}).
And we are plan to use this ${\bm{\rm G}}$-matrix as a rewriting rules that can construct the Adinkra in $p$-steps. If we express this is equation form, 

\begin{equation}
[{\bm{\rm L}}]_{i}{}^{\hat k}=[{\bm{\rm L}}_1]_{i}{}^{\hat k}+[{\bm{\rm L}}_2]_{i}{}^{\hat k}+[{\bm{\rm L}}_3]_{i}{}^{\hat k}
+[{\bm{\rm L}}_4]_{i}{}^{\hat k}=[{\bm{\rm G_{(L)}}}^p{\bm{\rm L}}_1]_{i}{}^{\hat k}
\label{eq8.1}
\end{equation}

Here $p=1$ is a trivial solution, and when $p$ is larger than 2 we need more rigorous proof that the available maximum values for $p$ is whether exist for the each supermultiplet. In phonomenological approach we define $p$ as 2, which we check that the value of ${\bm{\rm G}}$-matrix is exist.
According to the Mathematica code, the CLS $\bm\rL$- and $\bm\rR$-matrix exceed the calculation RAM amount usage so we could not check the whether there is solution for ${\bm{\rm G}}$-matrix at $p$=2.
In case of Chiral, Vector and Tensor case, for each supermultiplet we obtain 12 available different ${\bm{\rm G}}$-matrix value and we choose one of the matrix result in this chapter.\footnote{List of 12 available matrix and the calculation code is attached at Appendix A and B}

\begin{equation}
[{\bm{\rm L}}]_{i}{}^{\hat k}=[{\bm{\rm L}}_1]_{i}{}^{\hat k}+[{\bm{\rm L}}_2]+[{\bm{\rm L}}_3]_{i}{}^{\hat k}
+[{\bm{\rm L}}_4]_{i}{}^{\hat k}
=[{\bm{\rm G_{(L)}}}^2{\bm{\rm L}}_1]_{i}{}^{\hat k}
\label{eq8.2}
\end{equation}

Then we can find there is a ${\bm{\rm G}}$-matrix satisfy Eq. (\ref{eq8.2}).

\begin{equation}
[\bm{\rm G}_{(L)}]_{i}{}^{k}=
\begin{pmatrix}
1 & 0 & 1 & 0 \\
0 & -1 & 0 & 1 \\
0 & -1 & 0 & -1 \\
1 & 0 & -1 & 0 \\
\end{pmatrix}=
\begin{pmatrix}
\sigma_3 & \mathbf{I}_2 \\
-i\sigma_2 & -\sigma_1 \\
\end{pmatrix}
\end{equation}

When we only considering the absolute values for each matrix elements,

\begin{equation}
|[\bm{\rm G}_{(L)}]_i{}^{k}|=
\begin{pmatrix}
1 & 0 & 1 & 0 \\
0 & 1 & 0 & 1 \\
0 & 1 & 0 & 1 \\
1 & 0 & 1 & 0 \\
\end{pmatrix}=
\begin{pmatrix}
\mathbf{I}_2 & \mathbf{I}_2 \\
\mathbf{I}^A_2 & \mathbf{I}^A_2 \\
\end{pmatrix}
=\begin{pmatrix}
1 & 1 \\
\end{pmatrix}
\otimes
\begin{pmatrix}
\mathbf{I}_2 \\
\mathbf{I}^A_2 \\
\end{pmatrix},\quad\quad \mathbf{I}^A_2=
\begin{pmatrix}
0 & 1 \\
1 & 0 \\
\end{pmatrix}
\end{equation}

We can double check that the incidence matrix for the graph generated from multiplication of $[{\bm \rL}_1]_{i}{}^{\hat j}$ and ${\bm{\rm G}}$-matrix is given as

\begin{equation}
|[{\bm{\rm G_{(L)}L_1}}]_{i}{}^{\hat k}|=
\begin{pmatrix}
1 & 1 & 0 & 0 \\
0 & 0 & 1 & 1 \\
0 & 0 & 1 & 1 \\
1 & 1 & 0 & 0 \\
\end{pmatrix}
\label{eq8.5}
\end{equation}

\begin{figure}[H]
    \centering
    \includegraphics[scale=1.0]{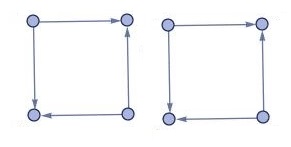}
    \caption{Graph based on Eq. (\ref{eq8.5}) as an incidence matrix}
    \label{fig8.6}
\end{figure}

\begin{equation}
|[{\bm{{\rm G_{(L)}}^2 \rm L_1}}]_{i}{}^{\hat k}|=
\begin{pmatrix}
1 & 1 & 1 & 1 \\
1 & 1 & 1 & 1 \\
1 & 1 & 1 & 1 \\
1 & 1 & 1 & 1 \\
\end{pmatrix}
\label{eq8.6}
\end{equation}

\begin{figure}[H]
    \centering
    \includegraphics[scale=1.0]{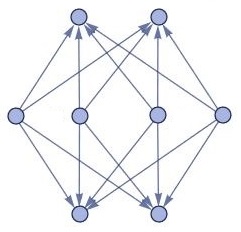}
    \caption{Graph based on Eq. (\ref{eq8.6}) as an incidence matrix}
    \label{fig8.7}
\end{figure}

And for $\bm\rR$-matrix, in the same way,

\begin{equation}
[{\bm{\rm R}}]_{\hat i}{}^{k}=[{\bm{\rm R}}_1]_{\hat i}{}^{k}+[{\bm{\rm R}}_2]_{\hat i}{}^{k}+[{\bm{\rm R}}_3]_{\hat i}{}^{k}+[{\bm{\rm R}}_4]_{\hat i}{}^{k}=[[{\bm{\rm G_{(R)}}}]^2{\bm{\rm R}}_1]_{\hat i}{}^{k}
\end{equation}

\begin{equation}
[{\bm{\rm G_{(R)}}}]_{\hat i}{}^{\hat k}=
\begin{pmatrix}
\sigma_1 & -\mathbf{I}_2 \\
i\sigma_2 & \sigma_3 \\
\end{pmatrix}
\end{equation}

\begin{equation}
|[{\bm{\rm G_{(R)}}}]_{\hat i}{}^{\hat k}|=
\begin{pmatrix}
\mathbf{I}^A_2 & \mathbf{I}_2 \\
\mathbf{I}^A_2 & \mathbf{I}_2\\
\end{pmatrix}
=
\begin{pmatrix}
1 \\
1 \\
\end{pmatrix}
\otimes
\begin{pmatrix}
\mathbf{I}^A_2 & \mathbf{I}_2 \\
\end{pmatrix}
\end{equation}

\begin{lcverbatim}
L1 = {{1, x}, {2, w}, {3, y}, {4, z}};
X1 = {{1, z}, {2, w}, {3, x}, {4, y}};
Y1 = {{1, w}, {2, z}, {3, y}, {4, x}};
Z1 = {{1, x}, {2, y}, {3, z}, {4, w}};
K = {{1, x}, {1, z}, {2, y}, {2, w}, {3, y}, {3, w}, {4, x}, {4, z}};
ResourceFunction["WolframModelPlot"][#, VertexLabels -> Automatic] & /@
  ResourceFunction["WolframModel"][{L1 -> K}, L1, 3, "StatesList"]
\end{lcverbatim}

\begin{figure}[H]
    \centering
    \includegraphics[scale=0.7]{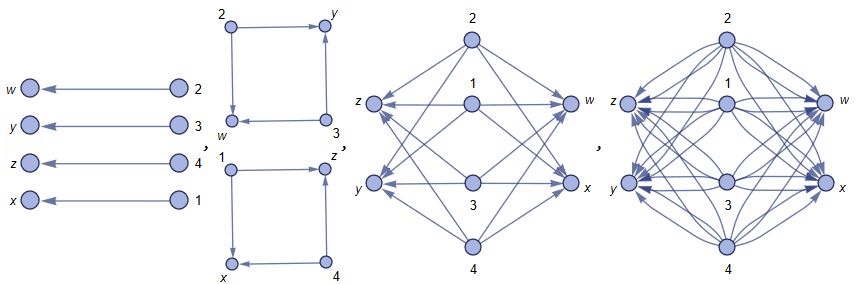}
    \caption{Recursive Chiral Supermultiplet Adinkra by uisng ${\bm \rL}_1$ and $\bm{\rm{G}}$. \protect\cite{37}}
    \label{fig8.8}
\end{figure}

\subsection{4D $\mathcal{N}=1$ Vector supermultiplets}

First for $\bm\rL$-matrix,

\begin{equation}
[{\bm{\rm L}}]_{i}{}^{\hat k}=[{\bm{\rm L}}_1]_{i}{}^{\hat k}+[{\bm{\rm L}}_2]_{i}{}^{\hat k}+[{\bm{\rm L}}_3]_{i}{}^{\hat k}
+[{\bm{\rm L}}_4]_{i}{}^{\hat k}
=[[{\bm{\rm G_{(L)}}}]^2{\bm{\rm L}}_1]_{i}{}^{\hat k}
\label{eq8.2}
\end{equation}

\begin{equation}
[\bm{\rm G_{(L)}}]_{i}{}^{k}=
\begin{pmatrix}
\sigma_3 & \mathbf{I}_2 \\
-i\sigma_2 & -\sigma_1 \\
\end{pmatrix}
\end{equation}

\begin{equation}
|[\bm{\rm G_{(L)}}]_{i}{}^{k}|=
\begin{pmatrix}
\mathbf{I}_2 & \mathbf{I}_2 \\
\mathbf{I}^A_2 & \mathbf{I}^A_2\\
\end{pmatrix}
=
\begin{pmatrix}
1 & 1\\
\end{pmatrix}
\otimes
\begin{pmatrix}
\mathbf{I}_2 \\
\mathbf{I}^A_2 \\
\end{pmatrix}
\end{equation}

Next for $\bm\rR$-matrix,

\begin{equation}
[{\bm{\rm R}}]_{\hat i}{}^{k}=[{\bm{\rm R}}_1]_{\hat i}{}^{k}+[{\bm{\rm R}}_2]_{\hat i}{}^{k}+[{\bm{\rm R}}_3]_{\hat i}{}^{k}+[{\bm{\rm R}}_4]_{\hat i}{}^{k}=[[{\bm{\rm G_{(R)}}}]^2{\bm{\rm R}}_1]_{\hat i}{}^{k}
\end{equation}

\begin{equation}
[{\bm{\rm G_{(R)}}}]_{\hat i}{}^{\hat k}=
\begin{pmatrix}
-\sigma_3 & \sigma_3 \\
\sigma_1 & \sigma_1 \\
\end{pmatrix}
\end{equation}

\begin{equation}
|[{\bm{\rm G_{(R)}}}]_{\hat i}{}^{\hat k}|=
\begin{pmatrix}
\mathbf{I}_2 & \mathbf{I}_2 \\
\mathbf{I}^A_2 & \mathbf{I}^A_2\\
\end{pmatrix}
=
\begin{pmatrix}
1 & 1\\
\end{pmatrix}
\otimes
\begin{pmatrix}
\mathbf{I}_2 \\
\mathbf{I}^A_2 \\
\end{pmatrix}
\end{equation}

\begin{lcverbatim}
L1 = {{1, y}, {2, w}, {3, x}, {4, z}};
X1 = {{1, z}, {2, w}, {3, x}, {4, y}};
Y1 = {{1, y}, {2, x}, {3, w}, {4, z}};
Z1 = {{1, x}, {2, y}, {3, z}, {4, w}};
K = {{1, x}, {1, z}, {2, y}, {2, w}, {3, y}, {3, w}, {4, x}, {4, z}};
ResourceFunction["WolframModelPlot"][#, VertexLabels -> Automatic] & /@
  ResourceFunction["WolframModel"][{L1 -> K}, L1, 3, "StatesList"]
\end{lcverbatim}

\begin{figure}[H]
    \centering
    \includegraphics[scale=0.7]{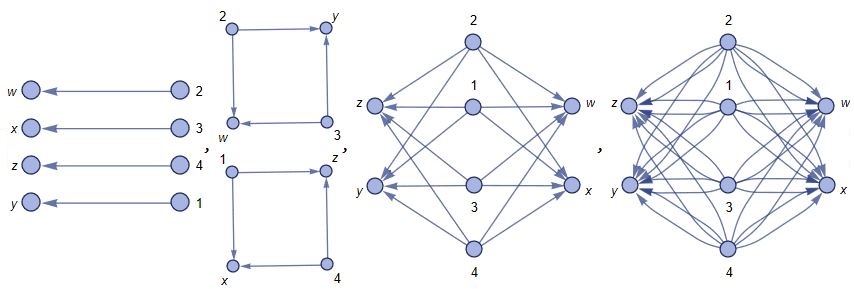}
    \caption{Recursive Vector Supermultiplet Adinkra by uisng ${\bm \rL}_1$ and $\bm{\rm G}$. \protect\cite{37}}
    \label{fig8.9}
\end{figure}

\subsection{4D $\mathcal{N}=1$ Tensor supermultiplets}
First for $\bm\rL$-matrix,
\begin{equation}
[{\bm{\rm L}}]_{i}{}^{\hat k}=[{\bm{\rm L}}_1]_{i}{}^{\hat k}+[{\bm{\rm L}}_2]_{i}{}^{\hat k}+[{\bm{\rm L}}_3]_{i}{}^{\hat k}+[{\bm{\rm L}}_4]_{i}{}^{\hat k}=[[{{\bm{\rm G_{(L)}}}}]^2{\bm{\rm L}}_1]_{i}{}^{\hat k}
\end{equation}

\begin{equation}
[{\bm{\rm G_{(L)}}}]_{i}{}^{k}=
\begin{pmatrix}
\sigma_1 & -\mathbf{I}_2 \\
i\sigma_2 & \sigma_3 \\
\end{pmatrix}
\end{equation}

\begin{equation}
|[{\bm{\rm G_{(L)}}}]_{i}{}^{k}|=
\begin{pmatrix}
\mathbf{I}^A_2 & \mathbf{I}_2 \\
\mathbf{I}^A_2 & \mathbf{I}_2\\
\end{pmatrix}
=
\begin{pmatrix}
1 \\
1 \\
\end{pmatrix}
\otimes
\begin{pmatrix}
\mathbf{I}^A_2 & \mathbf{I}_2 \\
\end{pmatrix}
\end{equation}

Next for $\bm\rR$-matrix,
\begin{equation}
[{\bm{\rm R}}]_{\hat i}{}^{k}=[{\bm{\rm R}}_1]_{\hat i}{}^{k}+[{\bm{\rm R}}_2]_{\hat i}{}^{k}+[{\bm{\rm R}}_3]_{\hat i}{}^{k}+[{\bm{\rm R}}_4]_{\hat i}{}^{k}=[[{\bm{\rm G_{(R)}}}]^2{\bm{\rm R}}_1]_{\hat i}{}^{k}
\end{equation}

\begin{equation}
[{\bm{\rm G_{(R)}}}]_{\hat i}{}^{\hat k}=
\begin{pmatrix}
-\sigma_1 & \sigma_1 \\
\sigma_3 & \sigma_3 \\
\end{pmatrix}
\end{equation}

\begin{equation}
|[{\bm{\rm G_{(R)}}}]_{\hat i}{}^{\hat k}|=
\begin{pmatrix}
\mathbf{I}^A_2 & \mathbf{I}^A_2\\
\mathbf{I}_2 & \mathbf{I}_2 \\
\end{pmatrix}
=
\begin{pmatrix}
1 & 1\\
\end{pmatrix}
\otimes
\begin{pmatrix}
\mathbf{I}^A_2 \\
\mathbf{I}_2 \\
\end{pmatrix}
\end{equation}

\begin{lcverbatim}
L1 = {{1, x}, {2, z}, {3, w}, {4, y}};
X1 = {{1, w}, {2, z}, {3, y}, {4, x}};
Y1 = {{1, y}, {2, x}, {3, w}, {4, z}};
Z1 = {{1, x}, {2, y}, {3, z}, {4, w}};
K = {{1, y}, {1, z}, {2, x}, {2, w}, {3, y}, {3, z}, {4, x}, {4, w}};
ResourceFunction["WolframModelPlot"][#, VertexLabels -> Automatic] & /@
  ResourceFunction["WolframModel"][{L1 -> K}, L1, 3, "StatesList"]
\end{lcverbatim}

\begin{figure}[H]
    \centering
\includegraphics[scale=0.7]{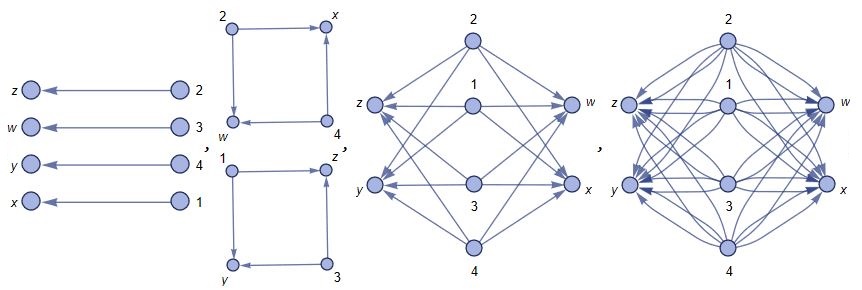}
   \caption{Recursive Tensor Supermultiplet Adinkra by uisng ${\bm \rL}_1$ and $\bm{\rm G}$. \protect\cite{37}}
    \label{fig8.10}
\end{figure}

\subsection{Summary Table}
So we can summarized the all result as a table,

\begin{table}[H]
\begin{center}
\begin{tabular}{|m{4.5em}|m{4cm}|m{4cm}|m{4cm}|m{4cm}|m{4cm}|m{1cm}|m{1cm}|m{1cm}|}
\hline
  $\bm\rL$-matrix  
  & Chiral 
  & Vector 
  & Tensor \\   
  \hline
  $[{\bm{\rm X}}_1]_{i}{}^{j}$
  & $i\sigma_2\otimes\sigma_3$ 
  & $i\sigma_2\otimes\sigma_3$
  & -$i\sigma_2\otimes\sigma_1$ \\ 
  \hline
  $[{\bm{\rm Y}}_1]_{i}{}^{j}$
  & $i\sigma_2\otimes \sigma_1$
  & $\mathbf{I}_2\otimes i\sigma_2$  
  & $\mathbf{I}_2\otimes i\sigma_2$ 
 \\ 
  \hline
  $[{\bm{\rm Z}}_1]_{i}{}^{j}$
  & $\mathbf{I}_4$
  & -$\mathbf{I}_4$ 
  & -$\mathbf{I}_4$
 \\ 
  \hline
  $[{\bm{\rm G_{(L)}}}]_{i}{}^{j}$
  & $\begin{pmatrix}
  \sigma_3&\mathbf{I}_2\\
  -i\sigma_2& -\sigma_1\\
  \end{pmatrix}$
  & $\begin{pmatrix}
  \sigma_3&\mathbf{I}_2\\
  -i\sigma_2& -\sigma_1\\
  \end{pmatrix}$ 
  & $\begin{pmatrix}
  \sigma_1&-\mathbf{I}_2\\
  i\sigma_2& \sigma_3\\
  \end{pmatrix}$ 
  \\ 
  \hline
$|[{\bm{\rm G_{(L)}}}]_{i}{}^{j}|$& 
$\begin{pmatrix}
\mathbf{I}_2&\mathbf{I}_2 \\
\mathbf{I}^A_2&\mathbf{I}^A_2 \\
\end{pmatrix}$&
$\begin{pmatrix}
\mathbf{I}_2&\mathbf{I}_2 \\
\mathbf{I}^A_2&\mathbf{I}^A_2 \\
\end{pmatrix}$&
$\begin{pmatrix}
\mathbf{I}^A_2 & \mathbf{I}_2 \\
\mathbf{I}^A_2 & \mathbf{I}_2 \\
\end{pmatrix}$
\\
\hline
\end{tabular}
\end{center}
\caption{$\bm\rL$-matrix summary table}
\end{table}

\begin{table}[H]
\begin{center}
\begin{tabular}{|m{4.5em}|m{4cm}|m{4cm}|m{4cm}|m{4cm}|m{4cm}|m{1cm}|m{1cm}|m{1cm}|}
  \hline
  $\bm\rR$-matrix & Chiral &  Vector & Tensor\\
  \hline
  $[{\bm{\rm X}}'_1]_{\hat i}{}^{\hat j}$& 
  -$\mathbf{I}_2\otimes i\sigma_2$& 
  $\sigma_3\otimes i\sigma_2$ & 
  -$\sigma_3\otimes i\sigma_2$ 
\\
  \hline
  $[{\bm{\rm Y}}'_1]_{\hat i}{}^{\hat j}$& 
  $i\sigma_2\otimes\sigma_3$  & 
  $i\sigma_2\otimes\mathbf{I}_2$ &
  $i\sigma_2\otimes\mathbf{I}_2$\\ 
  \hline
  $[{\bm{\rm Z}}'_1]_{\hat i}{}^{\hat j}$& 
  -$\mathbf{I}_4$ & 
  $\mathbf{I}_4$ &
  $\mathbf{I}_4$ \\ 
  \hline
  $[{\bm{\rm G_{(R)}}}]_{\hat i}{}^{\hat j}$&
  $\begin{pmatrix}
  \sigma_1&-\mathbf{I}_2\\
  i\sigma_2& \sigma_3\\
  \end{pmatrix}$  &
  $\begin{pmatrix}
  -\sigma_3&\sigma_3\\
  \sigma_1& \sigma_1\\
  \end{pmatrix}$  &
  $\begin{pmatrix}
  -\sigma_1&\sigma_1\\
  \sigma_3& \sigma_3\\
  \end{pmatrix}$  \\ 
  \hline
$|[{\bm{\rm G_{(R)}}}]_{\hat i}{}^{\hat j}|$&
$\begin{pmatrix}
\mathbf{I}^A_2 & \mathbf{I}_2 \\
\mathbf{I}^A_2 & \mathbf{I}_2 \\
\end{pmatrix}$&
$\begin{pmatrix}
\mathbf{I}_2 & \mathbf{I}_2 \\
\mathbf{I}^A_2 & \mathbf{I}^A_2 \\
\end{pmatrix}$&
$\begin{pmatrix}
\mathbf{I}^A_2 & \mathbf{I}^A_2 \\
\mathbf{I}_2 & \mathbf{I}_2 \\
\end{pmatrix}$
\\
\hline
\end{tabular}
\end{center}
\caption{$\bm\rR$-matrix summary table}
\label{tab:my_label}
\end{table}

The ${\bm{\rm G}}$-matrix and Valise Adinkra for the CLS valise is currently unavailable due to our current insufficient access to more robust computational capacities 
to run the code.
(Appendix A shows the mathematica code for calculate $\bm{\rm G}$-matrix. Appendix B shows the all available $\bm{\rm G}$-matrix values for different supermultiplets.)

\newpage
\section{Conclusion and Outlook}
In this paper we have studied a selection of adinkras from the perspective of the rewriting rules for the Chiral, Vector and Tensor, CLS multiplets.  The main reason is that the process at least for minimal four color ones converge rapidly. In further investigation we propose to develop an automatic Mathematica code that generates adinkra diagrams in one or two steps. Additionally, we need to find fundamental rewriting rules that can adjust tensor and vector supermultiplet operations and construct adinkra diagrams in a more efficient way. To accomplish this, we could explore existing libraries or packages in Mathematica that may be useful for this task, and adapt them to our needs. The development of such a code and rewriting rules would facilitate the analysis and computation of supersymmetric theories, and could have important applications in particle physics and beyond.
There are several outlooks for this research in further investigation such as, first Mathematical proof for available values for $p$ for ${\bm{\rm G}}$-matrix. We also need to solve the computational `bottleneck' issue of the simulation. Furthermore there is a need to develop an automatic simulation code for generating adinkras from the inputs of the $\bm\rL$- or $\bm\rR$- matrices.

We also wish to once more comment on the serendipitous emergence of the role played by the bosonic holoraumy matrices as providing the lowest order operators required at the first application of the rewriting relationships. The appearance of the hopping operators is most interesting as the latter act as translation operators on the permutahedron. The ${\bm{\rm X}}$-matrices that arise from the $\bm\rL$-matrices inherit this relation. This further implies that the ${\bm{\rm Y}}$-matrices correspond to ``accelerations," and the ${\bm{\rm Z}}$-matrices correspond to the ``jerks\footnote{See the discussion linked to URL https://www.desy.de/user/projects/Physics/General/jerk.htmlon-line.}." 
The fact that the ${\bm{\rm Z}}$-matrices correspond to identity matrices implies that they have no impact on the system in the permutahedra. This is tantalizing and curiously reminiscent to dynamical systems in Newtonian mechanics where the usual acceleration is constant in the absence of forces. Why this curious analogy should occur is mystifying.

\begin{center}
\parbox{4in}{{\it 
``While asleep, I had an unusual experience.'' \\ ${~}$ 
${~}$ 
\\ ${~}$ }\,\,-\,\, \, Srinivasa Ramanujan
$~~~~~~~~~$}
 \parbox{4in}{
$~~$}  
\end{center}

\noindent
{\bf {Author Affiliation}}\\[.1in] \indent
One of the co-authors, Youngik Lee, graduated from Brown University with an M.Sc. in Physics in 2023 and is currently an alumnus.

\noindent
{\bf {Acknowledgements}}\\[.1in] \indent
The research was supported during a portion of the time it was carried out by the endowment of the Ford Foundation Professorship of Physics at Brown 
University.  
We would like to thank Morgan Gillis, David Chester for the code and image provide, Lily Anderson, and Aleksander Cianciara for inspiring discussions, and also the members of Polytopic Representation for SUSY Research Group for the valuable comments.
All also gratefully acknowledge the support of the Brown Theoretical Physics Center. 
In addition, the research of S.J.G. is currently supported by the Clark Leadership Chair in Science endowment at the University of Maryland - College Park.  Finally, S.J.G. wishes to acknowledge Stephen Wolfram for the introduction to the
subject of his paradigm.

\newpage
\appendix
\section{Mathematica Code for G-matrix}

\begin{lcverbatim}
findMatricesG[A_, p_] := 
 Module[{}, Select[Tuples[{-1, 0, 1}, ConstantArray[Length[A], 2]], 
   MatrixPower[#, p] == A &]]

A = {{1, 0}, {0, 1}}; p = 2; 
{Map[MatrixForm, #], Length[#]} &@ findMatricesG[A, p]
\end{lcverbatim}

Here ${\bm{\rm A}}$ is the input array and the ${\bm{\rm G}}$ is the ${\bm{\rm G}}$-matrix we define in Chapter 7.
Sample code shows ${\bm{\rm G}}$-matrix calculation when  $p$=2, and iput matrix ${\bm{\rm A}}=\mathbf{I}_2$.

\section{${\bm{\rm G}}$-matrix Results for Chiral, Vector and Tensor supermultiplet}

\subsection{4D $\mathcal{N}=1$ Chiral, Vector supermultiplet, $p$=2}
\begin{equation}
\left(
\begin{array}{cccc}
 -1 & 0 & 0 & 1 \\
 0 & 1 & -1 & 0 \\
 -1 & 0 & 0 & -1 \\
 0 & -1 & -1 & 0 \\
\end{array}
\right)
,\quad
\left(
\begin{array}{cccc}
 -1 & 0 & 1 & 0 \\
 -1 & 0 & -1 & 0 \\
 0 & -1 & 0 & -1 \\
 0 & -1 & 0 & 1 \\
\end{array}
\right)
,\quad
\left(
\begin{array}{cccc}
 -1 & 1 & 0 & 0 \\
 0 & 0 & -1 & -1 \\
 0 & 0 & 1 & -1 \\
 -1 & -1 & 0 & 0 \\
\end{array}
\right)
\end{equation}

\begin{equation}
\left(
\begin{array}{cccc}
 0 & -1 & 0 & 1 \\
 0 & 1 & 0 & 1 \\
 -1 & 0 & -1 & 0 \\
 1 & 0 & -1 & 0 \\
\end{array}
\right)
,\quad
\left(
\begin{array}{cccc}
 0 & -1 & 1 & 0 \\
 -1 & 0 & 0 & 1 \\
 0 & -1 & -1 & 0 \\
 1 & 0 & 0 & 1 \\
\end{array}
\right)
,\quad
\left(
\begin{array}{cccc}
 0 & 0 & -1 & 1 \\
 1 & 1 & 0 & 0 \\
 -1 & 1 & 0 & 0 \\
 0 & 0 & -1 & -1 \\
\end{array}
\right)
\end{equation}

\begin{equation}
\left(
\begin{array}{cccc}
 0 & 0 & 1 & -1 \\
 -1 & -1 & 0 & 0 \\
 1 & -1 & 0 & 0 \\
 0 & 0 & 1 & 1 \\
\end{array}
\right)
,\quad
\left(
\begin{array}{cccc}
 0 & 1 & -1 & 0 \\
 1 & 0 & 0 & -1 \\
 0 & 1 & 1 & 0 \\
 -1 & 0 & 0 & -1 \\
\end{array}
\right)
,\quad
\left(
\begin{array}{cccc}
 0 & 1 & 0 & -1 \\
 0 & -1 & 0 & -1 \\
 1 & 0 & 1 & 0 \\
 -1 & 0 & 1 & 0 \\
\end{array}
\right)
\end{equation}

\begin{equation}
\left(
\begin{array}{cccc}
 1 & -1 & 0 & 0 \\
 0 & 0 & 1 & 1 \\
 0 & 0 & -1 & 1 \\
 1 & 1 & 0 & 0 \\
\end{array}
\right)
,\quad
\left(
\begin{array}{cccc}
 1 & 0 & -1 & 0 \\
 1 & 0 & 1 & 0 \\
 0 & 1 & 0 & 1 \\
 0 & 1 & 0 & -1 \\
\end{array}
\right)
,\quad
\left(
\begin{array}{cccc}
 1 & 0 & 0 & -1 \\
 0 & -1 & 1 & 0 \\
 1 & 0 & 0 & 1 \\
 0 & 1 & 1 & 0 \\
\end{array}
\right)
\end{equation}

\subsection{4D $\mathcal{N}=1$ Tensor supermultiplet, $p$=2}

\begin{equation}
\left(
\begin{array}{cccc}
 -1 & 0 & -1 & 0 \\
 0 & 1 & 0 & -1 \\
 0 & 1 & 0 & 1 \\
 -1 & 0 & 1 & 0 \\
\end{array}
\right)
,\quad
\left(
\begin{array}{cccc}
 -1 & 0 & 0 & 1 \\
 -1 & 0 & 0 & -1 \\
 0 & 1 & 1 & 0 \\
 0 & -1 & 1 & 0 \\
\end{array}
\right)
,\quad
\left(
\begin{array}{cccc}
 -1 & 1 & 0 & 0 \\
 0 & 0 & 1 & -1 \\
 1 & 1 & 0 & 0 \\
 0 & 0 & 1 & 1 \\
\end{array}
\right)
\end{equation}

\begin{equation}
\left(
\begin{array}{cccc}
 0 & -1 & -1 & 0 \\
 0 & 1 & -1 & 0 \\
 -1 & 0 & 0 & 1 \\
 -1 & 0 & 0 & -1 \\
\end{array}
\right)
,\quad
\left(
\begin{array}{cccc}
 0 & -1 & 0 & 1 \\
 -1 & 0 & -1 & 0 \\
 -1 & 0 & 1 & 0 \\
 0 & -1 & 0 & -1 \\
\end{array}
\right)
,\quad
\left(
\begin{array}{cccc}
 0 & 0 & -1 & -1 \\
 1 & 1 & 0 & 0 \\
 0 & 0 & -1 & 1 \\
 -1 & 1 & 0 & 0 \\
\end{array}
\right)
\end{equation}

\begin{equation}
\left(
\begin{array}{cccc}
 0 & 0 & 1 & 1 \\
 -1 & -1 & 0 & 0 \\
 0 & 0 & 1 & -1 \\
 1 & -1 & 0 & 0 \\
\end{array}
\right)
,\quad
\left(
\begin{array}{cccc}
 0 & 1 & 0 & -1 \\
 1 & 0 & 1 & 0 \\
 1 & 0 & -1 & 0 \\
 0 & 1 & 0 & 1 \\
\end{array}
\right)
,\quad
\left(
\begin{array}{cccc}
 0 & 1 & 1 & 0 \\
 0 & -1 & 1 & 0 \\
 1 & 0 & 0 & -1 \\
 1 & 0 & 0 & 1 \\
\end{array}
\right)
\end{equation}

\begin{equation}
\left(
\begin{array}{cccc}
 1 & -1 & 0 & 0 \\
 0 & 0 & -1 & 1 \\
 -1 & -1 & 0 & 0 \\
 0 & 0 & -1 & -1 \\
\end{array}
\right)
,\quad
\left(
\begin{array}{cccc}
 1 & 0 & 0 & -1 \\
 1 & 0 & 0 & 1 \\
 0 & -1 & -1 & 0 \\
 0 & 1 & -1 & 0 \\
\end{array}
\right)
,\quad
\left(
\begin{array}{cccc}
 1 & 0 & 1 & 0 \\
 0 & -1 & 0 & 1 \\
 0 & -1 & 0 & -1 \\
 1 & 0 & -1 & 0 \\
\end{array}
\right)
\end{equation}

\section{Diagonalized $\bm\rL$- and $\bm\rR$-matrix of CLS Field}

The complex linear supermultiplet (CLS) contains scalar $G$, pseudoscalar $L$, Majorana spinor $\zeta_a$ and auxiliary scalar $M$, auxiliary pseudoscalar $N$, auxiliary vector $V_\mu$, auxiliary axial-vector $U_\mu$, and auxiliary Majorana spinors $\rho_a$ and $\beta_a$. \cite{38}

\[
[{\rm L_1}]_{i}{}^{\hat k}  = \left[\begin{array}{cccc|cccc|cccc}
    1 & 0 & 0 & 0 & 0 & 0 & 0 & 0 & 0 & 0 & 0 & 0 \\ 
    0 & 1 & 0 & 0 & 0 & 0 & 0 & 0 & 0 & 0 & 0 & 0 \\ 
    0 & 0 & 1 & 0 & 0 & 0 & 0 & 0 & 0 & 0 & 0 & 0 \\ 
    0 & 0 & 0 & 1 & 0 & 0 & 0 & 0 & 0 & 0 & 0 & 0 \\ 
    \hline
    0 & 0 & 0 & 0 & 1 & 0 & 0 & 0 & 0 & 0 & 0 & 0 \\ 
    0 & 0 & 0 & 0 & 0 & 1 & 0 & 0 & 0 & 0 & 0 & 0 \\ 
    0 & 0 & 0 & 0 & 0 & 0 & 1 & 0 & 0 & 0 & 0 & 0 \\ 
    0 & 0 & 0 & 0 & 0 & 0 & 0 & 1 & 0 & 0 & 0 & 0 \\ 
    \hline
    0 & 0 & 0 & 0 & 0 & 0 & 0 & 0 & 1 & 0 & 0 & 0 \\ 
    0 & 0 & 0 & 0 & 0 & 0 & 0 & 0 & 0 & 1 & 0 & 0 \\ 
    0 & 0 & 0 & 0 & 0 & 0 & 0 & 0 & 0 & 0 & 1 & 0 \\ 
    0 & 0 & 0 & 0 & 0 & 0 & 0 & 0 & 0 & 0 & 0 & 1
\numberthis\end{array}\right] \]

\[
[{\rm L_2}]_{i}{}^{\hat k}  = \left[\begin{array}{cccc|cccc|cccc}
    0 & 1 & 0 & 0 & 0 & 0 & 0 & 0 & 0 & 0 & 0 & 0 \\  
    -1 & 0 & 0 & 0 & 0 & 0 & 0 & 0 & 0 & 0 & 0 & 0 \\ 
    0 & 0 & 0 & 1 & 0 & 0 & 0 & 0 & 0 & 0 & 0 & 0 \\  
    0 & 0 & -1 & 0 & 0 & 0 & 0 & 0 & 0 & 0 & 0 & 0 \\
    \hline
    0 & 0 & 0 & 0 & 0 & 1 & 0 & 0 & 0 & 0 & 0 & 0 \\
    0 & 0 & 0 & 0 & -1 & 0 & 0 & 0 & 0 & 0 & 0 & 0 \\
    0 & 0 & 0 & 0 & 0 & 0 & 0 & -1 & 0 & 0 & 0 & 0 \\
    0 & 0 & 0 & 0 & 0 & 0 & 1 & 0 & 0 & 0 & 0 & 0 \\
    \hline
    0 & 0 & 0 & 0 & 0 & 0 & 0 & 0 & 0 & -1 & 0 & 0 \\
    0 & 0 & 0 & 0 & 0 & 0 & 0 & 0 & 1 & 0 & 0 & 0 \\
    0 & 0 & 0 & 0 & 0 & 0 & 0 & 0 & 0 & 0 & 0 & 1 \\
    0 & 0 & 0 & 0 & 0 & 0 & 0 & 0 & 0 & 0 & -1 & 0
\numberthis\end{array}\right] \]

\[
[{\rm L_3}]_{i}{}^{\hat k}  = \left[\begin{array}{cccc|cccc|cccc}
    0 & 0 & 1 & 0 & 0 & 0 & 0 & 0 & 0 & 0 & 0 & 0 \\  
    0 & 0 & 0 & -1 & 0 & 0 & 0 & 0 & 0 & 0 & 0 & 0 \\ 
    -1 & 0 & 0 & 0 & 0 & 0 & 0 & 0 & 0 & 0 & 0 & 0 \\ 
    0 & 1 & 0 & 0 & 0 & 0 & 0 & 0 & 0 & 0 & 0 & 0 \\  
    \hline
    0 & 0 & 0 & 0 & 0 & 0 & -1 & 0 & 0 & 0 & 0 & 0 \\ 
    0 & 0 & 0 & 0 & 0 & 0 & 0 & -1 & 0 & 0 & 0 & 0 \\ 
    0 & 0 & 0 & 0 & 1 & 0 & 0 & 0 & 0 & 0 & 0 & 0 \\  
    0 & 0 & 0 & 0 & 0 & 1 & 0 & 0 & 0 & 0 & 0 & 0 \\  
    \hline
    0 & 0 & 0 & 0 & 0 & 0 & 0 & 0 & 0 & 0 & 1 & 0 \\  
    0 & 0 & 0 & 0 & 0 & 0 & 0 & 0 & 0 & 0 & 0 & 1 \\  
    0 & 0 & 0 & 0 & 0 & 0 & 0 & 0 & -1 & 0 & 0 & 0 \\ 
    0 & 0 & 0 & 0 & 0 & 0 & 0 & 0 & 0 & -1 & 0 & 0
\numberthis\end{array}\right] \]

\[
[{\rm L_4}]_{i}{}^{\hat k}  = \left[\begin{array}{cccc|cccc|cccc}
    0 & 0 & 0 & 1 & 0 & 0 & 0 & 0 & 0 & 0 & 0 & 0 \\
    0 & 0 & 1 & 0 & 0 & 0 & 0 & 0 & 0 & 0 & 0 & 0 \\
    0 & -1 & 0 & 0 & 0 & 0 & 0 & 0 & 0 & 0 & 0 & 0 \\
    -1 & 0 & 0 & 0 & 0 & 0 & 0 & 0 & 0 & 0 & 0 & 0 \\
    \hline
    0 & 0 & 0 & 0 & 0 & 0 & 0 & -1 & 0 & 0 & 0 & 0 \\
    0 & 0 & 0 & 0 & 0 & 0 & 1 & 0 & 0 & 0 & 0 & 0 \\
    0 & 0 & 0 & 0 & 0 & -1 & 0 & 0 & 0 & 0 & 0 & 0 \\
    0 & 0 & 0 & 0 & 1 & 0 & 0 & 0 & 0 & 0 & 0 & 0 \\
    \hline
    0 & 0 & 0 & 0 & 0 & 0 & 0 & 0 & 0 & 0 & 0 & -1 \\
    0 & 0 & 0 & 0 & 0 & 0 & 0 & 0 & 0 & 0 & 1 & 0 \\
    0 & 0 & 0 & 0 & 0 & 0 & 0 & 0 & 0 & -1 & 0 & 0 \\
    0 & 0 & 0 & 0 & 0 & 0 & 0 & 0 & 1 & 0 & 0 & 0
\numberthis\end{array}\right] \]

\[
[{\rm R_1}]_{\hat i}{}^{k} = \left[\begin{array}{cccc|cccc|cccc}
    1 & 0 & 0 & 0 & 0 & 0 & 0 & 0 & 0 & 0 & 0 & 0 \\ 
    0 & 1 & 0 & 0 & 0 & 0 & 0 & 0 & 0 & 0 & 0 & 0 \\ 
    0 & 0 & 1 & 0 & 0 & 0 & 0 & 0 & 0 & 0 & 0 & 0 \\ 
    0 & 0 & 0 & 1 & 0 & 0 & 0 & 0 & 0 & 0 & 0 & 0 \\ 
    \hline
    0 & 0 & 0 & 0 & 1 & 0 & 0 & 0 & 0 & 0 & 0 & 0 \\ 
    0 & 0 & 0 & 0 & 0 & 1 & 0 & 0 & 0 & 0 & 0 & 0 \\ 
    0 & 0 & 0 & 0 & 0 & 0 & 1 & 0 & 0 & 0 & 0 & 0 \\ 
    0 & 0 & 0 & 0 & 0 & 0 & 0 & 1 & 0 & 0 & 0 & 0 \\ 
    \hline
    0 & 0 & 0 & 0 & 0 & 0 & 0 & 0 & 1 & 0 & 0 & 0 \\ 
    0 & 0 & 0 & 0 & 0 & 0 & 0 & 0 & 0 & 1 & 0 & 0 \\ 
    0 & 0 & 0 & 0 & 0 & 0 & 0 & 0 & 0 & 0 & 1 & 0 \\ 
    0 & 0 & 0 & 0 & 0 & 0 & 0 & 0 & 0 & 0 & 0 & 1
\numberthis\end{array}\right] \]

\[
[{\rm R_2}]_{\hat i}{}^{k} = \left[\begin{array}{cccc|cccc|cccc}
    0 & -1 & 0 & 0 & 0 & 0 & 0 & 0 & 0 & 0 & 0 & 0 \\ 
    1 & 0 & 0 & 0 & 0 & 0 & 0 & 0 & 0 & 0 & 0 & 0 \\  
    0 & 0 & 0 & -1 & 0 & 0 & 0 & 0 & 0 & 0 & 0 & 0 \\ 
    0 & 0 & 1 & 0 & 0 & 0 & 0 & 0 & 0 & 0 & 0 & 0 \\  
    \hline
    0 & 0 & 0 & 0 & 0 & -1 & 0 & 0 & 0 & 0 & 0 & 0 \\ 
    0 & 0 & 0 & 0 & 1 & 0 & 0 & 0 & 0 & 0 & 0 & 0 \\  
    0 & 0 & 0 & 0 & 0 & 0 & 0 & 1 & 0 & 0 & 0 & 0 \\  
    0 & 0 & 0 & 0 & 0 & 0 & -1 & 0 & 0 & 0 & 0 & 0 \\ 
    \hline
    0 & 0 & 0 & 0 & 0 & 0 & 0 & 0 & 0 & 1 & 0 & 0 \\  
    0 & 0 & 0 & 0 & 0 & 0 & 0 & 0 & -1 & 0 & 0 & 0 \\ 
    0 & 0 & 0 & 0 & 0 & 0 & 0 & 0 & 0 & 0 & 0 & -1 \\ 
    0 & 0 & 0 & 0 & 0 & 0 & 0 & 0 & 0 & 0 & 1 & 0
\numberthis\end{array}\right] \]

\[
[{\rm R_3}]_{\hat i}{}^{k} = \left[\begin{array}{cccc|cccc|cccc}
    0 & 0 & -1 & 0 & 0 & 0 & 0 & 0 & 0 & 0 & 0 & 0 \\ 
    0 & 0 & 0 & 1 & 0 & 0 & 0 & 0 & 0 & 0 & 0 & 0 \\  
    1 & 0 & 0 & 0 & 0 & 0 & 0 & 0 & 0 & 0 & 0 & 0 \\  
    0 & -1 & 0 & 0 & 0 & 0 & 0 & 0 & 0 & 0 & 0 & 0 \\ 
    \hline
    0 & 0 & 0 & 0 & 0 & 0 & 1 & 0 & 0 & 0 & 0 & 0 \\  
    0 & 0 & 0 & 0 & 0 & 0 & 0 & 1 & 0 & 0 & 0 & 0 \\  
    0 & 0 & 0 & 0 & -1 & 0 & 0 & 0 & 0 & 0 & 0 & 0 \\ 
    0 & 0 & 0 & 0 & 0 & -1 & 0 & 0 & 0 & 0 & 0 & 0 \\ 
    \hline
    0 & 0 & 0 & 0 & 0 & 0 & 0 & 0 & 0 & 0 & -1 & 0 \\ 
    0 & 0 & 0 & 0 & 0 & 0 & 0 & 0 & 0 & 0 & 0 & -1 \\ 
    0 & 0 & 0 & 0 & 0 & 0 & 0 & 0 & 1 & 0 & 0 & 0 \\  
    0 & 0 & 0 & 0 & 0 & 0 & 0 & 0 & 0 & 1 & 0 & 0
\numberthis\end{array}\right] \]

\[
[{\rm R_4}]_{\hat i}{}^{k} = \left[\begin{array}{cccc|cccc|cccc}
    0 & 0 & 0 & -1 & 0 & 0 & 0 & 0 & 0 & 0 & 0 & 0 \\ 
    0 & 0 & -1 & 0 & 0 & 0 & 0 & 0 & 0 & 0 & 0 & 0 \\ 
    0 & 1 & 0 & 0 & 0 & 0 & 0 & 0 & 0 & 0 & 0 & 0 \\  
    1 & 0 & 0 & 0 & 0 & 0 & 0 & 0 & 0 & 0 & 0 & 0 \\  
    \hline
    0 & 0 & 0 & 0 & 0 & 0 & 0 & 1 & 0 & 0 & 0 & 0 \\  
    0 & 0 & 0 & 0 & 0 & 0 & -1 & 0 & 0 & 0 & 0 & 0 \\ 
    0 & 0 & 0 & 0 & 0 & 1 & 0 & 0 & 0 & 0 & 0 & 0 \\  
    0 & 0 & 0 & 0 & -1 & 0 & 0 & 0 & 0 & 0 & 0 & 0 \\ 
    \hline
    0 & 0 & 0 & 0 & 0 & 0 & 0 & 0 & 0 & 0 & 0 & 1 \\  
    0 & 0 & 0 & 0 & 0 & 0 & 0 & 0 & 0 & 0 & -1 & 0 \\ 
    0 & 0 & 0 & 0 & 0 & 0 & 0 & 0 & 0 & 1 & 0 & 0 \\  
    0 & 0 & 0 & 0 & 0 & 0 & 0 & 0 & -1 & 0 & 0 & 0
\numberthis\end{array}\right] \]

\newpage


\end{document}